\documentclass[doublecol]{epl2}
\usepackage{graphicx}
\usepackage{amsmath,amssymb}
\usepackage{here}
\usepackage{bm}
\usepackage{color}
\usepackage{framed}
\usepackage{ulem}
\usepackage{txfonts}

\def\agt{>\kern -9.2truept\lower 4.3truept\hbox{$\displaystyle\sim$}}
\def\alt{<\kern -9.2truept\lower 4.3truept\hbox{$\displaystyle\sim$}}
\title{Breakdown of the conventional spin-wave dynamics and its double-constraint modification
       in the spin-$\mathbf{\frac{1}{2}}$ triangular-prism Heisenberg antiferromagnet}
\shorttitle{Modified spin-wave dynamics of triangular-prism antiferromagnet}
\author{Shoji Yamamoto${}^{*}$ and Jun Ohara}
\shortauthor{S. Yamamoto and J. Ohara}
\institute{Department of Physics, Hokkaido University, Sapporo 060-0810, Japan}
\abstract
{Spontaneous magnon decays in an $S=\frac{1}{2}$ Heisenberg antiferromagnet on the equilateral
triangular prism are investigated in terms of modified magnon Green's functions.
In one dimension, the so-called infrared divergence prevents us from calculating
any---whether static or dynamic---structure factor within the conventional spin-wave theory
even at zero temperature.
The well-known modified spin-wave theory initiated by Takahashi completely fails to treat
anharmonicities to cause transverse-to-longitudinal coupling which are quite characteristic of
noncollinear antiferromagnets.
We propose imposing a double-constraint condition on spin waves to solve all these difficulties
and get a full view of the nonlinear spin-wave dynamics in one-dimensional frustrated noncollinear
antiferromagnets.
We reveal a novel instability of the single-particle spectrum in the absence of any well-defined
magnetically ordered ground state.}

\begin{document}

\maketitle

   There are such surprises on the way from one- to two-dimensional quantum magnets
\cite{R9235,G8901,A6233,D618,D964,H1607}
that the $S=\frac{1}{2}$ antiferromagnetic Heisenberg model on ladders exhibits an excitation
gap immediately above the ground state when consisting of an even number of legs
whereas
that reproduces the quantum critical Tomonaga-Luttinger liquid behavior \cite{S533}
of a single chain \cite{H1358,H4955} when consisting of an odd number of legs.
We can verify this theoretical scenario in various ladder materials such as
two-legged $\mathrm{LaCuO}_{2.5}$ \cite{H41},
two-legged $\mathrm{CaV}_2\mathrm{O}_5$ \cite{I2397}, and
$\frac{1}{2}(n+1)$-legged Sr$_{n-1}$Cu$_{n+1}$O$_{2n}$ ($n=3,5,\cdots$) \cite{A3463,K2812}.
There are more interesting approaches to form these ladder materials into a cylindrical shape.
Examples of such practices include
three-legged
$[(\mathrm{CuCl}_2\mathrm{tachH})_3\mathrm{Cl}]\mathrm{Cl}_2$
($\mathrm{tach}
 =\mathit{cis},\mathit{trans}\mbox{-}1,3,5\mbox{-}\mathrm{triamino}\mbox{-}\mathrm{cyclohexane}$)
\cite{S174420,S1580} and
$\mathrm{CsCrF}_4$ \cite{M093701,M084714},
four-legged
$\mathrm{Cu}_2\mathrm{Cl}_4\mbox{-}\mathrm{D}_8\mathrm{C}_4\mathrm{SO}_2$ \cite{G037206,G060404},
and nine-legged
$\mathrm{Na}_2\mathrm{V}_3\mathrm{O}_7$ \cite{M676,L060405,G064431}.

   Such chemical explorations motivated numerical investigations of the spin-$\frac{1}{2}$
nearest-neighbor antiferromagnetic Heisenberg model on a triangular prism
\cite{K4001,N054421,S184415,S403201}.
Surprisingly, its ground state is no longer critical, breaking a translational symmetry
in the leg direction \cite{K4001,N054421}.
$S=1$ \cite{C075108} and $S=\frac{3}{2}$ \cite{N224425} analogs also exhibit an excitation gap
immediately above their ground states with spontaneous dimerization.
With decreasing coupling between triangles, however, the $S=1$ ground state turns unique but
the $S=\frac{3}{2}$ one stays degenerate, though both remain gapped from their excitation
spectra.
Despite such theoretical findings, experimental observations of the excitation gap to
the lowest-lying triplet state are limited to more-than-three-legged tubes
\cite{G037206,G064431}.
$[(\mathrm{CuCl}_2\mathrm{tachH})_3\mathrm{Cl}]\mathrm{Cl}_2$ of $S=\frac{1}{2}$
is not a triangular prism \cite{S174420,S1580} but an alternating stack of triangles
\cite{O297,O1423,F014409}
to behave as an effective spin-$\frac{3}{2}$ antiferromagnetic Heisenberg chain
\cite{I037206,F014409}.
$\mathrm{CsCrF}_4$ of $S=\frac{3}{2}$ seems a three-legged straight tube \cite{M093701}
but substantial magnetic anisotropies \cite{K184405}
hinder our detecting so small a gap of intrinsic origin \cite{N224425}.

   A three-leg spin tube can give rise to a rich phase diagram \cite{C075108} and therefore
varied low-lying excitations, even if we limit our focus to an equilateral triangular prism
consisting of Heisenberg-exchange-coupled identical spins.
We are encouraged to calculate the dynamic spin structure factor, which can be accessed
directly by inelastic neutron scattering, for such a fascinating system, but it is far from
straightforward to do so in the thermodynamic limit.
Density-matrix renormalization group (DMRG) \cite{N054421,D205119} and
tensor network \cite{C1545}
are really powerful tools, but a direct access to the thermodynamic limit is
still hard even with them.
\textit{Magnetic frustration} generally demands an increasing number $M$ of states kept
in matrix renormalization \cite{N054421,M445} and several excited as well as the ground states
must be included as targets in \textit{dynamical} density-matrix renormalization at each frequency
$\omega$ \cite{N054421,J045114}, both causing an increase in computational cost.
We may consider a spin-wave (SW) theory in this context, but its naivest application to
one-dimensional $\mathrm{SU}(2)$ Heisenberg antiferromagnets ends in a fiasco.
In one dimension, the conventional SW (CSW) theory \cite{K568} has no way of
describing either static or dynamic structure factor even in the ground state, whether it is
collinear or noncollinear, unless hardening of soft antiferromagnons.
The staggered magnetization diverges logarithmically, to say nothing of the antiferromagnetic
structure factor as its fluctuation \cite{S075123},
in a Heisenberg antiferromagnet on the triangular prism
lattice within the CSW theory.
We can overcome this difficulty by modifying CSWs so as to keep the total staggered magnetization
zero \cite{T1524,T2494,H4769,T5000,Y064426}.
Hartree-Fock (HF) approximation of such modified SW (MSW) Hamiltonian truncated at the order of
$S^0$ yields an excellent low-temperature thermodynamics for collinear antiferromagnets
\cite{T1524,T2494}, which turns out essentially the same as a Schwinger-boson mean-field
description \cite{A617}, but it still ends in failure when applied to frustrated
noncollinear antiferromagnets.
Since the total uniform magnetization is no longer commutable with any SW Hamiltonian,
whether modified or not, MSWs conditioned as the above awfully overestimate the otherwise
moderate intratriangular ferromagnetic spin correlation.
Then we devise a new MSW scheme \cite{Y065004} capable of suppressing
such a ferromagnetic divergence and featuring cubic as well as quartic anharmonicities
peculiar to frustrated noncollinear antiferromagnets \cite{Y065004,C144416}.

   A SW language is characterized by its ``$1/S$ expansion" and is therefore suited for
visualizing quantum renormalization of the harmonic spectrum on the order of $S$.
$O(S^0)$ quantum corrections to harmonic antiferromagnons on the square and triangular lattices
contribute toward upward and downward renormalizations of their energies, respectively.
In the triangular-lattice antiferromagnet, this first-order renormalization causes such drastic
effects on SW excitations that their lifetimes no longer remain infinite in a major part of
the hexagonal Brillouin zone \cite{C144416,S180403,C207202} and rotonlike minima develop at
some points outside the decay region \cite{Z224420,M094407}.
In the square-lattice antiferromagnet, on the other hand, the lowest-order spectrum renormalization
brings about neither a finite lifetime nor a rotonlike minimum \cite{S216003,Z184440}.
Since the classical ground state of a triangular-prism antiferromagnet is a collinear
($180^\circ$-N\'eel) stack of coplanar ($120^\circ$-N\'eel) states,
quantum renormalization of its harmonic SW spectrum may have characteristics of both
one-dimensional collinear antiferromagnets \cite{H1358,H4955} in a Luttinger liquid state
\cite{S533} and two-dimensional coplanar antiferromagnets in an ordered state
\cite{H2531,J2727,S1766,B2590,A2483,B10048,C3899,W127004}.
We shall evaluate various MSW schemes in comparison with DMRG calculations, intending to find out
the best MSW language that has the potential to reveal such a subtle spectrum.
\begin{figure}
\centering
\includegraphics[width=85mm]{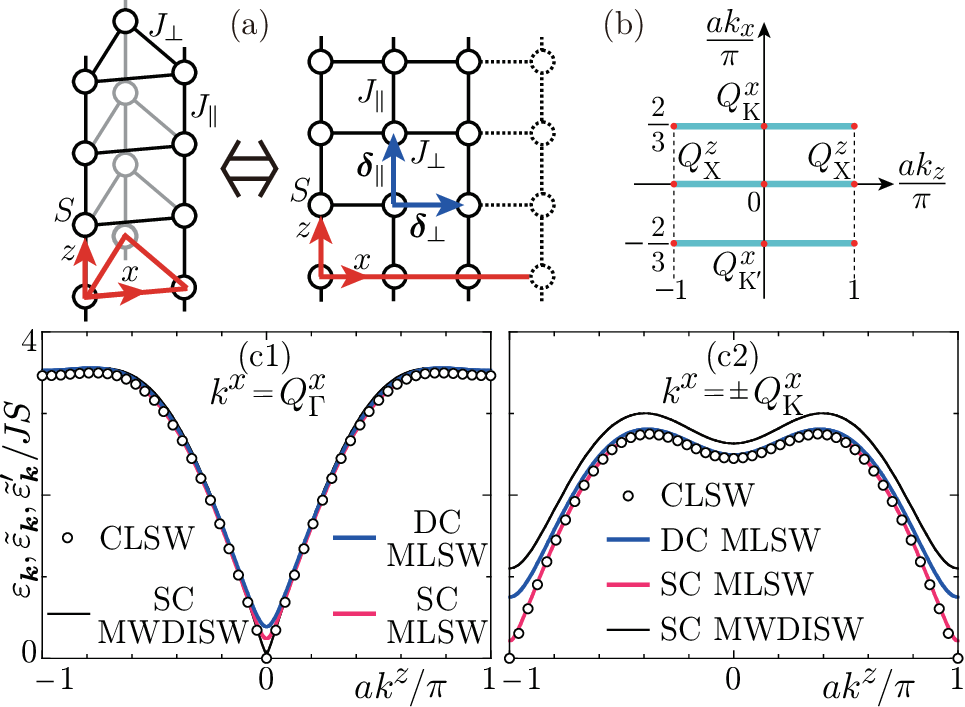}
\vspace*{-4mm}
\caption{An equilateral triangular prism may be identified with a planar three-leg ladder with
         the periodic boundary condition in the rung direction (a)
         whose first Brillouin zone consists of three parallel line segments
         at equal intervals in two dimensions (b).
         CLSW $\varepsilon_{\bm{k}}$ (\ref{E:ekCLSW}),
         SC-MLSW $\tilde{\varepsilon}_{\bm{k}}^{\mathrm{SC}}$ (\ref{E:ekSC-MLSW}),
         SC-MWDISW $\tilde{\varepsilon}'{}_{\!\!\bm{k}}^{\mathrm{SC}}$ (S24) \cite{Suppl_MSW},
         and
         DC-MLSW $\tilde{\varepsilon}_{\bm{k}}^{\mathrm{DC}}$ (\ref{E:ekDC-MLSW})
         dispersions as functions of $k^z$
         with $k^x=Q_\Gamma^x$ (c1) and $k^x=\pm Q_{\mathrm{K}}^x$ (c2)
         in the $N\rightarrow\infty$ limit.}
\label{F:Model&Disp}
\vspace*{-4mm}
\end{figure}

   We set the spin-$S$ nearest-neighbor antiferromagnetic Heisenberg model
$
   \mathcal{H}
  =\sum_{l=1}^L
   \sum_{\sigma=\parallel,\perp}
   J_\sigma
   \bm{S}_{\bm{r}_l}\cdot\bm{S}_{\bm{r}_l+\bm{\delta}_{\sigma}}
$
on a equilateral triangular prism of $L\equiv 3N$ sites
[cf. Figs. \ref{F:Model&Disp}(a) and \ref{F:Model&Disp}(b)].
Henceforth we set $J_\parallel$ and $J_\perp$ both equal to $J\,(>0)$ and accordingly suppose
$|\bm{\delta}_\parallel|=|\bm{\delta}_\perp|=a$.
In order to express the spin Hamiltonian in terms of Holstein-Primakoff bosons,
we first transform it into the rotating frame with its ``moving" $z$ axis
pointing along each local spin direction in the classical ground state designated by
the ordering vector
$(q^z,q^x,q^y)=\frac{\pi}{a}\left(1,\frac{2}{3},0\right)$.
With a tacit understanding of $q^y$ being $0$,
we henceforth use a short-hand notation in the two-dimensional reciprocal space,
\vspace*{-1.4mm}
\begin{align}
   \bm{Q}_{\Gamma,\Gamma}\!
  =\!(Q_\Gamma^z,Q_\Gamma^x)\!
  \equiv\!
   \frac{\pi}{a}
   \left(
    0,0
   \right),\ 
   \bm{Q}_{\mathrm{X},\mathrm{K}}\!
  =\!(Q_{\mathrm{X}}^z,Q_{\mathrm{K}}^x)\!
  \equiv\!
   \frac{\pi}{a}
   \left(
    1,\frac{2}{3}
   \right).\!\!
   \label{E:Vin2DBZ}
   \\[-7.5mm]\nonumber
\end{align}
We denote the local coordinate system by $(\tilde{z},\tilde{x},\tilde{y})$ distinguishably from
the laboratory frame $(z,x,y)$.
Note that every local $\tilde{z}$-$\tilde{x}$ plane coincides with the global $z$-$x$ ``plane",
which consists of the side faces of the triangular prism in practice.
For coplanar antiferromagnets with their spins lying in the thus-made $z$-$x$ plane,
the spin components in the laboratory and rotating frames are related with each other as
\vspace*{-1.4mm}
\begin{align}
   \left[
    \begin{array}{c}
     S_{\bm{r}_l}^z \\
     S_{\bm{r}_l}^x \\
     S_{\bm{r}_l}^y \\
    \end{array}
   \right]
  =\left[
    \begin{array}{ccc}
     \cos\phi_{\bm{r}_l} &-\sin\phi_{\bm{r}_l} & 0 \\
     \sin\phi_{\bm{r}_l} & \cos\phi_{\bm{r}_l} & 0 \\
               0           &           0           & 1 \\
    \end{array}
   \right]
   \left[
    \begin{array}{c}
     S_{\bm{r}_l}^{\tilde{z}} \\
     S_{\bm{r}_l}^{\tilde{x}} \\
     S_{\bm{r}_l}^{\tilde{y}} \\
    \end{array}
   \right],
   \label{E:Trans}
   \\[-7.5mm]\nonumber
\end{align}
where
$\phi_{\bm{r}_l}
\equiv
 \bm{Q}_{\mathrm{X},\mathrm{K}}\cdot\bm{r}_l$
is the angle formed by the axes $z$ ($x$) and $\tilde{z}$ ($\tilde{x}$) at $\bm{r}_l$.
Then we introduce Holstein-Primakoff bosons,
\vspace*{-1.4mm}
\begin{align}
   &
   S_{\bm{r}_l}^{\tilde{z}}
  =S-a_{\bm{r}_l}^\dagger a_{\bm{r}_l},\ 
   S_{\bm{r}_l}^{\tilde{x}}-iS_{\bm{r}_l}^{\tilde{y}}
   \equiv S_{\bm{r}_l}^{\tilde{+}\dagger}
   \equiv S_{\bm{r}_l}^{\tilde{-}}
  =\sqrt{2S}a_{\bm{r}_l}^\dagger\mathcal{R}_{\bm{r}_l}(S);
   \nonumber
   \allowdisplaybreaks \\
   &
   \mathcal{R}_{\bm{r}_l}(S)
  \equiv
   \sqrt{1-\frac{a_{\bm{r}_l}^\dagger a_{\bm{r}_l}}{2S}}
  =1
  -\sum_{l=1}^\infty
   \frac{(2l-3)!!}{l!}
   \left(
    \frac{a_{\bm{r}_l}^\dagger a_{\bm{r}_l}}{4S}
   \right)^l,
   \label{E:HPT}
   \\[-7.0mm]\nonumber
\end{align}
to expand the Hamiltonian as
$
   \mathcal{H}
  =\sum_{m=0}^4
   \mathcal{H}^{\left(\frac{m}{2}\right)}
  +O\left(S^{-\frac{1}{2}}\right),
$
where
$\mathcal{H}^{\left(\frac{m}{2}\right)}$, on the order of $S^{\frac{m}{2}}$, consist of $4-m$
boson operators \cite{Suppl_MSW} and those on the order of $S$ to a fractional power are
peculiar to noncollinear antiferromagnets.
$\mathcal{H}^{\left(\frac{3}{2}\right)}$ automatically vanishes because
the Holstein-Primakoff boson vacuum corresponds to a minimum of the classical energy
\cite{C144416}.

   We start from the naivest SW description of the triangular-prism antiferromagnet.
We move to the reciprocal space,
$
   a_{\bm{k}_{\nu,\sigma}}^\dagger
  =\sum_{l=1}^L
   e^{i\bm{k}_{\nu,\sigma}\cdot\bm{r}_l}
   a_{\bm{r}_l}^\dagger
   /\sqrt{L}
$,
and define the Bogoliubov bosons
$
   \alpha_{\bm{k}_{\nu,\sigma}}^\dagger
  =u_{\bm{k}_{\nu,\sigma}}a_{ \bm{k}_{\nu,\sigma}}^\dagger
  -v_{\bm{k}_{\nu,\sigma}}a_{-\bm{k}_{\nu,\sigma}}
$
so as to make the linear SW (LSW) Hamiltonian diagonal,
\vspace*{-1.4mm}
\begin{align}
   \sum_{m=1}^2
   \mathcal{H}^{(m)}
  \equiv
   \mathcal{H}_{\mathrm{harm}}
  =\sum_{l=1}^2
   E^{(m)}
  +\sum_{\sigma=0,\pm}\sum_{\nu=1}^N
   \varepsilon_{\bm{k}_{\nu,\sigma}}
   \alpha_{\bm{k}_{\nu,\sigma}}^\dagger
   \alpha_{\bm{k}_{\nu,\sigma}},
   \label{E:H(CLSW)}
   \\[-7.0mm]\nonumber
\end{align}
where
$
   E^{(2)}
  \equiv
  -\frac{3}{2}LJS^2
$
is the classical ground-state energy and
$
   E^{(1)}
  \equiv
  -\frac{3}{2}LJS
  +\sum_{\sigma=0,\pm}\sum_{\nu=1}^N
   \frac{\varepsilon_{\bm{k}_{\nu,\sigma}}}{2}
$
is its $O(S^1)$ quantum correction,
while $\alpha_{\bm{k}_{\nu,\sigma}}^\dagger$ creates a SW with wavevector
$
   \bm{k}_{\nu,\sigma}
  \equiv
   (k_\nu^z,k_\sigma^x)
  =\frac{\pi}{a}
   \left(
    \frac{2\nu}{N}-1,\frac{2}{3}\sigma
   \right)
$
and energy
\vspace*{-1.5mm}
\begin{align}
   &
   \varepsilon_{\left(k_\nu^z,k_0^x  \right)}
  =JS
   \sqrt{2(1-\cos{ak_\nu^z})(5+2\cos{ak_\nu^z})},
   \nonumber
   \allowdisplaybreaks \\
   &
   \varepsilon_{\left(k_\nu^z,k_\pm^x\right)}
  =JS
   \sqrt{2(1+\cos{ak_\nu^z})\left(\frac{7}{2}-2\cos{ak_\nu^z}\right)}.
   \label{E:ekCLSW}
   \\[-7.5mm]\nonumber
\end{align}
The complete breaking of the global $\mathrm{SO}(3)$ rotational symmetry results in three
Nambu-Goldstone modes \cite{C144416,B195124} of linear dispersion
[Figs. \ref{F:Model&Disp}(c1) and \ref{F:Model&Disp}(c2)],
one at the zone center $\bm{Q}_{\Gamma,\Gamma}$ with
velocity $v_{\bm{Q}_{\Gamma,\Gamma}}=\sqrt{7}JSa$
and two at the zone corners $\pm \bm{Q}_{\mathrm{X},\mathrm{K}}$
with velocity $v_{\bm{Q}_{\mathrm{X},\mathrm{K}}}=\sqrt{\frac{11}{2}}JSa$.
On the equilateral triangular prism \cite{K4001,N054421,N224425,C075108}, however,
these acoustic antiferromagnons are fictitious characters and make structure factors diverge
\cite{Suppl_MSW}.
\begin{figure*}[t]
\centering
\includegraphics[width=175mm]{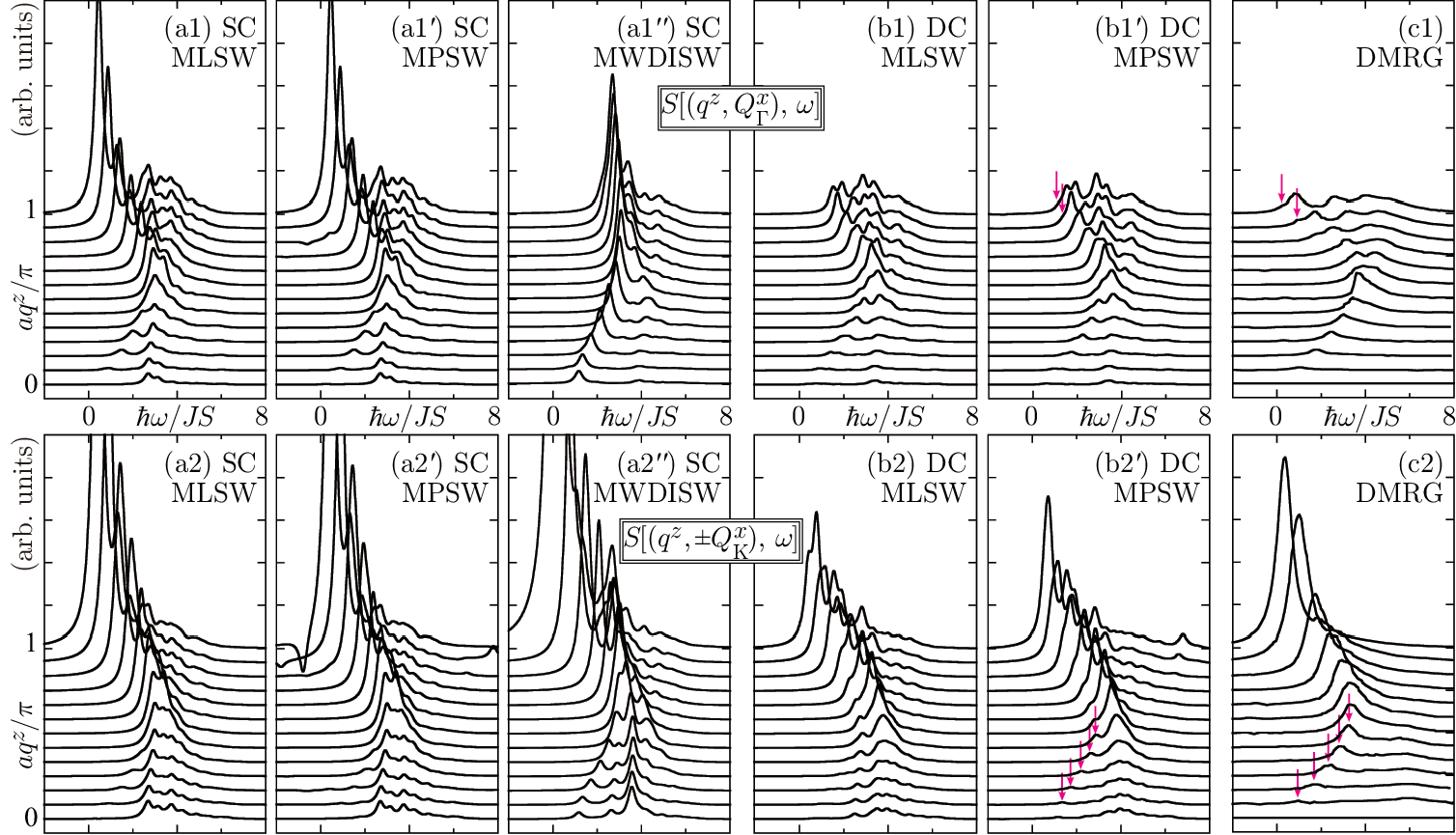}
\vspace*{-4mm}
\caption{Various MSW [($\mathrm{a}1$) to ($\mathrm{b}1'$) ($q^x=Q_\Gamma^x$) and
                      ($\mathrm{a}2$) to ($\mathrm{b}2'$) ($q^x=\pm Q_{\mathrm{K}}^x$)]
         versus DMRG [($\mathrm{c}1$) and ($\mathrm{c}2$)] \cite{N054421} calculations of
         the total dynamic structure factor
         $S(\bm{q},\omega)
         \equiv
          \sum_{\lambda=z,x,y}S^{\lambda\lambda}(\bm{q},\omega)$
         \{(S44), (S50), and (S51) \cite{Suppl_GF}\}
         for $S=\frac{1}{2}$ at $N=24$.
         Arrows to
         $S[(q^z,Q_\Gamma^x),\omega]$ at
         $\frac{Q_{\mathrm{X}}^z}{2} \ll |q^z|\alt Q_{\mathrm{X}}^z$
         [($\mathrm{b}1'$) and (c1)] and
         $S[(q^z,\pm Q_{\mathrm{K}}^x),\omega]$ at
         $Q_\Gamma^z \alt |q^z| \alt \frac{Q_{\mathrm{X}}^z}{2}$
         [($\mathrm{b}2'$) and (c2)]
         indicate particular spectral features describable only with pertinent single-magnon
         energies,
         $\tilde{\varepsilon}_{\bm{k}}^{\mathrm{DC}}
          \displaystyle
         +\mathop{\varSigma}^{\,\protect\substack{-\!\!\!-\protect\\[-1.6mm]}}\!{}_1^{(0)}(\bm{k})
         +\mathop{\varSigma}^{\,\protect\substack{-\!\!\!-\protect\\[-1.6mm]}}\!{}_2^{(0)}
          (\bm{k};\tilde{\varepsilon}_{\bm{k}}^{\mathrm{DC}}/\hbar)$
         \{(\ref{E:ekDC-MLSW}) and (S42) \cite{Suppl_GF}\}
         especially at $\bm{k}\simeq\pm\bm{Q}_{\Gamma,\mathrm{K}}$.}
\label{F:Sqw}
\vspace*{-4mm}
\end{figure*}

   Then we consider modifying CSW Hamiltonians.
The naivest MSW scheme \cite{T1524,T2494,H4769,T5000} consists of diagonalizing the conventional
LSW (CLSW) Hamiltonian (\ref{E:H(CLSW)}) subject to the single-constraint (SC) condition
$
   S
  -\langle
    a_{\bm{r}_l}^\dagger
    a_{\bm{r}_l}
   \rangle_T^{\mathrm{SC}}
  =0
$,
\vspace*{-1.5mm}
\begin{align}
   &
   \mathcal{H}_{\mathrm{harm}}
  +\sum_{l=1}^L
   \mu_l
   \left(
    S-a_{\bm{r}_l}^\dagger a_{\bm{r}_l}
   \right)
  \equiv
   \widetilde{\mathcal{H}}_{\mathrm{harm}}^{\mathrm{SC}}
   \nonumber
   \allowdisplaybreaks \\[-1mm]
   &\qquad
  =\sum_{m=1}^2
   E^{(m)}
  +\mu L
   \left(
    S+\frac{1}{2}
   \right)
  +\sum_{\sigma=0,\pm}\sum_{\nu=1}^N
   \tilde{\varepsilon}_{\bm{k}_{\nu,\sigma}}^{\rm{SC}}
   \alpha_{\bm{k}_{\nu,\sigma}}^\dagger
   \alpha_{\bm{k}_{\nu,\sigma}},
   \label{E:H(SC-MLSW)}
   \\[-7.5mm]\nonumber
\end{align}
where $\langle\cdots\rangle_T^{\mathrm{SC}}$ denotes a thermal average at temperature $T$
with respect to $\widetilde{\mathcal{H}}_{\mathrm{harm}}^{\mathrm{SC}}$,
the $L$ Lagrange multipliers $\mu_l$ degenerate into $\mu$, and then
$\alpha_{\bm{k}_{\nu,\sigma}}^\dagger$ creates a modified LSW (MLSW) with wavevector
$\bm{k}_{\nu,\sigma}$ and energy
\vspace*{-1.5mm}
\begin{align}
   &
   \frac{\tilde{\varepsilon}_{\left(k_\nu^z,k_0^x  \right)}^{\mathrm{SC}}}{JS}
  =\!\!\sqrt{2\left(1\!-\! \cos{ak_\nu^z}\!-\!\frac{\mu}{2JS}\right)\!
              \left(5\!+\!2\cos{ak_\nu^z}\!-\!\frac{\mu}{ JS}\right)},
   \nonumber \\
   &
   \frac{\tilde{\varepsilon}_{\left(k_\nu^z,k_\pm^x\right)}^{\mathrm{SC}}}{JS}\!
  =\!\!\sqrt{2\left(1\!+\! \cos{ak_\nu^z}\!-\!\frac{\mu}{2JS}\right)
              \left(\frac{7}{2}\!-\!2\cos{ak_\nu^z}\!-\!\frac{\mu}{ JS}\right)}.
   \label{E:ekSC-MLSW}
   \\[-7.5mm]\nonumber
\end{align}
Since its extension up to $O(S^0)$ within the HF approximation \cite{T1524,T2494},
which we call the SC-modified HF-decomposition-based-interacting-SW (SC-MHFISW) scheme,
yields a good thermodynamics of one-dimensional Haldane-gap antiferromagnets
\cite{R2589,W12805,Y769,Y822}, we try this too on the present system \cite{Suppl_MSW}.
Here we perform all calculations at $T=0$, and then, MHFISWs are no different from modified
Wick-decomposition-based-interacting SWs (MWDISWs) \cite{N034714,Y094412}.
Such calculated MSWs no longer have any soft mode
[see Figs. \ref{F:Model&Disp}(c1) and \ref{F:Model&Disp}(c2)]
to enable us to calculate the dynamic spin structure factor (S44) \cite{Suppl_GF}.

   We lead the argument off with the SC-MLSW findings.
A major portion of the spectral weight concentrates at $\bm{q}=\bm{Q}_{\mathrm{X},\mathrm{K}}$
[Figs. \ref{F:Sqw}(a2) and \ref{F:Sqw}(c2)], claiming that
while the coplanar $120^\circ$-N\'eel structure predominates in each cross-sectional triangle,
the collinear $180^\circ$-N\'eel structure remains significant in the leg direction.
Even so, SC MLSWs overestimate this single-magnon scattering intensity
$S(\bm{Q}_{\mathrm{X},\mathrm{K}},\omega)$ (S62) \cite{Suppl_GF}
with such a small excitation energy that
$\tilde{\varepsilon}_{\bm{Q}_{\Gamma,\Gamma}}^{\mathrm{SC}}\ll JS$.
Another serious fault of SC MLSWs is a spurious intensity at
$\bm{q}=\bm{Q}_{\mathrm{X},\Gamma}$ [Fig. \ref{F:Sqw}(a1)],
claiming a one-dimensional antiferromagnetic configuration of spins coupled ferromagnetically
within each triangle.
SC MLSWs overestimate this two-magnon scattering intensity
$S(\bm{Q}_{\mathrm{X},\Gamma},\omega)$ (S63) \cite{Suppl_GF} as well
with a still smaller excitation energy
$\tilde{\varepsilon}_{\pm\bm{Q}_{\mathrm{X},\mathrm{K}}}^{\mathrm{SC}}\ll JS$.
Indeed SC MLSWs at $\bm{Q}_{\Gamma,\Gamma}$ and $\pm\bm{Q}_{\mathrm{X},\mathrm{K}}$ are both
gapped, but their gaps seem to be still too small to reproduce the correct spectral weighting.
DMRG calculations show that the $S(\bm{q},\omega)$ spectral weighting is well dispersive
with $q^z$ moving from $Q_\Gamma^z$ to $Q_{\mathrm{X}}^z$ at every $q^x$
[Figs. \ref{F:Sqw}(c1) and \ref{F:Sqw}(c2)], but SC MLSWs can hardly reproduce such features
especially along the lines connecting
$(Q_\Gamma^z,\pm Q_{\mathrm{K}}^x)$ and $(\frac{1}{2}Q_{\mathrm{X}}^z,\pm Q_{\mathrm{K}}^x)$.
So how about the SC-MWDISW findings?
The increased gaps at $\pm\bm{Q}_{\mathrm{X},\mathrm{K}}$ suppress
the too strong SC-MLSW intensity
$S[\bm{Q}_{\mathrm{X},\Gamma},
   (\tilde{\varepsilon}_{ \bm{Q}_{\mathrm{X},\mathrm{K}}}^{\mathrm{SC}}
   +\tilde{\varepsilon}_{-\bm{Q}_{\mathrm{X},\mathrm{K}}}^{\mathrm{SC}})/\hbar]$
[Fig. \ref{F:Sqw}($\mathrm{a}1''$)],
whereas the diminished gap at $\bm{Q}_{\Gamma,\Gamma}$ even more enhances
the almost divergent SC-MLSW intensity
$S(\bm{Q}_{\mathrm{X},\mathrm{K}},
   \tilde{\varepsilon}_{\bm{Q}_{\Gamma,\Gamma}}^{\mathrm{SC}}/\hbar)$
[Fig. \ref{F:Sqw}($\mathrm{a}2''$)].
In any case, the SC-MWDISW findings are still far from the DMRG calculations or may even be
a change for the worse.

   In order to make cubic as well as quartic anharmonicities effective in sophisticating
harmonic SWs, we have to treat them perturbatively rather than variationally.
We correct the SC-MLSW Hamiltonian to first order in $1/S$ \cite{V134429,C144416}, i.e.,
to first order in $\mathcal{H}^{(0)}$ and
to second order in $\mathcal{H}^{\left(\frac{1}{2}\right)}$.
We employ magnon Green's functions \cite{C3264,M094407} to renormalize the spectrum and
accordingly correct the dynamic structure factor \cite{Suppl_GF}.
Since the longitudinal component of the structure factor, which probes the two-magnon continuum,
is a factor of $1/S$ smaller than the leading terms in the transverse correlation functions,
it may be calculated in terms of bare magnon Green's functions.
The effect of interactions between antiferromagnons is taken into its transverse components
through the lowest-order self energies $\varSigma^{(0)}(\bm{k};\omega+i\epsilon)$ \cite{Y065004,C144416}
consisting of the frequency-independent contribution $\varSigma_1^{(0)}(\bm{k})$ of first order
in $\mathcal{H}^{(0)}$ and the frequency-dependent contribution
$\varSigma_2^{(0)}(\bm{k};\omega+i\epsilon)$ of second order
in $\mathcal{H}^{\left(\frac{1}{2}\right)}$,
both on the order of $S^0$ \cite{Suppl_GF}.
Figures \ref{F:Sqw}($\mathrm{a}1'$) and \ref{F:Sqw}($\mathrm{a}2'$) show such corrected
calculations to (S44) in terms of modified perturbed SWs (MPSWs).
Even though any two-magnon correlation remains the same, yet some sort of change may be coming to
the one-magnon contributions.
However, the MLSW and MPSW findings look almost the same at least within the SC modification.
This is because the real parts of the two primary self-energies,
$\displaystyle
 \mathop{\varSigma}^{\,\protect\substack{-\!\!\!-\\[-1.6mm]}}\!{}_1^{(0)}(\bm{k})$ and
$\displaystyle
 \mathop{\varSigma}^{\,\protect\substack{-\!\!\!-\\[-1.6mm]}}\!{}_2^{(0)}
 (\bm{k};\tilde{\varepsilon}_{\bm{k}}^{\mathrm{SC}}/\hbar)$,
balance each other at $|k^z|\alt \frac{Q_{\mathrm{X}}^z}{2}$ \cite{Suppl_DCvsSC}.
Considering that neither SC MWDISWs nor SC MPSWs can hardly reproduce the DMRG calculations,
we are now in a position to reconsider the way of modifying CSWs.

   Then we should figure out two serious problems of SC MSWs in noncollinear antiferromagnets
\cite{Y065004,Suppl_MSW}.
First, they break the $\mathrm{U}(1)$ rotational symmetry about each local $\tilde{z}$ axis.
Quantum averages
$
   \langle
    a_{\bm{r}_l}^\dagger
    a_{\bm{r}_l}^\dagger
   +a_{\bm{r}_l}
    a_{\bm{r}_l}
   \rangle_0^{\mathrm{SC}}
$
in a noncollinear ground state are generally nonzero to cause
$\langle S_{\bm{r}_l}^{\tilde{x}} S_{\bm{r}_l}^{\tilde{x}} \rangle_0^{\mathrm{SC}}
\ne
 \langle S_{\bm{r}_l}^{\tilde{y}} S_{\bm{r}_l}^{\tilde{y}} \rangle_0^{\mathrm{SC}}$.
Demanding that MSWs should keep these unreasonable expectation values as well as
the total staggered magnetization zero causes to increase their otherwise stable
ground-state energy.
In order to solve this composite problem, we diagonalize the LSW Hamiltonian subject to
the double-constraint (DC) condition
$
   S
  -\langle
    a_{\bm{r}_l}^\dagger
    a_{\bm{r}_l}
   \rangle_0^{\mathrm{DC}}
  =\delta
$ and
$
   \langle
    a_{\bm{r}_l}^\dagger
    a_{\bm{r}_l}^\dagger
   +a_{\bm{r}_l}
    a_{\bm{r}_l}
   \rangle_0^{\mathrm{DC}}
  =0
$, 
\vspace*{-1.4mm}
\begin{align}
   &\!\!\!
   \mathcal{H}_{\mathrm{harm}}
  +\sum_{l=1}^L
   \left[
    \mu_l
    \left(
     S-\delta-a_{\bm{r}_l}^\dagger a_{\bm{r}_l}
    \right)
   +\eta_l
    \frac{a_{\bm{r}_l}^\dagger a_{\bm{r}_l}^\dagger
         +a_{\bm{r}_l}         a_{\bm{r}_l}        }
         {2}
   \right]
  \equiv
   \widetilde{\mathcal{H}}_{\mathrm{harm}}^{\mathrm{DC}}
   \nonumber
   \allowdisplaybreaks \\
   &\!\!\!
  =\sum_{m=1}^2
   E^{(m)}
  +\mu L
   \left(
    S-\delta+\frac{1}{2}
   \right)
  +\sum_{\sigma=0,\pm}\sum_{\nu=1}^N
   \tilde{\varepsilon}_{\bm{k}_{\nu,\sigma}}^{\rm{DC}}
   \alpha_{\bm{k}_{\nu,\sigma}}^\dagger
   \alpha_{\bm{k}_{\nu,\sigma}},
   \label{E:HDC-MLSW}
   \\[-7.5mm]\nonumber
\end{align}
where $\mu_l$'s and $\eta_l$'s both degenerate to be set to $\mu$ and $\eta$, respectively,
$\delta$ is determined so as to minimize the ground-state energy,
and then $\alpha_{\bm{k}_{\nu,\sigma}}^\dagger$ creates an MSW with energy
\vspace*{-1.4mm}
\begin{align}
   &
   \frac{\tilde{\varepsilon}_{\left(k_\nu^z,k_0^x  \right)}^{\mathrm{DC}}}{JS}
  =\!\!\sqrt{2\left(1\!-\! \cos{ak_\nu^z}\!-\!\frac{\mu\!-\!\eta}{2JS}\right)\!
              \left(5\!+\!2\cos{ak_\nu^z}\!-\!\frac{\mu\!+\!\eta}{ JS}\right)},
   \nonumber \\
   &
   \frac{\tilde{\varepsilon}_{\left(k_\nu^z,k_\pm^x\right)}^{\mathrm{DC}}}{JS}\!
  =\!\!\sqrt{2\left(1\!+\! \cos{ak_\nu^z}\!-\!\frac{\mu\!+\!\eta}{2JS}\right)
              \left(\frac{7}{2}\!-\!2\cos{ak_\nu^z}\!-\!\frac{\mu\!-\!\eta}{ JS}\right)}.
   \label{E:ekDC-MLSW}
   \\[-7.5mm]\nonumber
\end{align}
Note that the DC condition reduces to the original SC condition for collinear antiferromagnets
without any frustration.
Such DC-MLSW calculations
$S[\bm{Q}_{\mathrm{X},\Gamma},
   (\tilde{\varepsilon}_{ \bm{Q}_{\mathrm{X},\mathrm{K}}}^{\mathrm{DC}}
   +\tilde{\varepsilon}_{-\bm{Q}_{\mathrm{X},\mathrm{K}}}^{\mathrm{DC}})/\hbar]$
[Fig. \ref{F:Sqw}($\mathrm{b}1$)]
and
$S(\bm{Q}_{\mathrm{X},\mathrm{K}},
   \tilde{\varepsilon}_{\bm{Q}_{\Gamma,\Gamma}}^{\mathrm{DC}}/\hbar)$
[Fig. \ref{F:Sqw}($\mathrm{b}2$)]
are remarkable improvements compared to any of the SC-MSW overestimates.
This is more or less thanks to the enhanced gaps at
$\pm\bm{Q}_{\mathrm{X},\mathrm{K}}$ [Fig. \ref{F:Model&Disp}(c2)] and
$\bm{Q}_{\Gamma,\Gamma}$ [Fig. \ref{F:Model&Disp}(c1)], indeed,
but we should be reminded of the spurious peak in $S(\bm{Q}_{\mathrm{X},\Gamma},\omega)$
fabricated by SC MWDISWs with even larger gaps at $\pm\bm{Q}_{\mathrm{X},\mathrm{K}}$
[Fig. \ref{F:Model&Disp}(c2)]
to learn that the DC modification scheme---restoration of the local $\mathrm{U}(1)$ rotational
symmetry and then required ground-state optimization---must be more essential.

   DC MLSWs are no doubt much better than any of SC MSWs but still lack subtlety in reproducing
the DMRG calculations.
They are totally ignorant of
the subtle shoulder [Fig. \ref{F:Sqw}($\mathrm{c}1$)] and
curved ridge [Fig. \ref{F:Sqw}($\mathrm{c}2$)]
on the lower-energy side of the two-magnon scattering continuum,
both originating in the renormalized one-magnon spectrum.
There is no such observation without sufficiently strong renormalization of magnon energies
via the cubic interaction $\mathcal{H}^{\left(\frac{1}{2}\right)}$ which serves to couple
one-magnon transverse and two-magnon longitudinal fluctuations.
The $\mathcal{H}^{\left(\frac{1}{2}\right)}$-driven $O(S^0)$ self-energy
$\displaystyle
 \mathop{\varSigma}^{\,\protect\substack{-\!\!\!-\\[-1.6mm]}}\!{}_2^{(0)}
 (\bm{k};\tilde{\varepsilon}_{\bm{k}}^{\mathrm{DC}}/\hbar)$
causes a strong downward renormalization of the bare magnon energies
$\tilde{\varepsilon}_{\bm{k}}^{\mathrm{DC}}$
for $k^x=\pm Q_{\mathrm{K}}^x$ and $|k^z|\alt\frac{Q_{\mathrm{X}}^z}{2}$.
The thus-obtained DC MPSWs (S66) and (S67) \cite{Suppl_GF}
successfully reproduce the DMRG calculations,
as are illustrated with vertical arrows in
Figs. \ref{F:Sqw}($\mathrm{b}1'$) and \ref{F:Sqw}($\mathrm{b}2'$), respectively.
\begin{figure*}
\centering
\includegraphics[width=175mm]{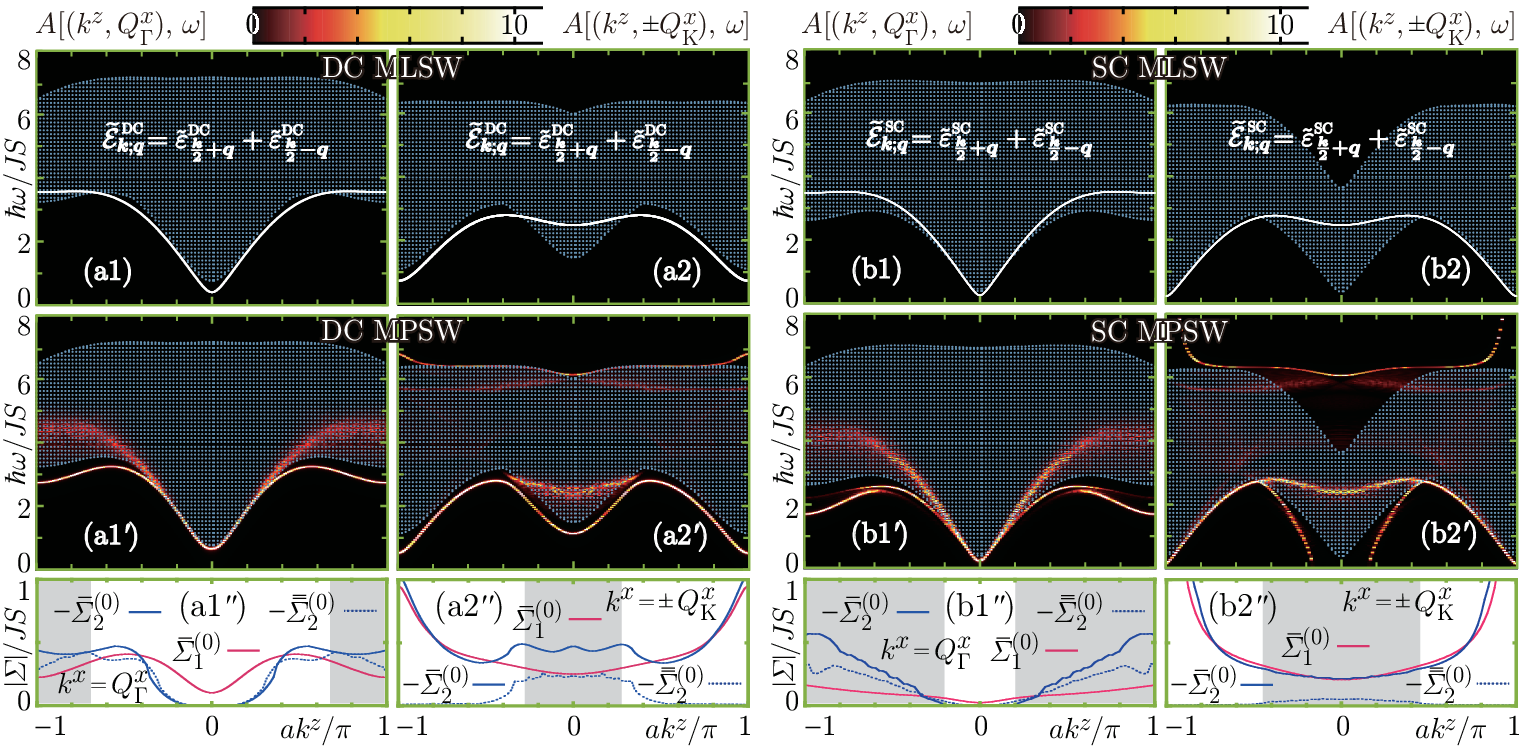}
\vspace*{-4mm}
\caption{DC-MLSW [($\mathrm{a}1$) and ($\mathrm{a}2$)],
         DC-MPSW [($\mathrm{a}1'$) and ($\mathrm{a}2'$)],
         SC-MLSW [($\mathrm{b}1$) and ($\mathrm{b}2$)], and
         SC-MPSW [($\mathrm{b}1'$) and ($\mathrm{b}2'$)]
         calculations of the spectral function
         $A(\bm{k},\omega)$ (\ref{E:Aqw}) plotted against the background of
         the two-magnon continuum
         $\widetilde{\mathcal{E}}_{\bm{k};\bm{q}}
         =\tilde{\varepsilon}_{\frac{\bm{k}}{2}+\bm{q}}
         +\tilde{\varepsilon}_{\frac{\bm{k}}{2}-\bm{q}}$
         with running $\bm{q}$ dotted finely and homogeneously
         for $S=\frac{1}{2}$ in the $N\rightarrow\infty$ limit.
         An excitation with wavevector $\bm{k}$ is unstable when the minimum of
         $\widetilde{\mathcal{E}}_{\bm{k};\bm{q}}$, denoted by
         $\widetilde{\mathcal{E}}_{\bm{k};\bm{q}_{\mathrm{min}}(\bm{k})}$,
         is lower than
         $\tilde{\varepsilon}_{\bm{k}}$
         [see ($\mathrm{a}1$) to ($\mathrm{b}2$)],
         explicitly shown by shading in
         ($\mathrm{a}1''$) to ($\mathrm{b}2''$),
         and then the quasiparticle dispersion
         $\tilde{\varepsilon}_{\bm{k}}
          \displaystyle
         +\mathop{\varSigma}^{\,\protect\substack{-\!\!\!-\protect\\[-1.6mm]}}\!{}_1^{(0)}(\bm{k})
         +\mathop{\varSigma}^{\,\protect\substack{-\!\!\!-\protect\\[-1.6mm]}}\!{}_2^{(0)}
          (\bm{k};\tilde{\varepsilon}_{\bm{k}}/\hbar)$
         is pushed out below the lower edge of
         $\widetilde{\mathcal{E}}_{\bm{k};\bm{q}}$
         [see ($\mathrm{a}1'$) to ($\mathrm{b}2'$)]
         with emergent imaginary self-energy correction
         $\displaystyle
          \mathop{\varSigma}
                ^{\,\protect\substack
                  {\protect\\[-0.6mm]-\!\!\!-\protect\\[-1.4mm] -\!\!\!-\protect\\[-1.4mm]}}
            \!{}_2^{(0)}
          (\bm{k};\tilde{\varepsilon}_{\bm{k}}/\hbar)$.
         Note that in the triangular-prism antiferromagnet
         $\displaystyle
          \mathop{\varSigma}
                ^{\,\protect\substack
                  {\protect\\[-0.6mm]-\!\!\!-\protect\\[-1.4mm] -\!\!\!-\protect\\[-1.4mm]}}
            \!{}_2^{(0)}
          (\bm{k};\tilde{\varepsilon}_{\bm{k}}/\hbar)$
         can be finite even if the inequality
         $\widetilde{\mathcal{E}}_{\bm{k};\bm{q}_{\mathrm{min}}(\bm{k})}
         <\tilde{\varepsilon}_{\bm{k}}$
         is not met [see ($\mathrm{a}1''$) in particular].}
\label{F:Aqw}
\vspace*{-4mm}
\end{figure*}

   Noncollinear magnets generally have a propensity of their elementary excitations
to spontaneous decays into pairs of other magnons \cite{S180403,C207202,Z219}.
The present system is also the case and hence the complicated dynamic structure factor.
In order to understand such decay-induced magnon dynamics in more detail,
we calculate a spectral function,
which is an analytic continuation of the one-magnon Green's function \cite{Suppl_GF}
and explicitly written as
\vspace*{-1.4mm}
\begin{align}
   \!
   A(\bm{k},\omega)\!
  =\!
   \left\{\!
   \begin{array}{ll}
   \displaystyle
   {\mathrm{Im}}
   \frac{-1/\pi}
        {\hbar\omega\!-\!\tilde{\varepsilon}_{\bm{k}}\!+\!i\hbar\epsilon}
  =\delta(\hbar\omega\!-\!\tilde{\varepsilon}_{\bm{k}})
   & \mathrm{(MLSW)} \\
   \displaystyle
   {\mathrm{Im}}
   \frac{-1/\pi}
        {\hbar\omega\!-\!\tilde{\varepsilon}_{\bm{k}}\!
        -\!\varSigma^{(0)}(\bm{k};\omega\!+\!i\epsilon)\!+\!i\hbar\epsilon}
   \!
   & \mathrm{(MPSW)} \\
   \end{array}
   \right.\!\!,
   \label{E:Aqw}
   \\[-5.5mm]\nonumber
\end{align}
where $\tilde{\varepsilon}_{\bm{k}}$ is assumed to be
either $\tilde{\varepsilon}_{\bm{k}}^{\mathrm{SC}}$ (\ref{E:ekSC-MLSW})
or $\tilde{\varepsilon}_{\bm{k}}^{\mathrm{DC}}$ (\ref{E:ekDC-MLSW}).
Figure \ref{F:Aqw} shows $A(\bm{k},\omega)$
in the DC-MLSW [($\mathrm{a}1$) and ($\mathrm{a}2$)],
DC-MPSW [($\mathrm{a}1'$) and ($\mathrm{a}2'$)],
SC-MLSW [($\mathrm{b}1$) and ($\mathrm{b}2$)], and
SC-MPSW [($\mathrm{b}1'$) and ($\mathrm{b}2'$)]
ground states.
MLSWs have an infinite lifetime without damping and therefore their energy spectrum consists of
delta functions.
If we denote the energy sum of two free MLSWs with wavevectors $\bm{k}_1$ and $\bm{k}_2$ by
$\widetilde{\mathcal{E}}_{\bm{k};\bm{q}}
\equiv
 \tilde{\varepsilon}_{\frac{\bm{k}}{2}+\bm{q}}
+\tilde{\varepsilon}_{\frac{\bm{k}}{2}-\bm{q}}$
and plot it as a function of the conserved quantity $\bm{k}\equiv\bm{k}_1+\bm{k}_2$ with
the free parameter $2\bm{q}\equiv\bm{k}_1-\bm{k}_2$ running over the full Brillouin zone,
$\widetilde{\mathcal{E}}_{\bm{k};\bm{q}}$ becomes a continuum band,
as is dotted finely and homogeneously in Fig. \ref{F:Aqw}.
If a single-magnon energy exceeds the bottom of the two-magnon continuum at each value of $\bm{k}$,
$\widetilde{\mathcal{E}}_{\bm{k};\bm{q}_{\mathrm{min}}(\bm{k})}<\tilde{\varepsilon}_{\bm{k}}$,
such a quasiparticle is no longer stable but tends to decay into a pair of quasiparticles with
their relative wavevector being $2\bm{q}_{\mathrm{min}}(\bm{k})$.
Figures \ref{F:Aqw}(a1) and \ref{F:Aqw}(a2) demonstrate that the single DC-MLSW delta-function
spectrum $\tilde{\varepsilon}_{\bm{k}}^{\mathrm{DC}}$ indeed penetrates their pair-excitation
continuum $\widetilde{\mathcal{E}}_{\bm{k};\bm{q}}^{\mathrm{DC}}$.
The CLSW spectrum contains three Nambu-Goldstone modes to be buried in its pair excitation
continuum, as is shown in Figs. S3($\mathrm{a}1$) and S3($\mathrm{a}2$) \cite{Suppl_DCvsSC}.
Modifying CLSWs causes a hardening of their dispersion as
$0
<\tilde{\varepsilon}_{\pm\bm{Q}_{\mathrm{X},\mathrm{K}}}^{\mathrm{SC}}
<\tilde{\varepsilon}_{\bm{Q}_{\Gamma,\Gamma}}^{\mathrm{SC}}
<\tilde{\varepsilon}_{\bm{Q}_{\Gamma,\Gamma}}^{\mathrm{DC}}
<\tilde{\varepsilon}_{\pm\bm{Q}_{\mathrm{X},\mathrm{K}}}^{\mathrm{DC}}$
and pushes out the single-magnon branch below the two-magnon continuum according to
the modification scheme (Fig. \ref{F:Aqw}).
The sufficient hardening of the DC-MLSW mode
$\tilde{\varepsilon}_{\pm\bm{Q}_{\mathrm{X},\mathrm{K}}}^{\mathrm{DC}}$ results in
substantially lifting up its pair excitation energy
$\tilde{\varepsilon}_{\pm\bm{Q}_{\mathrm{X},\mathrm{K}}}^{\mathrm{DC}}
+\tilde{\varepsilon}_{\pm\bm{Q}_{\mathrm{X},\mathrm{K}}}^{\mathrm{DC}}
=\widetilde{\mathcal{E}}_{\mp\bm{Q}_{\Gamma,\mathrm{K}};
                          \bm{q}_{\mathrm{min}}(\mp\bm{Q}_{\Gamma,\mathrm{K}})}
                        ^{\mathrm{DC}}$
so that the perturbed single-magnon branch
$\tilde{\varepsilon}_{\mp\bm{Q}_{\Gamma,\mathrm{K}}}^{\mathrm{DC}}
+{\displaystyle
  \bigl|
   \mathop{\varSigma}^{\,\substack{-\!\!\!-\\[-1.8mm]}}\!{}_1^{(0)}
   (\mp\bm{Q}_{\Gamma,\mathrm{K}})
  \bigr|}
-{\displaystyle
  \bigl|
   \mathop{\varSigma}^{\,\substack{-\!\!\!-\\[-1.8mm]}}\!{}_2^{(0)}
   (\mp\bm{Q}_{\Gamma,\mathrm{K}};
    \tilde{\varepsilon}_{\mp\bm{Q}_{\Gamma,\mathrm{K}}}^{\mathrm{DC}}/\hbar)
  \bigr|}$
can appear bellow that [Fig. \ref{F:Aqw}($\mathrm{a}2'$)].
However, an overlap of the single-magnon branch with the two-magnon continuum does not
necessarily lead to a strong downward renormalization of the bare magnon energies and
a rotonlike dip in the single-magnon dispersion emergent below the two-magnon continuum.
DC MSWs owe their such behaviors at $\bm{k}\simeq\bm{Q}_{\mathrm{X},\Gamma}$ and
$\bm{k}\simeq\pm\bm{Q}_{\Gamma,\mathrm{K}}$ to cubic dominant anharmonicities,
$\displaystyle
 \bigl|
  \mathop{\varSigma}^{\,\substack{-\!\!\!-\\[-1.8mm]}}\!{}_1^{(0)}(\bm{k})
 \bigr|
<\bigl|
  \mathop{\varSigma}^{\,\substack{-\!\!\!-\\[-1.8mm]}}\!{}_2^{(0)}
  (\bm{k};\tilde{\varepsilon}_{\bm{k}}^{\mathrm{DC}}/\hbar)
 \bigr|$.
Note that at $\bm{k}\simeq\bm{Q}_{\Gamma,\Gamma}$,
$\displaystyle
 \bigl|
  \mathop{\varSigma}^{\,\substack{-\!\!\!-\\[-1.8mm]}}\!{}_1^{(0)}(\bm{k})
 \bigr|$
predominates over
$\displaystyle
 \bigl|
  \mathop{\varSigma}^{\,\substack{-\!\!\!-\\[-1.8mm]}}\!{}_2^{(0)}(\bm{k};
  \tilde{\varepsilon}_{\bm{k}}^{\mathrm{DC}}/\hbar)
 \bigr|$
to renormalize the single-magnon branch upward, which is the case with single-chain \cite{N034714}
and square-lattice \cite{S216003} collinear antiferromagnets, for instance.
SC MLSWs with $\bm{k}\simeq\pm\bm{Q}_{\Gamma,\mathrm{K}}$ fail to enjoy a quantum rotonlike
stabilization [Fig. \ref{F:Aqw}($\mathrm{b}2'$)]
with their balanced self-energy corrections,
$\displaystyle
 \bigl|
  \mathop{\varSigma}^{\,\substack{-\!\!\!-\\[-1.8mm]}}\!{}_1^{(0)}(\bm{k})
 \bigr|
\simeq
 \bigl|
  \mathop{\varSigma}^{\,\substack{-\!\!\!-\\[-1.8mm]}}\!{}_2^{(0)}
  (\bm{k};\tilde{\varepsilon}_{\bm{k}}^{\mathrm{SC}}/\hbar)
 \bigr|$
[Fig. \ref{F:Aqw}($\mathrm{b}2''$)].

   The inequality
$\widetilde{\mathcal{E}}_{\bm{k};\bm{q}_{\mathrm{min}}(\bm{k})}<\tilde{\varepsilon}_{\bm{k}}$,
which is shown by shading in Figs. \ref{F:Aqw}($\mathrm{a}1''$) to \ref{F:Aqw}($\mathrm{b}2''$),
guarantees the single-magnon branch to overlap with the two-magnon continuum and then
the cubic interaction makes its lifetime finite to cause one-to-two-magnon decays even at
zero temperature.
At $\bm{k}\simeq\pm\bm{Q}_{\Gamma,\mathrm{K}}$,
the decay rate
$\displaystyle
-\mathop{\varSigma}^{\,\substack{\\[-0.6mm]-\!\!\!- \\[-1.4mm] -\!\!\!-\\[-1.4mm]}}\!{}_2^{(0)}
 (\bm{k};\tilde{\varepsilon}_{\bm{k}}^{\mathrm{DC}}/\hbar)$
stays finite well consistently with the shaded region,
whereas on the way from $\bm{Q}_{\mathrm{X},\Gamma}$ to $\bm{Q}_{\Gamma,\Gamma}$,
the imaginary part survives far beyond the shaded region possibly with a different decay scenario
such as fractionalization of spin excitations into spinons.
The present wavevector $\bm{Q}_{\mathrm{X},\Gamma}$ represents such a collinear spin configuration
as ferromagnetic and antiferromagnetic in the rung and leg directions, respectively,
while
the midedges of the hexagonal Brillouin zone of the triangular-lattice antiferromagnet,
which are often referred to as the $\mathrm{M}$ points, also represent such collinear
configurations as ferromagnetic and antiferromagnetic in the directions of certain two primitive
translation vectors with a relative angle of $60^\circ$.
These two ordering vectors thus correspond well to each other,
and interestingly enough, the renormalized antiferromagnon spectra in one and two dimensions
have similar appearances in their vicinities \cite{M094407}.
Series-expansion calculations for the triangular-lattice antiferromagnet
\cite{Z224420,Z057201} reveal that the bare magnon energies are strongly renormalized
downward by quantum fluctuations so as to form rotonlike minima near $\mathrm{M}$.
Bosonic \cite{G184403} and fermionic \cite{Z075108} spinon-language analyses both claim that
these rotonlike minima should be collective modes of a spinon condensate, i.e.,
two-spinon bound states that spontaneously break the $\mathrm{SU}(2)$ symmetry.
In the triangular-lattice antiferromagnet,
the inequality condition of bare magnons being kinematically unstable towards decays into pairs of
other magnons and the region of the lowest-order self-energy corrections to bare magnons having
an imaginary part coincide to form a hexagram, which is inscribed in the hexagonal Brillouin zone
occupying two thirds of it but not including the $\mathrm{M}$ points
\cite{C144416,S180403,C207202}.
In the triangular-prism antiferromagnet, the magnon decay rate
$\displaystyle
-\mathop{\varSigma}^{\,\substack{\\[-0.6mm]-\!\!\!- \\[-1.4mm] -\!\!\!-\\[-1.4mm]}}\!{}_2^{(0)}
 (\bm{k};\tilde{\varepsilon}_{\bm{k}}^{\mathrm{DC}}/\hbar)$
remains finite far beyond the spontaneous two-magnon decay region on the way from
$\bm{Q}_{\mathrm{X},\Gamma}$ to $\bm{Q}_{\Gamma,\Gamma}$.
Then, there is a possibility of bare magnons decaying into two spinons instead, but if so
why is the magnon decay rate vanishing near $\mathrm{M}$ on the hexagonal Brillouin zone?
We find a probable key to this mystery in general arguments on frustrated quantum antiferromagnets
in terms of lattice gauge theories \cite{F3682} that spinons are necessarily confined in their
ordered phases \cite{C14729(R)} but should be deconfined in their disordered phases \cite{R1773}.
Calculating the dynamic structure factor of the triangular-lattice antiferromagnet in terms of
spinon Green's functions \cite{G184403,Z075108} indeed shows that when coming closer to
the $\mathrm{M}$ points, single magnons are no longer allowed to decay into two spinons but
stabilize below the lower edge of the two-spinon continuum with their line width disappearing.
The present system on the way from $\bm{Q}_{\mathrm{X},\Gamma}$ to $\bm{Q}_{\Gamma,\Gamma}$
behaves as an effective spin-$\frac{3}{2}$ chain rather than spin-$\frac{1}{2}$ frustrated
triangles, wherein the instability of its single-magnon spectrum may change in character,
from spontaneous two-magnon decays to two-spinon decays.
Unlike the collective antiferromagnon modes with their decay rates vanishing at the $\mathrm{M}$
points on the triangular lattice, here one-dimensional antiferromagnon modes are persistently
accompanied by nonvanishing decay rates on the way from $\bm{Q}_{\mathrm{X},\Gamma}$ to
$\bm{Q}_{\Gamma,\Gamma}$, suggesting their possible decays into some other elementary excitations.
Deconfined spinons are indeed the case with the spin-$\frac{1}{2}$ antiferromagnetic Heisenberg
chain \cite{F375}, whose eigenstates necessarily contain an even number of kinks without forming
any bound state of them, each with quasimomentum running through half the Brillouin zone.
Though its $O(S^0)$ self-energy is independent of frequency to have no imaginary part,
higher-order self-energy corrections necessarily have an imaginary part without
cubic interactions.
Note that
the well-known rotonlike dip in the single-magnon dispersion of the spin-$\frac{1}{2}$
square-lattice Heisenberg antiferromagnet \cite{Z184440} is never describable
without $O(S^{-1})$ self-energy corrections \cite{U282,S216003}.
It is worth noting further that a recently developed bond-operator technique is
quite efficient in reproducing such a quantum rotonlike stabilization \cite{S184421}.
Considering that spinon observations are qualitatively common to all fractional-spin chains
\cite{H4955,A5291}, two-spinon bound states as stable triplet magnons are unlikely
in the present system.
It should also be noted that a strong-rung-coupling effective Hamiltonian \cite{C6241,W174410}
for the triangular-prism antiferromagnet yields massive spinonlike as well as magnonlike
excitation spectra with varying interchain and intrachain couplings.
Ordered but noncollinear magnets have cubic anharmonicities,
i.e., coupling of transverse and longitudinal fluctuations,
hence the ubiquitous propensity of their excitations to spontaneous decays.
The ground states of fractional-spin antiferromagnetic Heisenberg chains are neither ordered nor
disordered completely, hence deconfined spinons.
The spin-$\frac{1}{2}$ triangular-prism antiferromagnet is possessed of both characteristics
and may exhibit a novel \textit{mixed instability} of the single-particle spectrum.

\vspace{-3mm}
\acknowledgments
\vspace{-4mm}
This work is supported by JSPS KAKENHI Grant No. 22K03502.


\vspace{-5mm}
\begin{thebibliography}{99}
\makeatletter%comments in bibliography
\@newlistfalse%comments in bibliography
\makeatother%comments in bibliography
\vspace{-3mm}
\noindent\par
{${}^{*}$ {Corresponding author: yamamoto@phys.sci.hokudai.ac.jp}}

\bibitem{R9235}
   Reigrotzki M., Tsunetsugu H. and Rice T. M.,
%      Strong-coupling expansions for antiferromagnetic Heisenberg spin-one-half ladders,
      \textit{J. Phys.: Condens. Matter}, \textbf{6} (1994) 9235.

\bibitem{G8901}
   Gopalan S., Rice T. M. and Sigrist M.,
%      Spin ladders with spin gaps: A description of a class of cuprates,
      \textit{Phys. Rev. B}, \textbf{49} (1994) 8901.

\bibitem{A6233}
   Azzouz M., Chen Liang and Moukouri S.,
%      Calculation of the singlet-triplet gap of the antiferromagnetic Heisenberg model on
%      a ladder,
      \textit{Phys. Rev. B}, \textbf{50} (1994) 6233.

\bibitem{D618}
   Dagotto E. and Rice T. M.,
%      Surprises on the way from one- to two-dimensional quantum magnets: The ladder materials,
      \textit{Science}, \textbf{271} (1996) 618.

\bibitem{D964}
   Dai X. and Su Z.,
%   Mean-field theory for the spin-ladder system,
      \textit{Phys. Rev. B}, \textbf{57} (1998) 964.

\bibitem{H1607}
   Hori H. and Yamamoto S.,
%      Fermionic description of spin-gap states of antiferromagnetic Heisenberg ladders
%      in a magnetic field,
      \textit{J. Phys. Soc. Jpn.}, \textbf{71} (2002) 1607.

\bibitem{S533}
   Schulz H. J.,
      \textit{Fermi liquids and non-Fermi liquids} in
      \textit{Mesoscopic Quantum Physics, Les Houches, Session LXI, 1994}
      edited by Akkermans E., Montambaux G., Pichard J. L. and Zinn-Justin J.
      (Elsevier, Amsterdam, 1995), p. 533.

\bibitem{H1358}
   Haldane F. D. M.,
%      General relation of correlation exponents and spectral properties of one-dimensional
%      Fermi systems: Application to the anisotropic $S=\frac{1}{2}$ Heisenberg chain,
      \textit{Phys. Rev. Lett.}, \textbf{45} (1980) 1358.

\bibitem{H4955}
   Hallberg K., Wang X. Q. G., Horsch P. and Moreo A.,
%      Critical behavior of the $S=3/2$ antiferromagnetic Heisenberg chain,
      \textit{Phys. Rev. Lett.}, \textbf{76} (1996) 4955.

\bibitem{H41}
   Hiroi Z. and Takano M.,
%      Absence of superconductivity in the doped antiferromagnetic spin-ladder compound
%      $(\mathrm{La},\mathrm{Sr})\mathrm{CuO}_{2.5}$,
      \textit{Nature}, \textbf{377} (1995) 41.

\bibitem{I2397}
   Iwase H., Isobe M., Ueda Y. and Yasuoka H.,
%      Observation of spin gap in $\mathrm{CaV}_2\mathrm{O}_5$ by NMR,
      \textit{J. Phys. Soc. Jpn.}, \textbf{65} (1996) 2397.

\bibitem{A3463}
   Azuma M., Hiroi Z., Takano M., Ishida K. and Kitaoka Y.,
%      Observation of a spin gap in $\mathrm{SrCu}_2\mathrm{O}_3$ comprising
%      spin-$\frac{1}{2}$ quasi-1D two-leg ladders,
      \textit{Phys. Rev. Lett.}, \textbf{73} (1994) 3463.

\bibitem{K2812}
   Kojima K., Keren A., Luke G. M., Nachumi B., Wu W. D., Uemura Y. J., Azuma M. and Takano M.,
%      Magnetic behavior of the 2-leg and 3-leg spin ladder cuprates
%      $\mathrm{Sr}_{n-1}\mathrm{Cu}_{n+1}\mathrm{O}_{2n}$,
      \textit{Phys. Rev. Lett.}, \textbf{74} (1995) 2812.

\bibitem{S174420}
   Schnack J., Nojiri H. and K\"ogerleri P.,
%      Magnetic characterization of the frustrated three-leg ladder compound
%      $[(\mathrm{CuCl}_2\mathrm{tachH})_3\mathrm{Cl}]\mathrm{Cl}_2$,
      \textit{Phys. Rev. B}, \textbf{70} (2004) 174420.

\bibitem{S1580}
   Seeber G., K\"ogerler P., Kariukic B. M. and Cronin L.,
%      Supramolecular assembly of ligand-directed triangular
%      $\{\mathrm{Cu}_3^{\mathrm{II}}\mathrm{Cl}\}$ clusters
%      with spin frustration and spin-chain behaviour,
      \textit{Chem. Commun.}, \textbf{2004} (2004) 1580.

\bibitem{M093701}
   Manaka H., Hirai Y., Hachigo Y., Mitsunaga M., Ito M. and Terada N.,
%      Spin-liquid state study of equilateral triangle $S=3/2$ spin tubes formed in
%      $\mathrm{CsCrF}_4$,
      \textit{J. Phys. Soc. Jpn.}, \textbf{78} (2009) 093701.

\bibitem{M084714}
   Manaka H., Etoh T., Honda Y., Iwashita N., Oagata K., Terada N.,
   Hisamatsu T., Ito M., Narumi Y., Kondo A., Kindo K. and Miura Y.,
%      Effects of geometrical spin frustration on triangular spin tubes formed
%      in $\mathrm{CsCrF}_4$ and $\alpha\mbox{-}\mathrm{KCrF}_4$,
      \textit{J. Phys. Soc. Jpn.}, \textbf{80} (2011) 084714.

\bibitem{G037206}
   Garlea V. O., Zheludev A., Regnault L.-P., Chung J.-H., Qiu Y., Boehm M., Habicht K. and
   Meissner M.,
%      Excitations in a four-leg antiferromagnetic Heisenberg spin tube,
      \textit{Phys. Rev. Lett.}, \textbf{100} (2008) 037206.

\bibitem{G060404}
   Garlea V. O., Zheludev A., Habicht K., Meissner M., Grenier B., Regnault L.-P. and
   Ressouche E.,
%      Dimensional crossover in a spin-liquid-to-helimagnet quantum phase transition,
      \textit{Phys. Rev. B}, \textbf{79} (2009) 060404(R).

\bibitem{M676}
   Millet P., Henry J. Y., Mila F. and Galy J.,
%      Vanadium(IV)-oxide nanotubes: Crystal structure of
%      the low-dimensional quantum magnet $\mathrm{Na}_2\mathrm{V}_3\mathrm{O}_7$,
      \textit{J. Solid State Chem.}, \textbf{147} (1999) 676.

\bibitem{L060405}
   L\"uscher A., Noack R. M., Misguich G., Kotov V. N. and Mila F.,
%      Soliton binding and low-lying singlets in frustrated odd-legged $S=\frac{1}{2}$ spin tubes,
      \textit{Phys. Rev. B}, \textbf{70} (2004) 060405(R).

\bibitem{G064431}
   Gavilano J. L., Felder E., Rau D., Ott H. R., Millet P., Mila F., Cichorek T. and
   Mota A. C.,
%      Unusual magnetic properties of the low-dimensional quantum magnet
%      $\mathrm{Na}_2\mathrm{V}_3\mathrm{O}_7$,
      \textit{Phys. Rev. B}, \textbf{72} (2005) 064431.

\bibitem{K4001}
   Kawano K. and Takahashi M.,
%      Three-leg antiferromagnetic Heisenberg ladder with frustrated boundary condition;
%      Ground state properties,
      \textit{J. Phys. Soc. Jpn.}, \textbf{66} (1997) 4001.

\bibitem{N054421}
   Nishimoto S. and Arikawa M.,
%      Low-lying excitations of the three-leg spin tube:
%      A density-matrix renormalization group study,
      \textit{Phys. Rev. B}, \textbf{78} (2008) 054421.

\bibitem{S184415}
   Sakai T., Sato M., Okunishi K., Otsuka Y., Okamoto K. and Itoi C.,
%      Quantum phase transitions of the asymmetric three-leg spin tube,
      \textit{Phys. Rev. B}, \textbf{78} (2008) 184415.

\bibitem{S403201}
   Sakai T., Sato M., Okamoto K., Okunishi K. and Itoi C.,
%      Quantum spin nanotubes-frustration, competing orders and criticalities,
      \textit{J. Phys. Condens. Matter}, \textbf{22} (2010) 403201.

\bibitem{C075108}
   Charrier D., Capponi S., Oshikawa M. and Pujol P.,
%      Quantum phase transitions in three-leg spin tubes,
      \textit{Phys. Rev. B}, \textbf{82} (2010) 075108.

\bibitem{N224425}
   Nishimoto S., Fuji Y. and Ohta Y.,
%      Spin gap of the three-leg $S=\frac{3}{2}$ Heisenberg tube,
      \textit{Phys. Rev. B}, \textbf{83} (2011) 224425.

\bibitem{O297}
   Okunishi K., Yoshikawa S., Sakai T. and Miyashita S.,
%      Low-energy excitations of the $S=1/2$ quantum spin tube with
%      the triangular lattice structure,
      \textit{Prog. Theor. Phys. Suppl.}, \textbf{159} (2005) 297.

\bibitem{O1423}
   Okunishi K., Yoshikawa S., Sakai T. and Miyashita S.,
%      Quantum phase transition of a triangular lattice spin tube
%      and edge spin effects,
      \textit{Int. J. Mod. Phys. C}, \textbf{20} (2009) 1423.

\bibitem{F014409}
   Fouet J.-B., L\"auchli A., Pilgram S., Noack R. M. and Mila F.,
%      Frustrated three-leg spin tubes: From spin $1/2$ with chirality to spin $3/2$,
      \textit{Phys. Rev. B}, \textbf{73} (2006) 014409.

\bibitem{I037206}
   Ivanov N. B., Schnack J., Schnalle R., Richter J., K\"ogerler P., Newton G. N., Cronin L.,
   Oshima Y. and Nojiri H.,
%      Heat capacity reveals the physics of a frustrated spin tube,
      \textit{Phys. Rev. Lett.}, \textbf{105} (2010) 037206.

\bibitem{K184405}
   Kikuchi H., Asai S., Manaka H., Hagihala M., Itoh S. and Masuda T.,
%      Inelastic neutron scattering in the weakly coupled triangular spin tube candidate
%      $\mathrm{CsCrF}_4$,
      \textit{Phys. Rev. B}, \textbf{107} (2023) 184405.

\bibitem{D205119}
   Dargel P. E., W\"ollert A., Honecker A., McCulloch I. P.,
   Schollw\"ock U. and Pruschke T.,
%      Lanczos algorithm with matrix product states for dynamical correlation functions,
      \textit{Phys. Rev. B}, \textbf{85} (2012) 205119.

\bibitem{C1545}
   Chen X., Ran S.-J., Liu T., Peng C., Huang Y.-Z. and Su G.,
%      Thermodynamics of spin-$1/2$ Kagom\'{e} Heisenberg antiferromagnet:
%      algebraic paramagnetic liquid and finite-temperature phase diagram
      \textit{Science Bull.}, \textbf{63} (2018) 1545.

\bibitem{M445}
   Maisinger K. and Schollw\"ock U.,
%      Thermodynamics of Frustrated Quantum Spin Chains,
      \textit{Phys. Rev. Lett.}, \textbf{81} (1998) 445.

\bibitem{J045114}
   Jeckelmann E.,
%      Dynamical density-matrix renormalization-group method,
      \textit{Phys. Rev. B}, \textbf{66} (2002) 045114.

\bibitem{K568}
   Kubo R.,
%      The spin-wave theory of antiferromagnetics,
      \textit{Phys. Rev.}, \textbf{87} (1952) 568.

\bibitem{S075123}
   Schmidt B., Siahatgar M. and Thalmeier P.,
%      Ordered moment in the anisotropic and frustrated square lattice Heisenberg model,
      \textit{Phys. Rev. B}, \textbf{83} (2011) 075123.

\bibitem{T1524}
   Takahashi M.,
%      A spin-wave theory for $\mathrm{La}_2\mathrm{CuO}_4$,
      \textit{J. Phys. Soc. Jpn.}, \textbf{58} (1989) 1524.

\bibitem{T2494}
   Takahashi M.,
%      Modified spin-wave theory of a square-lattice antiferromagnet,
      \textit{Phys. Rev. B}, \textbf{40} (1989) 2494.

\bibitem{H4769}
   Hirsch J. E. and Tang S.,
%      Spin-wave theory of the quantum antiferromagnet with unbroken sublattice symmetry,
      \textit{Phys. Rev. B}, \textbf{40} (1989) 4769.

\bibitem{T5000}
   Tang S., Lazzouni M. E. and Hirsch J. E.,
%      Sublattice-symmetric spin-wave theory for the Heisenberg antiferromagnet,
      \textit{Phys. Rev. B}, \textbf{40} (1989) 5000.

\bibitem{Y064426}
   Yamamoto S.,
%      Bosonic representation of one-dimensional Heisenberg ferrimagnets,
      \textit{Phys. Rev. B}, \textbf{69} (2004) 064426.

\bibitem{A617}
   Auerbach A. and Arovas D. P.,
%      Spin dynamics in the square-lattice antiferromagnet,
      \textit{Phys. Rev. Lett.}, \textbf{61} (1988) 617.

\bibitem{Y065004}
   Yamamoto S. and Ohara J.,
%      Thermal features of Heisenberg antiferromagnets on edge- versus corner-sharing
%      triangular-based lattices: A message from spin waves,
      \textit{J. Phys. Commun.}, \textbf{7} (2023) 065004.

\bibitem{C144416}
   Chernyshev A. L. and Zhitomirsky M. E.,
%      Spin waves in a triangular lattice antiferromagnet:
%      Decays, spectrum renormalization, and singularities,
      \textit{Phys. Rev. B}, \textbf{79} (2009) 144416.

\bibitem{S180403}
   Starykh O. A., Chubukov A. V. and Abanov A. G.,
%      Flat spin-wave dispersion in a triangular antiferromagnet,
      \textit{Phys. Rev. B}, \textbf{74} (2006) 180403(R).

\bibitem{C207202}
   Chernyshev A. L. and Zhitomirsky M. E.,
%      Magnon decay in noncollinear quantum antiferromagnets,
      \textit{Phys. Rev. Lett.}, \textbf{97} (2006) 207202.

\bibitem{Z224420}
   Zheng W., Fj{\ae}restad J. O., Singh R. R. P., McKenzie R. H. and Coldea R.,
%      Excitation spectra of the spin-$\frac{1}{2}$ triangular-lattice Heisenberg antiferromagnet,
      \textit{Phys. Rev. B}, \textbf{74} (2006) 224420.

\bibitem{M094407}
   Mourigal M., Fuhrman W. T., Chernyshev A. L. and Zhitomirsky M. E.,
%      Dynamical structure factor of the triangular-lattice antiferromagnet,
      \textit{Phys. Rev. B}, \textbf{88} (2013) 094407.

\bibitem{S216003}
   Syromyatnikov A. V.,
%      Spectrum of short-wavelength magnons in a two-dimensional quantum Heisenberg antiferromagnet
%      on a square lattice: Third-order expansion in $1/S$,
      \textit{J. Phys.: Condens. Matter}, \textbf{22} (2010) 216003.

\bibitem{Z184440}
   Zheng W., Oitmaa J. and Hamer C. J.,
%      Series studies of the spin-$\frac{1}{2}$ Heisenberg antiferromagnet at $T=0$:
%      Magnon dispersion and structure factors,
      \textit{Phys. Rev. B}, \textbf{71} (2005) 184440.

%%%%% LRO on S=1/2 triangular-lattice Heisenberg AFM %%%%%
\bibitem{H2531}
   Huse D. A. and Elser V.,
%      Simple variational wave functions for two-dimensional Heisenberg spin-$\frac{1}{2}$
%      antiferromagnets,
      \textit{Phys. Rev. Lett.}, \textbf{60} (1988) 2531.

\bibitem{J2727}
   Jolicoeur Th. and Guillou J. C. L.,
%      Spin-wave results for the triangular Heisenberg antiferromagnet,
      \textit{Phys. Rev. B}, \textbf{40} (1989) 2727(R).

\bibitem{S1766}
   Singh R. R. P. and Huse D. A.,
%      Three-sublattice order in triangular- and Kagom\'e-lattice spin-half antiferromagnets,
      \textit{Phys. Rev. Lett.}, \textbf{68} (1992) 1766.

\bibitem{B2590}
   Bernu B., Lhuillier C. and Pierre L.,
%      Signature of N\'eel order in exact spectra of quantum antiferromagnets on finite lattices,
      \textit{Phys. Rev. Lett.}, \textbf{69} (1992) 2590.

\bibitem{A2483}
   Azaria P., Delamotte B. and Mouhanna D.,
%      Spontaneous symmetry breaking in quantum frustrated antiferromagnets,
      \textit{Phys. Rev. Lett.}, \textbf{70} (1993) 2483.

\bibitem{B10048}
   Bernu B., Lecheminant P., Lhuillier C. and Pierre L.,
%      Exact spectra, spin susceptibilities, and order parameter of the quantum Heisenberg
%      antiferromagnet on the triangular lattice,
      \textit{Phys. Rev. B}, \textbf{50} (1994) 10048.

\bibitem{C3899}
   Capriotti L., Trumper A. E. and Sorella S.,
%      Long-range N\'eel order in the triangular Heisenberg model,
      \textit{Phys. Rev. Lett.}, \textbf{82} (1999) 3899.

\bibitem{W127004}
   White S. R. and Chernyshev A. L.,
%      N\'eel order in square and triangular lattice Heisenberg models,
      \textit{Phys. Rev. Lett.}, \textbf{99} (2007) 127004.
%%%%% LRO on S=1/2 triangular-lattice Heisenberg AFM %%%%%

\bibitem{Suppl_MSW}
   Supplemental Material Sect. S1.
      A Holstein-Primakoff bosonic Hamiltonian for the spin-$S$ nearest-neighbor Heisenberg
      antiferromagnet on the equilateral triangular prism is introduced in detail and
      its various modification schemes are explicitly explained within and beyond
      the harmonic approximation.

\bibitem{B195124}
   Bauer D.-V. and Fj{\ae}restad J. O.,
%      Spin-wave study of entanglement and R\'enyi entropy for coplanar and collinear magnetic
%      orders in two-dimensional quantum Heisenberg antiferromagnets,
      \textit{Phys. Rev. B}, \textbf{101} (2020) 195124.

%%%%%%%% SC-MSW for Haldane-gap AFM %%%%%%%%%%%%%%%%%%%%%%%%%%%%%%%%%%%%%%%%%%%%%%%%%%%%%%%%%
\bibitem{R2589}%DM transformation MSW exp. observ. for S=1 HG AFM $\mathrm{CsNiCl}_3$.
   Rezende S. M.,
%      Spin-wave theory of the Haldane gap in one-dimensional antiferromagnets,
      \textit{Phys. Rev. B}, \textbf{42} (1990) 2589.

\bibitem{W12805}%NO expansion MSW for the ground-state energy and the gap at S=1
   Wang H., Shen J., Li K. and Su Z.,
%      Self-consistent mean-field theory of a spin-1 antiferromagnetic chain,
      \textit{Phys. Rev. B}, \textbf{42} (1994) 12805.

\bibitem{Y769}%ThD properties (chi)
   Yamamoto S. and Hori H.,
%      Spin-wave description of Haldane-gap antiferromagnets,
      \textit{J. Phys. Soc. Jpn.}, \textbf{72} (2003) 769.

\bibitem{Y822}%1/T_1 NENP
   Yamamoto S. and Hori H.,
%      Nuclear magnetic relaxation in the Haldane-gap antiferromagnet
%      $\mathrm{Ni}(\mathrm{C}_2\mathrm{H}_8\mathrm{N}_2)_2\mathrm{NO}_2(\mathrm{ClO}_4)$,
      \textit{J. Phys. Soc. Jpn.}, \textbf{72} (2004) 822.
%%%%%%%% SC-MSW for Haldane-gap AFM %%%%%%%%%%%%%%%%%%%%%%%%%%%%%%%%%%%%%%%%%%%%%%%%%%%%%%%%%

\bibitem{N034714}
   Noriki Y. and Yamamoto S.,
%      Modified spin-wave theory on low-dimensional Heisenberg ferrimagnets:
%      A new robust formulation,
      \textit{J. Phys. Soc. Jpn.}, \textbf{86} (2017) 034714.

\bibitem{Y094412}
   Yamamoto S. and Noriki Y.,
%      Spin-wave thermodynamics of square-lattice antiferromagnets revisited,
      \textit{Phys. Rev. B}, \textbf{99} (2019) 094412.

\bibitem{Suppl_GF}
   Supplemental Material Sect. S2.
      Magnon Green's functions are introduced to formulate the dynamic spin structure factors
      $S^{\lambda\lambda}(\bm{q},\omega)\,(\lambda=\tilde{x},\tilde{y},\tilde{z})$
      in the bare and correlated magnon ground states.

\bibitem{V134429}
   Veillette M. Y., James A. J. A. and Essler F. H. L.,
%      Spin dynamics of the quasi-two-dimensional spin-$\frac{1}{2}$ quantum magnet
%      $\mathrm{Cs}_2\mathrm{CuCl}_4$,
      \textit{Phys. Rev. B}, \textbf{72} (2005) 134429.

\bibitem{C3264}
   Canali C. M. and Wallin M.,
%      Spin-spin correlation functions for the square-lattice Heisenberg antiferromagnet
%      at zero temperature,
      \textit{Phys. Rev. B}, \textbf{48} (1993) 3264.

\bibitem{Suppl_DCvsSC}
   Supplemental Material Sect. S3.
      Manners of modifying conventional spin waves are explained in more detail with particular
      emphasis on similarities and differences between the single-constraint and
      double-constraint conditions.
      A comparative analysis on the spectral function $A(\bm{k},\omega)$ is also presented.

\bibitem{Z219}
   Zhitomirsky M. E. and Chernyshev A. L.,
%      \textit{Colloquium}: Spontaneous magnon decays,
      \textit{Rev. Mod. Phys.}, \textbf{85} (2013) 219.

\bibitem{Z057201}
   Zheng W., Fj{\ae}restad J. O., Singh R. R. P., McKenzie R. H. and Coldea R.,
%      Anomalous excitation spectra of frustrated quantum antiferromagnets,
      \textit{Phys. Rev. Lett.}, \textbf{96} (2006) 057201.

\bibitem{G184403}
   Ghioldi E. A., Gonzalez M. G., Zhang S.-S., Kamiya Y., Manuel L. O., Trumper A. E. and
   Batista C. D.,
%      Dynamical structure factor of the triangular antiferromagnet:
%      Schwinger boson theory beyond mean field,
      \textit{Phys. Rev. B}, \textbf{98} (2018) 184403.

\bibitem{Z075108}
   Zhang C. and Li T.,
%      Resonating valence bond theory of anomalous spin dynamics of spin-$\frac{1}{2}$
%      triangular lattice Heisenberg antiferromagnet and its application to
%      $\mathrm{Ba}_3\mathrm{CoSb}_2\mathrm{O}_9$,
      \textit{Phys. Rev. B}, \textbf{102} (2020) 075108.

\bibitem{F3682}
   Fradkin E. and Shenker S. H.,
%      Phase diagrams of lattice gauge theories with Higgs fields,
      \textit{Phys. Rev. D}, \textbf{19} (1979) 3682.

\bibitem{C14729(R)}
   Chubukov A. V. and Starykh O. A.,
%      Crossover from O(3) to O(4) behavior in weakly frustrated antiferromagnets,
      \textit{Phys. Rev. B}, \textbf{53} (1996) 14729(R).

\bibitem{R1773}
   Read N. and Sachdev S.,
%      Large-$N$ expansion for frustrated quantum antiferromagnets,
      \textit{Phys. Rev. Lett.}, \textbf{66} (1991) 1773.

\bibitem{F375}
   Faddeev L. D. and Takhtajan L. A.,
%      What is the spin of a spin wave?
      \textit{Phys. Lett. A}, \textbf{85} (1981) 375.

\bibitem{U282}
   Uhrig G. S. and Majumdar K.,
%      Varied perturbation theory for the dispersion dip
%      in the two-dimensional Heisenberg quantum antiferromagnet
      \textit{Eur. Phys. J. B}, \textbf{86} (2013) 282.

\bibitem{S184421}
  Syromyatnikov A. V.,
%  Collective excitations in spin-1/2 magnets through bond-operator formalism
%  designed both for paramagnetic and ordered phases
      \textit{Phys. Rev. B}, \textbf{98} (2018) 184421.

\bibitem{A5291}
   Affleck I. and Haldane F. D. M.,
%      Critical theory of quantum spin chains,
      \textit{Phys. Rev. B}, \textbf{36} (1987) 5291.

\bibitem{C6241}
   Cabra D. C., Honecker A. and Pujol P.,
%      Magnetization plateaux in $N$-leg spin ladders,
      \textit{Phys. Rev. B}, \textbf{58} (1998) 6241.

\bibitem{W174410}
   Wang H.-T.,
%      Elementary excitations in the spin-tube and spin-orbit models,
      \textit{Phys. Rev. B}, \textbf{64} (2001) 174410.
\end{thebibliography}
\end{document}

\title{\textit{Supplemental Material for}\\
Breakdown of the conventional spin-wave dynamics and its double-constraint modification\\
in the spin-$\mathbf{\frac{1}{2}}$ triangular-prism Heisenberg antiferromagnet}
\author{Shoji Yamamoto and Jun Ohara}
\address{Department of Physics, Hokkaido University, Sapporo 060-0810, Japan}
\date{\today}

\maketitle

\section{Spin Waves in Noncollinear Antiferromagnets and Their Modification}
\label{S:SW}
\subsection{Bosonic Hamiltonian}

   We define the nearest-neighbor antiferromagnetic Heisenberg model on the equilateral triangular
prism (three-legged ladder with the periodic boundary condition in the rung direction) of
$L\equiv 3\times N$ lattice sites [Fig. 1(a)],
\begin{align}
   \mathcal{H}
  =\sum_{l=1}^L
   \sum_{\sigma=\parallel,\perp}
   J_\sigma
   \bm{S}_{\bm{r}_l}\cdot\bm{S}_{\bm{r}_l+\bm{\delta}_\sigma},
   \label{SE:H}
\end{align}
where
$\bm{S}_{\bm{r}_l}$ and $\bm{S}_{\bm{r}_l+\bm{\delta}_\sigma}$ are the vector spin operators of
magnitude $S$ attached to the nearest-neighbor sites at $\bm{r}_l$ and
$\bm{r}_l+\bm{\delta}_\sigma$, respectively, with $\bm{\delta}_\sigma$ pointing to
a nearest-neighbor site to any site $\bm{r}_l$ in the leg ($\sigma=\parallel$) or
rung ($\sigma=\perp$) direction.
Note that $\bm{\delta}_\perp$, as well as $\bm{\delta}_\parallel$, no longer depends on
the site index $l$ when unfolding the prism into a two-dimensional plane
[Fig. 1(a)] and then $\bm{r}_l$ reads, say,
$n_\parallel\bm{\delta}_\parallel+n_\perp\bm{\delta}_\perp
=a(n_\parallel,n_\perp,0)$
with integers $n_\parallel=0,1,\cdots,N$ and $n_\perp=0,1,2$.
Henceforth we set $J_\parallel$ and $J_\perp$ both equal to $J\,(>0)$ and accordingly suppose
$|\bm{\delta}_\parallel|=|\bm{\delta}_\perp|=a$.
Then the classical ground state of (\ref{SE:H}) is designated by the ordering vector
$(q^z,q^x,q^y)=\frac{\pi}{a}\left(1,\frac{2}{3},0\right)$.
As long as $q^y=0$, we use a short-hand notation in the two-dimensional reciprocal space such as
\begin{align}
   \bm{Q}_{\Gamma,\Gamma}\!
  =\!(Q_\Gamma^z,Q_\Gamma^x)\!
  \equiv\!
   \frac{\pi}{a}
   \left(
    0,0
   \right),\ 
   \bm{Q}_{\mathrm{X},\mathrm{K}}\!
  =\!(Q_{\mathrm{X}}^z,Q_{\mathrm{K}}^x)\!
  \equiv\!
   \frac{\pi}{a}
   \left(
    1,\frac{2}{3}
   \right).\!\!
   \tag{1}
   \label{E:Vin2DBZ}
\end{align}
%\begin{figure}[t]
%\centering
%\includegraphics[width=86mm]{FigS1}
%\vspace*{-5mm}
%\caption{An equilateral triangular prism may be identified with a planar three-leg ladder with
%         the periodic boundary condition in the rung direction (a)
%         whose first Brillouin zone consists of three parallel line segments
%         at equal intervals in two dimensions (b).}
%\label{SF:PRISM}
%\end{figure}

   It is useful to define the rotating frame $(\tilde{z},\tilde{x},\tilde{y})$ with its
$\tilde{z}$ axis pointing along each local spin direction in the classical ground state.
The spin components in the coordinate system at rest and the rotating one are related
with each other as
\begin{align}
   \left[
    \begin{array}{c}
     S_{\bm{r}_l}^z \\
     S_{\bm{r}_l}^x \\
     S_{\bm{r}_l}^y \\
    \end{array}
   \right]
  =\left[
    \begin{array}{ccc}
     \cos\bm{Q}_{\mathrm{X},\mathrm{K}}\cdot\bm{r}_l &
    -\sin\bm{Q}_{\mathrm{X},\mathrm{K}}\cdot\bm{r}_l &
     0                                               \\
     \sin\bm{Q}_{\mathrm{X},\mathrm{K}}\cdot\bm{r}_l &
     \cos\bm{Q}_{\mathrm{X},\mathrm{K}}\cdot\bm{r}_l &
     0                                               \\
     0                                               &
     0                                               &
     1                                               \\
    \end{array}
   \right]
   \left[
    \begin{array}{c}
     S_{\bm{r}_l}^{\tilde{z}} \\
     S_{\bm{r}_l}^{\tilde{x}} \\
     S_{\bm{r}_l}^{\tilde{y}} \\
    \end{array}
   \right],
   \tag{2}
   \label{E:Trans}
\end{align}
Let us introduce Holstein-Primakoff bosons in the rotating frame as
\begin{align}
   &
   S_{\bm{r}_l}^{\tilde{z}}
  =S-a_{\bm{r}_l}^\dagger a_{\bm{r}_l},
   \nonumber
   \allowdisplaybreaks \\
   &
   S_{\bm{r}_l}^{\tilde{x}}+iS_{\bm{r}_l}^{\tilde{y}}
  \equiv
   S_{\bm{r}_l}^{\tilde{+}}
  =\sqrt{2S}\mathcal{R}_{\bm{r}_l}(S)
   a_{\bm{r}_l};
   \nonumber
   \allowdisplaybreaks \\
   &
   S_{\bm{r}_l}^{\tilde{x}}-iS_{\bm{r}_l}^{\tilde{y}}
  \equiv
   S_{\bm{r}_l}^{\tilde{-}}
  =\sqrt{2S} a_{\bm{r}_l}^\dagger
   \mathcal{R}_{\bm{r}_l}(S),
   \nonumber
   \allowdisplaybreaks \\
   &
   \mathcal{R}_{\bm{r}_l}(S)
  \equiv
   \sqrt{1-\frac{a_{\bm{r}_l}^\dagger a_{\bm{r}_l}}{2S}}
  =1
  -\sum_{l=1}^\infty
   \frac{(2l-3)!!}{l!}
   \left(
    \frac{a_{\bm{r}_l}^\dagger a_{\bm{r}_l}}{4S}
   \right)^l
   \tag{3}
   \label{E:HPT}
\end{align}
to obtain a spin-wave (SW) Hamiltonian,
\begin{align}
   \mathcal{H}
  =J
   \sum_{l=1}^{L}
   \sum_{\sigma=\parallel,\perp}
   \bm{S}_{\bm{r}_l}\cdot\bm{S}_{\bm{r}_l+\bm{\delta}_{\sigma}}
  =\sum_{m=0}^{4}
   \mathcal{H}^{\left(\frac{m}{2}\right)}
  +O\left(S^{-\frac{1}{2}}\right),
   \label{SE:Hexpanded}
\end{align}
where $\mathcal{H}^{\left(\frac{m}{2}\right)}$, on the order of $S^{\frac{m}{2}}$, read
\begin{align}
   &\!\!
   \mathcal{H}^{(2)}
 =-\frac{3}{2}JS^{2}L,
   \label{SE:H(2)}
   \allowdisplaybreaks \\
   &\!\!
   \mathcal{H}^{\left(\frac{3}{2}\right)}
  =0,
   \label{SE:H(3/2)}
   \allowdisplaybreaks \\
   &\!\!
   \mathcal{H}^{(1)}
 =-JS
   \sum_{l=1}^{L}
   \sum_{\sigma=\parallel,\perp}
   \left[
    \left(
     a_{\bm{r}_l}^\dagger
     a_{\bm{r}_l}
    +a_{\bm{r}_l+\bm{\delta}_\sigma}^\dagger
     a_{\bm{r}_l+\bm{\delta}_\sigma}
    \right)
    \cos\bm{Q}_{\mathrm{X},\mathrm{K}}\cdot{\bm{\delta}_\sigma}
    \vphantom{\frac{\cos\bm{Q}_{\mathrm{X},\mathrm{K}}\cdot{\bm{\delta}_\sigma}-1}{2}}
   \right.
   \nonumber 
   \allowdisplaybreaks \\
   &\!\!\qquad\ 
   -\frac{\cos\bm{Q}_{\mathrm{X},\mathrm{K}}\cdot{\bm{\delta}_\sigma}-1}{2}
    \left(
     a_{\bm{r}_l}^\dagger
     a_{\bm{r}_l+\bm{\delta}_\sigma}^\dagger
    +a_{\bm{r}_l}
     a_{\bm{r}_l+\bm{\delta}_\sigma}
    \right)
   \nonumber 
   \allowdisplaybreaks \\
   &\!\!\qquad\ 
   \left.
   -\frac{\cos\bm{Q}_{\mathrm{X},\mathrm{K}}\cdot{\bm{\delta}_\sigma}+1}{2}
    \left(
     a_{\bm{r}_l}^\dagger
     a_{\bm{r}_l+\bm{\delta}_\sigma}
    +a_{\bm{r}_l+\bm{\delta}_\sigma}^\dagger
     a_{\bm{r}_l}
    \right)
   \right]
   \label{SE:H(1)}
   \allowdisplaybreaks \\
   &\!\!
   \mathcal{H}^{\left(\frac{1}{2}\right)}
  =J\sqrt{\frac{S}{2}}
   \sum_{l=1}^{L}
   \sum_{\sigma=\parallel,\perp}
   \left[
    a_{\bm{r}_l}^\dagger
    a_{\bm{r}_l}
    \left(
     a_{{\bm{r}_l}+\bm{\delta}_\sigma}^\dagger
    +a_{{\bm{r}_l}+\bm{\delta}_\sigma}
    \right) 
   \right.
   \nonumber
   \allowdisplaybreaks \\
   &\!\!\qquad\ 
   \left.
   -\left(
     a_{\bm{r}_l}^\dagger
    +a_{\bm{r}_l}
    \right)
    a_{{\bm{r}_l}+\bm{\delta}_\sigma}^\dagger
    a_{{\bm{r}_l}+\bm{\delta}_\sigma}
   \right]
   \sin\bm{Q}_{\mathrm{X},\mathrm{K}}\cdot{\bm{\delta}_\sigma},
   \label{SE:H(1/2)}
   \allowdisplaybreaks \\
   &\!\!
   \mathcal{H}^{(0)}
  =J
   \sum_{l=1}^{L}
   \sum_{\sigma=\parallel,\perp}
   \left\{
    \vphantom{\frac{\cos\bm{Q}_{\mathrm{X},\mathrm{K}}\cdot{\bm{\delta}_\sigma}-1}{8}}
    a_{\bm{r}_l}^\dagger
    a_{\bm{r}_l}
    a_{{\bm{r}_l}+\bm{\delta}_\sigma}^\dagger
    a_{{\bm{r}_l}+\bm{\delta}_\sigma}
    \cos\bm{Q}_{\mathrm{X},\mathrm{K}}\cdot{\bm{\delta}_\sigma}
    \right.
    \nonumber 
    \allowdisplaybreaks \\
    &\!\!\qquad\ 
    \left.
   -\frac{\cos\bm{Q}_{\mathrm{X},\mathrm{K}}\cdot{\bm{\delta}_\sigma}-1}{8}
    \left[
     a_{\bm{r}_l}^\dagger
     a_{{\bm{r}_l}+\bm{\delta}_\sigma}^\dagger
     \left(
      a_{{\bm{r}_l}}^\dagger
      a_{{\bm{r}_l}}
     +a_{{\bm{r}_l}+\bm{\delta}_\sigma}^\dagger
      a_{{\bm{r}_l}+\bm{\delta}_\sigma}
     \right)
     \right.
    \right.
    \nonumber 
    \allowdisplaybreaks \\
    &\!\!\qquad\qquad\qquad\ 
     \left.
     +\left(
       a_{{\bm{r}_l}}^\dagger
       a_{{\bm{r}_l}}
      +a_{{\bm{r}_l}+\bm{\delta}_\sigma}^\dagger
       a_{{\bm{r}_l}+\bm{\delta}_\sigma}
      \right)
      a_{\bm{r}_l}
      a_{{\bm{r}_l}+\bm{\delta}_\sigma}
     \right]
    \nonumber 
    \allowdisplaybreaks \\
    &\!\!\qquad\ 
    -\frac{\cos\bm{Q}_{\mathrm{X},\mathrm{K}}\cdot{\bm{\delta}_\sigma}+1}{8}
     \left[
      a_{\bm{r}_l}^\dagger
      \left(
       a_{{\bm{r}_l}}^\dagger
       a_{{\bm{r}_l}}
      +a_{{\bm{r}_l}+\bm{\delta}_\sigma}^\dagger
       a_{{\bm{r}_l}+\bm{\delta}_\sigma}
      \right)
      a_{{\bm{r}_l}+\bm{\delta}_\sigma}
     \right.
     \nonumber 
     \allowdisplaybreaks \\
   &\!\!\qquad\qquad\qquad\ 
   \left.
    \left.
    +a_{{\bm{r}_l}+\bm{\delta}_\sigma}^\dagger
     \left(
      a_{{\bm{r}_l}}^\dagger
      a_{{\bm{r}_l}}
     +a_{{\bm{r}_l}+\bm{\delta}_\sigma}^\dagger
      a_{{\bm{r}_l}+\bm{\delta}_\sigma}
     \right)
     a_{\bm{r}_l}
    \right]
    \vphantom{\frac{\cos\bm{Q}_{\mathrm{X},\mathrm{K}}\cdot{\bm{\delta}_\sigma}-1}{8}}
   \right\}.
   \label{SE:H(0)}
\end{align}
We divide the conventional SW (CSW) Hamiltonian into the harmonic part
$\mathcal{H}_{\mathrm{harm}}
\equiv
 \mathcal{H}^{(2)}+\mathcal{H}^{(1)}$
to yield linear SWs (LSWs) and higher-order anharmonicities
$\mathcal{V}
\equiv
 \mathcal{H}^{\left(\frac{1}{2}\right)}+\mathcal{H}^{(0)}$
consisting of cubic and quartic terms.
We treat $\mathcal{V}$ in two ways, variationally on one hand, which is usual with collinear
magnets, and perturbatively on the other hand, which is indispensable for noncollinear magnets,
to obtain what we distinguishably call interacting SWs (ISWs) and perturbed SWs (PSWs),
respectively.
Starting from diagonalizing $\mathcal{H}_{\mathrm{harm}}$ to obtain conventional LSWs (CLSWs),
we consider various manners of modifying them and further bringing them into interaction.
We introduce a wavevector running over the first Brillouin zone [Fig. 1(b)],
\begin{align}
   \bm{k}_{\nu,\sigma}
  \equiv
   (k_\nu^z,k_\sigma^x)
  =\frac{\pi}{a}
   \left(
    \frac{2\nu}{N}-1,
    \frac{2}{3}\sigma
   \right),
   \label{SE:WV}
\end{align}
to define Fourier transforms of spin operators and Holstein-Primakoff bosons,
\begin{align}
   &
   S_{\bm{k}_{\nu,\sigma}}^\lambda
  =\sum_{l=1}^L
   \frac{e^{ i\bm{k}_{\nu,\sigma}\cdot\bm{r}_l}}{\sqrt{L}}
   S_{\bm{r}_l}^\lambda,\ 
   S_{\bm{r}_l}^\lambda
  =\sum_{\sigma=0,\pm}
   \sum_{\nu=1}^N
   \frac{e^{-i\bm{k}_{\nu,\sigma}\cdot\bm{r}_l}}{\sqrt{L}}
   S_{\bm{k}_{\nu,\sigma}}^\lambda,
   \label{SE:FTspin}
   \allowdisplaybreaks \\
   &
   a_{\bm{k}_{\nu,\sigma}}^\dagger
  =\sum_{l=1}^L
   \frac{e^{ i\bm{k}_{\nu,\sigma}\cdot\bm{r}_l}}{\sqrt{L}}
   a_{\bm{r}_l}^\dagger,\ 
   a_{\bm{r}_l}^\dagger
  =\sum_{\sigma=0,\pm}
   \sum_{\nu=1}^N
   \frac{e^{-i\bm{k}_{\nu,\sigma}\cdot\bm{r}_l}}{\sqrt{L}}
   a_{\bm{k}_{\nu,\sigma}}^\dagger.
   \label{SE:FTboson}
\end{align}
At each momentum, a Bogoliubov quasiparticle is created by
\begin{align}
   \alpha_{\bm{k}_{\nu,\sigma}}^\dagger
  =u_{\bm{k}_{\nu,\sigma}}a_{ \bm{k}_{\nu,\sigma}}^\dagger
  -v_{\bm{k}_{\nu,\sigma}}a_{-\bm{k}_{\nu,\sigma}}
   \label{SE:BTboson}
\end{align}
with $u_{\bm{k}_{\nu,\sigma}}$ and $v_{\bm{k}_{\nu,\sigma}}$ diagonalizing
$\mathcal{H}_{\mathrm{harm}}$ or any modified-SW (MSW) Hamiltonian in question.
Here we perform all calculations at $T=0$.

\subsection{Harmonic spin waves}

   The harmonic Hamiltonian is diagonalized as
\begin{align}
   &
   \mathcal{H}_{\mathrm{harm}}
  =\sum_{m=1}^2
   E^{(m)}
  +\sum_{\sigma=0,\pm}\sum_{\nu=1}^N
   \varepsilon_{\bm{k}_{\nu,\sigma}}
   \alpha_{\bm{k}_{\nu,\sigma}}^\dagger
   \alpha_{\bm{k}_{\nu,\sigma}};
   \tag{4}
   \label{E:H(CLSW)}
   \allowdisplaybreaks \\
   &
   E^{(2)}
 =-\frac{3}{2}LJS^2,\ 
   E^{(1)}
 =-\frac{3}{2}LJS
  +\sum_{\sigma=0,\pm}\sum_{\nu=1}^N
   \frac{\varepsilon_{\bm{k}_{\nu,\sigma}}}{2},
   \nonumber
   \allowdisplaybreaks \\
   &
   \varepsilon_{\left(k_\nu^z,k_0^x\right)}
  =JS\sqrt{2(1-\cos{ak_\nu^z})(5+2\cos{ak_\nu^z})},
   \nonumber
   \allowdisplaybreaks \\
   &
   \varepsilon_{\left(k_\nu^z,k_\pm^x\right)}
  =JS\sqrt{2(1+\cos{ak_\nu^z})\left(\frac{7}{2}-2\cos{ak_\nu^z}\right)}
   \tag{5}
   \label{E:ekCLSW}
\end{align}
via the Bogoliubov transformation
\begin{align}
   &
   u_{\left(k_\nu^z,k_0^x\right)}
  =\sqrt{\frac{2\varepsilon_{\left(k_\nu^z,k_0^x\right)}+7JS}
              {4\varepsilon_{\left(k_\nu^z,k_0^x\right)}}
        },
   \nonumber
   \allowdisplaybreaks \\
   &
   v_{\left(k_\nu^z,k_0^x\right)}
  =\frac{JS\left(4\cos{ak_\nu^z}+3\right)}
        {\sqrt{4\varepsilon_{\left(k_\nu^z,k_0^x\right)}
               \left(
                2\varepsilon_{\left(k_\nu^z,k_0^x\right)}+7JS
               \right)
              }
        },
   \nonumber
   \allowdisplaybreaks \\[-3mm]
   &
   \nonumber
   \allowdisplaybreaks \\
   &
   u_{\left(k_\nu^z,k_\pm^x\right)}
  =\sqrt{\frac{4\varepsilon_{\left(k_\nu^z,k_\pm^x\right)}+11JS}
              {8\varepsilon_{\left(k_\nu^z,k_\pm^x\right)}}
        },
   \nonumber
   \allowdisplaybreaks \\
   &
   v_{\left(k_\nu^z,k_\pm^x\right)}
  =\frac{JS\left(8\cos{ak_\nu^z}-3\right)}
        {\sqrt{8\varepsilon_{\left(k_\nu^z,k_\pm^x\right)}
               \left(
                4\varepsilon_{\left(k_\nu^z,k_\pm^x\right)}+11JS
               \right)
              }
        }.
   \label{SE:BT-CLSW}
\end{align}
\begin{widetext}
\noindent
Local spin expectation values in the CLSW vacuum state are given by
\begin{align}
   &
   \langle S_{\bm{r}_l}^{\tilde{z}}\rangle_0
  =S+1
  -\frac{JS}{24\pi}
   \int_{-\frac{\pi}{a}}^{\frac{\pi}{a}}
   \left(
    \frac{ 9}{\varepsilon_{\left(k^z,k_0^x\right)}}
   +\frac{11}{\varepsilon_{\left(k^z,k_\pm^x\right)}}
   \right)
   dk^z,\ 
   \langle S_{\bm{r}_l}^{\tilde{x}}\rangle_0
  =0,\ 
   \langle S_{\bm{r}_l}^{\tilde{y}}\rangle_0
  =0.
   \label{SE:SUBMCLSW}
\end{align}
The static structure factors are defined in terms of spatial spin fluctuations
$
   \delta S_{\bm{r}_l}^\lambda
  \equiv
   S_{\bm{r}_l}^\lambda
  -\langle
    S_{\bm{r}_l}^\lambda
   \rangle_0
$ as
\begin{align}
   &
   S^{\lambda\lambda'}(\bm{q})
  \equiv
   \sum_{l,l'=1}^{L}
   \frac{e^{i\bm{q}\cdot(\bm{r}_l-\bm{r}_{l'})}}{L}
   \langle
    \delta S_{\bm{r}_l   }^{\lambda }
    \delta S_{\bm{r}_{l'}}^{\lambda'}
   \rangle_0
  =\sum_{l,l'=1}^{L}
   \frac{e^{i\bm{q}\cdot(\bm{r}_l-\bm{r}_{l'})}}{L}
   \left(
    \langle
     S_{\bm{r}_l   }^{\lambda }
     S_{\bm{r}_{l'}}^{\lambda'}
    \rangle_0
   -\langle
     S_{\bm{r}_l}^\lambda
    \rangle_0
    \langle
     S_{\bm{r}_{l'}}^{\lambda'}
    \rangle_0
   \right)
   \label{SE:SSF}
\end{align}
and written in the rotating frame as
\begin{align}
   &
   S^{\tilde{z}\tilde{z}}(\bm{q})
  =\frac{(JS)^{2}}{96\pi}
   \int_{-\frac{\pi}{a}}^{\frac{\pi}{a}}
   \left\{
     \frac{
      42+7\cos{aq^x}+(3+4\cos{ak^z})
      \left[3\cos{aq^x}+4\cos{a(k^z+q^z)}\right]
          }
     {\varepsilon_{(k^z,k_0^x)}\varepsilon_{(k^z+q^z,k_0^x+q^x)}}
     \right.
     \nonumber
     \allowdisplaybreaks \\
     &\qquad\qquad
     \left.
    +\frac{
      132-11\cos{aq^x}+(6-16\cos{ak^z})
      \left[3\cos{aq^x}-8\cos{a(k^z+q^z)}\right]
          }
     {\varepsilon_{(k^z,k_\pm^x)}\varepsilon_{(k^z+q^z,k_\pm^x+q^x)}}
   \right\}
   dk^z
  -\frac{1}{4},
   \nonumber
   \allowdisplaybreaks \\
   &
   S^{\tilde{x}\tilde{x}}(\bm{q})
  =\frac{JS(3+2\cos{aq^x}+2\cos{aq^z})}{\varepsilon_{(q^z,q^x)}}
   \left[
    \frac{S}{2}+\frac{1}{4}
   -\frac{JS}{48\pi}
    \int_{-\frac{\pi}{a}}^{\frac{\pi}{a}}
    \left(
     \frac{13+ 8\cos{ak^z}}{\varepsilon_{(k^z,k_0^x)}}
    +\frac{ 5+16\cos{ak^z}}{\varepsilon_{(k^z,k_{\pm}^x)}}
    \right)
    dk^z
   \right]
   +O(S^{-1}),
   \nonumber
   \allowdisplaybreaks \\
   &
   S^{\tilde{y}\tilde{y}}(\bm{q})
  =\frac{JS(6-\cos{aq^x}-4\cos{aq^z})}{2\varepsilon_{(q^z,q^x)}}
   \left[
    \frac{S}{2}+\frac{1}{4}
   -\frac{JS}{48\pi}
    \int_{-\frac{\pi}{a}}^{\frac{\pi}{a}}
    \left(
     \frac{ 1- 4\cos{ak^z}}{\varepsilon_{(k^z,k_0^x)}}
    +\frac{17-16\cos{ak^z}}{\varepsilon_{(k^z,k_{\pm}^x)}}
    \right)
    dk^z
   \right]
   +O(S^{-1}),
   \nonumber
   \allowdisplaybreaks \\
   &
   S^{\tilde{z}\tilde{x}}(\bm{q})
  =\left[S^{\tilde{x}\tilde{z}}(\bm{q})\right]^*
  =0,
   \ 
   S^{\tilde{y}\tilde{z}}(\bm{q})
  =\left[S^{\tilde{z}\tilde{y}}(\bm{q})\right]^*
  =0,
   \ 
   S^{\tilde{x}\tilde{y}}(\bm{q})
  =\left[S^{\tilde{y}\tilde{x}}(\bm{q})\right]^*
  =\frac{1}{4i}
   \left[
   \vphantom{\frac{JS}{384\pi}
             \int_{-\frac{\pi}{a}}^{\frac{\pi}{a}}
             \left(
              \frac{65+12\cos{ak^z}}{\varepsilon_{(k^z,k_0^x)}}
             +\frac{79+24\cos{ak^z}}{\varepsilon_{(k^z,k_{\pm}^x)}}
             \right)
             dk^z
            }
   -\frac{S}{2}-\frac{1}{4}
   +\frac{JS(6+\cos{aq^x})}{2\varepsilon_{(q^z,q^x)}}
   \right.
   \nonumber
   \allowdisplaybreaks \\
   &
   \left.
    \times
    \frac{JS}{384\pi}
    \int_{-\frac{\pi}{a}}^{\frac{\pi}{a}}
    \left(
     \frac{3+4\cos{ak^z}}{\varepsilon_{(k^z,k_0^x)}}
    -\frac{3-8\cos{ak^z}}{\varepsilon_{(k^z,k_{\pm}^x)}}
    \right)
    dk^z
   +\frac{JS}{384\pi}
    \int_{-\frac{\pi}{a}}^{\frac{\pi}{a}}
    \left(
     \frac{65+12\cos{ak^z}}{\varepsilon_{(k^z,k_0^x)}}
    +\frac{79+24\cos{ak^z}}{\varepsilon_{(k^z,k_{\pm}^x)}}
    \right)
    dk^z
   \right]
   +O(S^{-1}).
   \label{SE:SqLocCLSW}
\end{align}
Having in mind that
\begin{align}
   &
   S_{\bm{q}}^{z}
  =\frac{1}{2}
   \sum_{\tau=\pm}
   \left(
         S_{\tau\bm{Q}_{\mathrm{X},\mathrm{K}}+\bm{q}}^{\tilde{z}}
  -i\tau S_{\tau\bm{Q}_{\mathrm{X},\mathrm{K}}+\bm{q}}^{\tilde{x}}
   \right),
   \ 
   S_{\bm{q}}^{x}
  =\frac{1}{2}
   \sum_{\tau=\pm}
   \left(
         S_{\tau\bm{Q}_{\mathrm{X},\mathrm{K}}+\bm{q}}^{\tilde{x}}
  +i\tau S_{\tau\bm{Q}_{\mathrm{X},\mathrm{K}}+\bm{q}}^{\tilde{z}}
   \right),
   \ 
   S_{\bm{q}}^{y}
  =S_{\bm{q}}^{\tilde{y}}
   \label{SE:LocS(q)toLabS(q)}
\end{align}
together with
$
   \left\langle
    \delta S_{ \bm{q} }^{\lambda}
    \delta S_{-\bm{q}'}^{\lambda'}
   \right\rangle_0
  =\delta_{\bm{q}\bm{q}'}
   S^{\lambda\lambda'}(\bm{q})
$,
we can readily construct the structure factors in the laboratory frame from (\ref{SE:SqLocCLSW}),
\begin{align}
   \!\!
   S^{zz}(\bm{q})=S^{xx}(\bm{q})
  =\frac{1}{4}
   \sum_{\tau=\pm}
   \left\{
    S^{\tilde{z}\tilde{z}}(\tau\bm{Q}_{\mathrm{X},\mathrm{K}}+\bm{q})
   +S^{\tilde{x}\tilde{x}}(\tau\bm{Q}_{\mathrm{X},\mathrm{K}}+\bm{q})
   +i\tau
   \left[
    S^{\tilde{z}\tilde{x}}(\tau\bm{Q}_{\mathrm{X},\mathrm{K}}+\bm{q})
   -S^{\tilde{x}\tilde{z}}(\tau\bm{Q}_{\mathrm{X},\mathrm{K}}+\bm{q})
   \right]
   \right\},
   \ 
   S^{yy}(\bm{q})
  =S^{\tilde{y}\tilde{y}}(\bm{q}).
   \label{E:TransSq}
\end{align}
Since the CLSW spectrum (\ref{E:ekCLSW}) contain fictitious soft modes of linear dispersion,
one-dimensional integrals in (\ref{SE:SUBMCLSW}) and (\ref{SE:SqLocCLSW}) logarithmically diverge,
not to mention any such calculations in the laboratory frame.
\end{widetext}

\subsection{Single-constraint modification}

   MSWs constrained to keep their total staggered magnetization zero, which were originally
designed for square-lattice antiferromagnets \cite{T1524,T2494,H4769,T5000} and are
good at describing collinear antiferromagnets \cite{Y064426,Y074703} in general,
can be applied to noncollinear antiferromagnets \cite{Y065004} as well.
The single-constraint (SC)-modified-LSW (MLSW) Hamiltonian is diagonalized as
\vspace{-5mm}
\begin{align}
   &\!\!\!
   \mathcal{H}_{\mathrm{harm}}
  +\sum_{l=1}^{L}
   \mu_l(S-a_{\bm{r}_l}^\dagger a_{\bm{r}_l})
  \equiv
   \widetilde{\mathcal{H}}_{\mathrm{harm}}^{\mathrm{SC}}
   \nonumber 
   \allowdisplaybreaks \\
   &\!\!\!
  =\sum_{m=1}^2 E^{(m)}
  +\mu L\left(S+\frac{1}{2}\right)
  +\sum_{\sigma=0,\pm}\sum_{\nu=1}^N
   \tilde{\varepsilon}_{\bm{k}_{\nu,\sigma}}^{\mathrm{SC}}
   \alpha_{\bm{k}_{\nu,\sigma}}^\dagger
   \alpha_{\bm{k}_{\nu,\sigma}};
   \tag{6}
   \label{E:H(SC-MLSW)}
   \allowdisplaybreaks \\
   &\!\!\!
   E^{(2)}
 =-\frac{3}{2}LJS^2,\ 
   E^{(1)}
 =-\frac{3}{2}LJS
  +\sum_{\sigma=0,\pm}\sum_{\nu=1}^N
   \frac{\tilde{\varepsilon}_{\bm{k}_{\nu,\sigma}}^{\mathrm{SC}}}{2},
   \nonumber
   \allowdisplaybreaks \\
   &\!\!\!
   \frac{\tilde{\varepsilon}_{\left(k_\nu^z,k_0^x\right)}^{\mathrm{SC}}}{JS}\!
  =\sqrt{2\left(1- \cos{ak_\nu^z}-\frac{\mu}{2JS}\right)
          \left(5+2\cos{ak_\nu^z}-\frac{\mu}{ JS}\right)},
   \nonumber
   \allowdisplaybreaks \\
   &\!\!\!
   \frac{\tilde{\varepsilon}_{\left(k_\nu^z,k_\pm^x\right)}^{\mathrm{SC}}}{JS}\!
  =\sqrt{2\left(          1+ \cos{ak_\nu^z}-\frac{\mu}{2JS}\right)
          \left(\frac{7}{2}-2\cos{ak_\nu^z}-\frac{\mu}{ JS}\right)}
   \tag{7}
   \label{E:ekSC-MLSW}
\end{align}
via the Bogoliubov transformation
\begin{align}
   &
   u_{\left(k_\nu^z,k_0^x\right)}
  =\sqrt{\frac{2\tilde{\varepsilon}_{\left(k_\nu^z,k_0^x\right)}^{\mathrm{SC}}+7JS-2\mu}
              {4\tilde{\varepsilon}_{\left(k_\nu^z,k_0^x\right)}^{\mathrm{SC}}}
        },
   \nonumber
   \allowdisplaybreaks \\
   &
   v_{\left(k_\nu^z,k_0^x\right)}
  =\frac{JS\left(4\cos{ak_\nu^z}+3\right)}
        {\sqrt{4\tilde{\varepsilon}_{\left(k_\nu^z,k_0^x\right)}^{\mathrm{SC}}
               \left(
                2\tilde{\varepsilon}_{\left(k_\nu^z,k_0^x\right)}^{\mathrm{SC}}+7JS-2\mu
               \right)
              }
        },
   \nonumber
   \allowdisplaybreaks \\
   &
   u_{\left(k_\nu^z,k_\pm^x\right)}
  =\sqrt{\frac{4\tilde{\varepsilon}_{\left(k_\nu^z,k_\pm^x\right)}^{\mathrm{SC}}+11JS-4\mu}
              {8\tilde{\varepsilon}_{\left(k_\nu^z,k_\pm^x\right)}^{\mathrm{SC}}}
        },
   \nonumber
   \allowdisplaybreaks \\[-3mm]
   &
   \nonumber 
   \allowdisplaybreaks \\
   &
   v_{\left(k_\nu^z,k_\pm^x\right)}
  =\frac{JS\left(8\cos{ak_\nu^z}-3\right)}
        {\sqrt{8\tilde{\varepsilon}_{\left(k_\nu^z,k_\pm^x\right)}^{\mathrm{SC}}
               \left(
                4\tilde{\varepsilon}_{\left(k_\nu^z,k_\pm^x\right)}^{\mathrm{SC}}+11JS-4\mu
               \right)
              }
        }
   \label{SE:BT-SC-MLSW}
\end{align}
on condition that
\begin{align}
   \langle
    S_{\bm{r}_l}^{\tilde{z}}
   \rangle_0^{\mathrm{SC}}
  =S
  -\langle
    a_{\bm{r}_l}^\dagger
    a_{\bm{r}_l}
   \rangle_0^{\mathrm{SC}}
  =0
   \label{SE:SC-MLSW}
\end{align}
with
$\langle \cdots \rangle_0^{\mathrm{SC}}$ denoting a quantum average in the SC-MLSW ground state.

   We can bring such MLSWs into interaction in a variational manner \cite{N034714,Y094412}.
Denoting some quadratic terms by
\begin{align}
   &
   \mathcal{A}
  \equiv
   a_{\bm{r}_l}^\dagger a_{\bm{r}_l},\ 
   \mathcal{B}_\sigma
  \equiv
   \frac{a_{\bm{r}_l}^\dagger
         a_{\bm{r}_l+\bm{\delta}_\sigma}^\dagger
        +a_{\bm{r}_l}
         a_{\bm{r}_l+\bm{\delta}_\sigma}}
        {2},
   \nonumber
   \allowdisplaybreaks \\
   &
   \mathcal{C}
  \equiv
   \frac{a_{\bm{r}_l}^\dagger
         a_{\bm{r}_l}^\dagger
        +a_{\bm{r}_l}
         a_{\bm{r}_l}}
        {2},\ 
   \mathcal{D}_\sigma
  \equiv
   \frac{a_{\bm{r}_l}^\dagger
         a_{\bm{r}_l+\bm{\delta}_\sigma}
        +a_{\bm{r}_l+\bm{\delta}_\sigma}^\dagger
         a_{\bm{r}_l}}
        {2}
   \label{SE:WICKMF}
\end{align}
and their thermal averages at temperature $T$ with respect to a certain MSW Hamiltonian by
$\langle\cdots\rangle_T$,
we apply the Hartree-Fock decomposition to the $O(S^0)$ quartic interaction (\ref{SE:H(0)})
\begin{widetext}
\vspace{-5mm}
\begin{align}
   &
   \mathcal{H}^{(0)}
  \simeq
   J
   \sum_{l=1}^{L}
   \sum_{\sigma=\parallel,\perp}
   \left\{
    \left[
     \left(
      \langle\mathcal{A}\rangle_T
     -\frac{\langle\mathcal{B}_\sigma\rangle_T
           +\langle\mathcal{D}_\sigma\rangle_T}{2}
     \right)
     \cos\bm{Q}_{\mathrm{X},\mathrm{K}}\cdot{\bm{\delta}_\sigma}
   \right.
   \left.
    +\frac{\langle\mathcal{B}_\sigma\rangle_T
          -\langle\mathcal{D}_\sigma\rangle_T}{2}
    \right]
   \right.
    \left(
     a_{\bm{r}_{l}}^\dagger
     a_{\bm{r}_{l}}
    +a_{\bm{r}_l+\bm{\delta}_{\sigma}}^\dagger
     a_{\bm{r}_l+\bm{\delta}_{\sigma}}
    \right)
   \nonumber
   \allowdisplaybreaks \\
   &\qquad
   +\left[
     \vphantom{ \frac{2\langle\mathcal{A}\rangle_T
           + \langle\mathcal{C}\rangle_T}{4}}
     \left(
      \langle\mathcal{B}_\sigma\rangle_T
     -\frac{2\langle\mathcal{A}\rangle_T
           + \langle\mathcal{C}\rangle_T}{4}
     \right)
     \cos\bm{Q}_{\mathrm{X},\mathrm{K}}\cdot{\bm{\delta}_\sigma}
    +\frac{2\langle\mathcal{A}\rangle_T
          - \langle\mathcal{C}\rangle_T}{4}
    \right]
    \left(
     a_{\bm{r}_{l}}^\dagger
     a_{\bm{r}_l+\bm{\delta}_{\sigma}}^\dagger
    +a_{\bm{r}_{l}}
     a_{\bm{r}_l+\bm{\delta}_{\sigma}}
    \right)
   \nonumber
   \allowdisplaybreaks \\
   &\qquad
   +\left[
     \left(
      \langle\mathcal{D}_\sigma\rangle_T
     -\frac{2\langle\mathcal{A}\rangle_T
     +\langle\mathcal{C}\rangle_T}{4}
     \right)
     \cos\bm{Q}_{\mathrm{X},\mathrm{K}}\cdot{\bm{\delta}_\sigma}
    -\frac{2\langle\mathcal{A}\rangle_T
    -\langle\mathcal{C}\rangle_T}{4}
    \right]
    \left(
     a_{\bm{r}_{l}}^\dagger
     a_{\bm{r}_l+\bm{\delta}_{\sigma}}
    +a_{\bm{r}_l+\bm{\delta}_{\sigma}}^\dagger
     a_{\bm{r}_{l}}
    \right)
   \nonumber
   \allowdisplaybreaks \\
   &\qquad
   -\left[
     \frac{\langle\mathcal{B}_\sigma\rangle_T
    +\langle\mathcal{D}_\sigma\rangle_T}{8}
     \cos\bm{Q}_{\mathrm{X},\mathrm{K}}\cdot{\bm{\delta}_\sigma}
    +\frac{\langle\mathcal{B}_\sigma\rangle_T
    -\langle\mathcal{D}_\sigma\rangle_T}{8}
    \right]
    \left(
     a_{\bm{r}_{l}}^\dagger
     a_{\bm{r}_{l}}^\dagger
    +a_{\bm{r}_l+\bm{\delta}_{\sigma}}^\dagger
     a_{\bm{r}_l+\bm{\delta}_{\sigma}}^\dagger
    +a_{\bm{r}_{l}}
     a_{\bm{r}_{l}}
    +a_{\bm{r}_l+\bm{\delta}_{\sigma}}
     a_{\bm{r}_l+\bm{\delta}_{\sigma}}
    \right)
   \nonumber
   \allowdisplaybreaks \\
   &\qquad
   -\left[
     \vphantom{
     \frac{\langle\mathcal{B}_\sigma\rangle_T
          +\langle\mathcal{D}_\sigma\rangle_T}{2}
     }
     \left(\langle\mathcal{A}\rangle_T\right)^2
    +\left(\langle\mathcal{B}_\sigma\rangle_T\right)^2
    +\left(\langle\mathcal{D}_\sigma\rangle_T\right)^2
    -\left(
      2\langle\mathcal{A}\rangle_T
     + \langle\mathcal{C}\rangle_T
     \right)
     \frac{\langle\mathcal{B}_\sigma\rangle_T
          +\langle\mathcal{D}_\sigma\rangle_T}{2}
    \right]
     \cos\bm{Q}_{\mathrm{X},\mathrm{K}}\cdot{\bm{\delta}_\sigma}
   \nonumber
   \allowdisplaybreaks \\
   &\qquad
    \left.
   -\left(
     2\langle\mathcal{A}\rangle_T
    - \langle\mathcal{C}\rangle_T
    \right)
    \frac{\langle\mathcal{B}_\sigma\rangle_T
         -\langle\mathcal{D}_\sigma\rangle_T}{2}
   \right\}
  \equiv
   \mathcal{H}^{(0)}_{\mathrm{quad}}
   \label{SE:H(0)quad2D}
   \\[-7.5mm]\nonumber
\end{align}
to have the tractable SW Hamiltonian
$\mathcal{H}_{\mathrm{quad}}
=\mathcal{H}_{\mathrm{harm}}
+\mathcal{H}_{\mathrm{quad}}^{(0)}$.
Its SC-MSW extension, i.e., the SC-modified Hartree-Fock-decomposition-based-ISW (SC-MHFISW)
Hamiltonian, where we suppose $\langle\cdots\rangle_T$ in (\ref{SE:H(0)quad2D}) to be
$\langle\cdots\rangle'{}_{\!\!T}^{\mathrm{SC}}$, is diagonalized as
\begin{align}
   &
   \mathcal{H}_{\mathrm{quad}}
  +\sum_{l=1}^{L}
   \mu_l (S-a_{\bm{r}_l}^\dagger a_{\bm{r}_l})
  \equiv
   \widetilde{\mathcal{H}}'{}^{\mathrm{SC}}
  =\sum_{m=0}^2 E^{(m)}
   +\mu L\left(S+\frac{1}{2}\right)
   +\sum_{\sigma=0,\pm}\sum_{\nu=1}^N
   \tilde{\varepsilon}'{}_{\!\!\bm{k}_{\nu,\sigma}}^{\mathrm{SC}}
   \alpha_{\bm{k}_{\nu,\sigma}}^\dagger
   \alpha_{\bm{k}_{\nu,\sigma}};
   \label{SE:DH(SC-MWDISW)}
   \allowdisplaybreaks \\
   &
   E^{(2)}
 =-\frac{3}{2}LJS^2,\ 
   E^{(1)}
 =-\frac{3}{2}LJS
  +\sum_{\sigma=0,\pm}\sum_{\nu=1}^N
   \frac{\tilde{\varepsilon}'{}_{\!\!\bm{k}_{\nu,\sigma}}^{\mathrm{SC}}}{2},\ 
   E^{(0)}
  =\langle\mathcal{H}^{(0)}\rangle'{}_{\!\!T}^{\mathrm{SC}}
 =-JL
   \nonumber
   \allowdisplaybreaks \\
   &\times
   \sum_{\sigma=\parallel,\perp}
   \left\{
    \left[
     \vphantom{\frac{\langle\mathcal{B}_\sigma\rangle'{}_{\!\!T}^{\mathrm{SC}}
                    +\langle\mathcal{D}_\sigma\rangle'{}_{\!\!T}^{\mathrm{SC}}}{2}}
     \left(\langle\mathcal{A}\rangle'{}_{\!\!T}^{\mathrm{SC}}\right)^2
    +\left(\langle\mathcal{B}_\sigma\rangle'{}_{\!\!T}^{\mathrm{SC}}\right)^2
    +\left(\langle\mathcal{D}_\sigma\rangle'{}_{\!\!T}^{\mathrm{SC}}\right)^2
    -\left(
      2\langle\mathcal{A}\rangle'{}_{\!\!T}^{\mathrm{SC}}
     + \langle\mathcal{C}\rangle'{}_{\!\!T}^{\mathrm{SC}}
     \right)
     \frac{\langle\mathcal{B}_\sigma\rangle'{}_{\!\!T}^{\mathrm{SC}}
          +\langle\mathcal{D}_\sigma\rangle'{}_{\!\!T}^{\mathrm{SC}}}{2}
    \right]
    \cos\bm{Q}_{\mathrm{X},\mathrm{K}}\cdot{\bm{\delta}_\sigma}
    \right.
    \nonumber
    \allowdisplaybreaks \\
    &
    \left.
    -\left(
      2\langle\mathcal{A}\rangle'{}_{\!\!T}^{\mathrm{SC}}
     - \langle\mathcal{C}\rangle'{}_{\!\!T}^{\mathrm{SC}}
     \right)
     \frac{\langle\mathcal{B}_\sigma\rangle'{}_{\!\!T}^{\mathrm{SC}}
          -\langle\mathcal{D}_\sigma\rangle'{}_{\!\!T}^{\mathrm{SC}}}{2}
   -\left(
     \frac{\langle\mathcal{A}\rangle'{}_{\!\!T}^{\mathrm{SC}}}{2}
    -\frac{\langle\mathcal{B}_\sigma\rangle'{}_{\!\!T}^{\mathrm{SC}}
           +\langle\mathcal{D}_\sigma\rangle'{}_{\!\!T}^{\mathrm{SC}}}{4}
    \right)
    \cos\bm{Q}_{\mathrm{X},\mathrm{K}}\cdot{\bm{\delta}_\sigma}
   -\frac{\langle\mathcal{B}_\sigma\rangle'{}_{\!\!T}^{\mathrm{SC}}
         -\langle\mathcal{D}_\sigma\rangle'{}_{\!\!T}^{\mathrm{SC}}}{4}
   \right\},
   \label{SE:E0SC-MWDISW}
   \allowdisplaybreaks \\[4mm]
   &
   \frac{\tilde{\varepsilon}'{}_{\!\!\left(k_\nu^z,k_0^x\right)}^{\mathrm{SC}}}{JS}
  =\sqrt{2\left[
           1+\frac{F_{0}-G_{0}}{2}
            -\left(1-\frac{F_{0}'-G_{0}'}{2}\right)\cos{ak_\nu^z}
            -\frac{\mu}{2JS}
          \right]
          \left[
            5+F_{0}+G_{0}
             +\left(2+F_{0}'+G_{0}'\right)\cos{ak_\nu^z}
             -\frac{\mu}{JS}
          \right]
        },
   \nonumber
   \allowdisplaybreaks \\
   &
   \frac{\tilde{\varepsilon}'{}_{\!\!\left(k_\nu^z,k_\pm^x\right)}^{\mathrm{SC}}}{JS}
  =\sqrt{2\left[
            1+\frac{F_{\pm}-G_{\pm}}{2}
             +\left(1+\frac{F_{\pm}'-G_{\pm}'}{2}\right)\cos{ak_\nu^z}
             -\frac{\mu}{2JS}
          \right]
          \left[
            \frac{7}{2}+F_{\pm}+G_{\pm}
           -\left(2-F_{\pm}'-G_{\pm}'\right)\cos{ak_\nu^z}-\frac{\mu}{JS}
          \right]
        },
   \label{SE:ekSC-MWDISW}
   \allowdisplaybreaks \\[4mm]
   &
   F_{0}
  =-\frac{7\langle \mathcal{A} \rangle'{}_{\!\!T}^{\mathrm{SC}}}{2S}
   +\frac{3\langle \mathcal{C} \rangle'{}_{\!\!T}^{\mathrm{SC}}}{4S}
   +\frac{3\langle \mathcal{B}_\perp \rangle'{}_{\!\!T}^{\mathrm{SC}}}{2S}
   +\frac{2\langle \mathcal{B}_\parallel \rangle'{}_{\!\!T}^{\mathrm{SC}}}{S}
   -\frac{3\langle \mathcal{D}_\perp \rangle'{}_{\!\!T}^{\mathrm{SC}}}{2S},\ 
   F_{0}'
  =\frac{\langle \mathcal{C} \rangle'{}_{\!\!T}^{\mathrm{SC}}
       -2\langle \mathcal{D}_\parallel \rangle'{}_{\!\!T}^{\mathrm{SC}}}{S},
   \nonumber
   \allowdisplaybreaks \\
   &
   G_{0}
  = \frac{3\langle \mathcal{A} \rangle'{}_{\!\!T}^{\mathrm{SC}}}{2S}
   -\frac{ \langle \mathcal{B}_\perp \rangle'{}_{\!\!T}^{\mathrm{SC}}}{S}
   -\frac{ \langle \mathcal{C} \rangle'{}_{\!\!T}^{\mathrm{SC}}}{2S}
   +\frac{3\langle \mathcal{D}_\perp \rangle'{}_{\!\!T}^{\mathrm{SC}}}{4S}
   +\frac{ \langle \mathcal{D}_\parallel \rangle'{}_{\!\!T}^{\mathrm{SC}}}{S},\ 
   G_{0}'
  =\frac{2\left(\langle \mathcal{A} \rangle'{}_{\!\!T}^{\mathrm{SC}}
         -\langle \mathcal{B}_\parallel \rangle'{}_{\!\!T}^{\mathrm{SC}}\right)}{S},
   \nonumber
   \allowdisplaybreaks \\
   &
   F_{\pm}
  =-\frac{11\langle \mathcal{A} \rangle'{}_{\!\!T}^{\mathrm{SC}}}{4S}
   +\frac{ 3\langle \mathcal{B}_\perp \rangle'{}_{\!\!T}^{\mathrm{SC}}}{2S}
   +\frac{ 2\langle \mathcal{B}_\parallel \rangle'{}_{\!\!T}^{\mathrm{SC}}}{S}
   -\frac{ 3\langle \mathcal{C} \rangle'{}_{\!\!T}^{\mathrm{SC}}}{8S},\ 
   F_{\pm}'
  =\frac{2\left(\langle \mathcal{C} \rangle'{}_{\!\!T}^{\mathrm{SC}}
               -\langle \mathcal{D}_\parallel \rangle'{}_{\!\!T}^{\mathrm{SC}}
          \right)}{S},
   \nonumber
   \allowdisplaybreaks \\
   &
   G_{\pm}
  =-\frac{3\langle \mathcal{A} \rangle'{}_{\!\!T}^{\mathrm{SC}}}{4S}
   +\frac{ \langle \mathcal{B}_\perp \rangle'{}_{\!\!T}^{\mathrm{SC}}}{2S}
   +\frac{ \langle \mathcal{C} \rangle'{}_{\!\!T}^{\mathrm{SC}}}{8S}
   +\frac{3\langle \mathcal{D}_\perp \rangle'{}_{\!\!T}^{\mathrm{SC}}}{4S}
   +\frac{ \langle \mathcal{D}_\parallel \rangle'{}_{\!\!T}^{\mathrm{SC}}}{S},\ 
   G_{\pm}'
  =\frac{2\left(\langle \mathcal{A} \rangle'{}_{\!\!T}^{\mathrm{SC}}
               -\langle \mathcal{B}_\parallel \rangle'{}_{\!\!T}^{\mathrm{SC}}
          \right)}{S}
   \label{SE:WDORDER2}
\end{align}
via the Bogoliubov transformation
\begin{align}
   &
   u_{\left(k_\nu^z,k_0^x\right)}
  =\sqrt{\frac{2\tilde{\varepsilon}'{}_{\!\!\left(k_\nu^z,k_0^x\right)}^{\mathrm{SC}}
               +JS\left(7+2F_0+2F_0'\cos{ak_\nu^z}\right)-2\mu}
              {4\tilde{\varepsilon}'{}_{\!\!\left(k_\nu^z,k_0^x\right)}^{\mathrm{SC}}}
        },
   \nonumber
   \allowdisplaybreaks \\
   &
   v_{\left(k_\nu^z,k_0^x\right)}
  =\frac{JS\left(4\cos{ak_\nu^z}-2G_0'\cos{ak_\nu^z}+3-2G_0\right)}
   {\sqrt{
    4\tilde{\varepsilon}'{}_{\!\!\left(k_\nu^z,k_0^x\right)}^{\mathrm{SC}}
   \left[
    2\tilde{\varepsilon}'{}_{\!\!\left(k_\nu^z,k_0^x\right)}^{\mathrm{SC}}
    +JS\left(7+2F_0+2F_0'\cos{ak_\nu^z}\right)-2\mu
   \right]
         }
   },
   \nonumber 
   \allowdisplaybreaks \\[2mm]
   &
   u_{\left(k_\nu^z,k_\pm^x\right)}
  =\sqrt{\frac{4\tilde{\varepsilon}'{}_{\!\!\left(k_\nu^z,k_\pm^x\right)}^{\mathrm{SC}}
               +JS\left(11+4F_0+4F_0'\cos{ak_\nu^z}\right)-4\mu}
              {8\tilde{\varepsilon}'{}_{\!\!\left(k_\nu^z,k_\pm^x\right)}^{\mathrm{SC}}}
        },
   \nonumber 
   \allowdisplaybreaks \\
   &
   v_{\left(k_\nu^z,k_\pm^x\right)}
  =\frac{JS\left(8\cos{ak_\nu^z}-4G_0'\cos{ak_\nu^z}+3-2G_0\right)}
   {\sqrt{8\tilde{\varepsilon}'{}_{\!\!\left(k_\nu^z,k_\pm^x\right)}^{\mathrm{SC}}
   \left[
    4\tilde{\varepsilon}'{}_{\!\!\left(k_\nu^z,k_\pm^x\right)}^{\mathrm{SC}}
    +JS\left(11+4F_0+4F_0'\cos{ak_\nu^z}\right)-4\mu
   \right]
         }
   }
   \label{SE:BT-SC-MWDISW}
\end{align}
on condition that
\begin{align}
   S
  -\langle
    \mathcal{A}
   \rangle'{}_{\!\!T}^{\mathrm{SC}}
  =S
  -\langle
    a_{\bm{r}_l}^\dagger
    a_{\bm{r}_l}
   \rangle'{}_{\!\!T}^{\mathrm{SC}}
  =0
   \label{SE:SC-MHFISW}
\end{align}
with
$
   \langle
    \cdots
   \rangle'{}_{\!\!T}^{\mathrm{SC}}
$
denoting a thermal average at temperature $T$ with respect to
$\widetilde{\mathcal{H}}'{}^{\mathrm{SC}}$.
In terms of the thermal distribution function
\begin{align}
   \langle
    \alpha_{\bm{k}_{\nu,\sigma}}^\dagger \alpha_{\bm{k}_{\nu,\sigma}}
   \rangle'_T{\vphantom{\rangle}}^{\!\!\mathrm{SC}}
  =\frac
   {\mathrm{Tr}
    \left[
     e^{-\tilde{\varepsilon}'{}_{\!\!\!\!\bm{k}_{\nu,\sigma}}^{\mathrm{SC}}
        \alpha_{\bm{k}_{\nu,\sigma}}^\dagger \alpha_{\bm{k}_{\nu,\sigma}}
       /k_{\mathrm{B}}T}
     \alpha_{\bm{k}_{\nu,\sigma}}^\dagger \alpha_{\bm{k}_{\nu,\sigma}}
    \right]}
   {\mathrm{Tr}
    \left[
     e^{-\tilde{\varepsilon}'{}_{\!\!\!\!\bm{k}_{\nu,\sigma}}^{\mathrm{SC}}
         \alpha_{\bm{k}_{\nu,\sigma}}^\dagger \alpha_{\bm{k}_{\nu,\sigma}}
       /k_{\mathrm{B}}T}
    \right]}
  =\frac{1}
   {e^{\tilde{\varepsilon}'{}_{\!\!\!\!\bm{k}_{\nu,\sigma}}^{\mathrm{SC}}/k_{\mathrm{B}}T}-1}
  \equiv
   \tilde{n}'{}_{\!\!\bm{k}_{\nu,\sigma}}^{\mathrm{SC}},
   \label{E:ThAvSC-MHFISW}
\end{align}
such self-consistent fields as to diagonalize $\widetilde{\mathcal{H}}'{}^{\mathrm{SC}}$ are
expressed as
\begin{align}
   &
   \langle \mathcal{A} \rangle'{}_{\!\!T}^{\mathrm{SC}}
  =\frac{JS}{N}
   \sum_{\nu=1}^{N}
   \left[
    \frac{ 7+2F_{0}+2F_{0}'\cos{ak_\nu^z}-2\mu/JS}
    {\tilde{\varepsilon}'{}_{\!\!\left(k_\nu^z,k_0^x\right)}^{\mathrm{SC}}}
    \left(
     \tilde{n}'{}_{\!\!(k_\nu^z,k_0^x)}^{\mathrm{SC}}+\frac{1}{2}
    \right)
   +\frac{11+4F_{\pm}+4F_{\pm}'\cos{ak_\nu^z}-4\mu/JS}
    {\tilde{\varepsilon}'{}_{\!\!\left(k_\nu^z,k_\pm^x\right)}^{\mathrm{SC}}}
    \left(
     \tilde{n}'{}_{\!\!(k_\nu^z,k_\pm^x)}^{\mathrm{SC}}+\frac{1}{2}
    \right)
   -\frac{1}{2JS}
   \right],
   \nonumber
   \allowdisplaybreaks \\
   &
   \langle \mathcal{B}_{\sigma} \rangle'{}_{\!\!T}^{\mathrm{SC}}
  =\frac{JS}{N}
   \sum_{\nu=1}^{N}
   \left[
    \frac{-3+2G_{0}-2(2-G_{0}')\cos{ak_\nu^z}}
    {\tilde{\varepsilon}'{}_{\!\!\left(k_\nu^z,k_0^x\right)}^{\mathrm{SC}}}
    \left(
     \tilde{n}'{}_{\!\!(k_\nu^z,k_0^x)}^{\mathrm{SC}}+\frac{1}{2}
    \right)
    \cos{a(k_\nu^z\delta_{\sigma,\parallel}+k_0^x\delta_{\sigma,\perp})}
   +\frac{ 3+4G_{0}-4(2-G_{0}')\cos{ak_\nu^z}}
    {\tilde{\varepsilon}'{}_{\!\!\left(k_\nu^z,k_\pm^x\right)}^{\mathrm{SC}}}
   \right.
   \nonumber
   \allowdisplaybreaks \\
   &
   \left.
    \qquad\qquad\qquad\quad\times
    \left(
     \tilde{n}'{}_{\!\!(k_\nu^z,k_\pm^x)}^{\mathrm{SC}}+\frac{1}{2}
    \right)
    \cos{a(k_\nu^z\delta_{\sigma,\parallel}+k_{\pm}^x\delta_{\sigma,\perp})}
   \right],
   \nonumber
   \allowdisplaybreaks \\
   &
   \langle \mathcal{C} \rangle'{}_{\!\!T}^{\mathrm{SC}}
  =\frac{JS}{N}
   \sum_{\nu=1}^{N}
   \left[
    \frac{-3+2G_{0}-2(2-G_{0}')\cos{ak_\nu^z}}
    {\tilde{\varepsilon}'{}_{\!\!\left(k_\nu^z,k_0^x\right)}^{\mathrm{SC}}}
    \left(
     \tilde{n}'{}_{\!\!(k_\nu^z,k_0^x)}^{\mathrm{SC}}+\frac{1}{2}
    \right)
   +\frac{ 3+4G_{0}-4(2-G_{0}')\cos{ak_\nu^z}}
    {\tilde{\varepsilon}'{}_{\!\!\left(k_\nu^z,k_\pm^x\right)}^{\mathrm{SC}}}
    \left(
     \tilde{n}'{}_{\!\!(k_\nu^z,k_\pm^x)}^{\mathrm{SC}}+\frac{1}{2}
    \right)
   \right],
   \nonumber
   \allowdisplaybreaks \\
   &
   \langle \mathcal{D}_{\sigma} \rangle'{}_{\!\!T}^{\mathrm{SC}}
  =\frac{JS}{N}
   \sum_{\nu=1}^{N}
   \left[
    \frac{ 7+2F_{0}+2F_{0}'\cos{ak_\nu^z}-2\mu/JS}
    {\tilde{\varepsilon}'{}_{\!\!\left(k_\nu^z,k_0^x\right)}^{\mathrm{SC}}}
    \left(
     \tilde{n}'{}_{\!\!(k_\nu^z,k_0^x)}^{\mathrm{SC}}+\frac{1}{2}
    \right)
    \cos{a(k_\nu^z\delta_{\sigma,\parallel}+k_0^x\delta_{\sigma,\perp})}
   +\frac{11+4F_{\pm}+4F_{\pm}'\cos{ak_\nu^z}-4\mu/JS}
    {\tilde{\varepsilon}'{}_{\!\!\left(k_\nu^z,k_\pm^x\right)}^{\mathrm{SC}}}
   \right.
   \nonumber
   \allowdisplaybreaks \\
   &
   \left.
    \qquad\qquad\qquad\quad\times
    \left(
     \tilde{n}'{}_{\!\!(k_\nu^z,k_\pm^x)}^{\mathrm{SC}}+\frac{1}{2}
    \right)
    \cos{a(k_\nu^z\delta_{\sigma,\parallel}+k_{\pm}^x\delta_{\sigma,\perp})}
   -\frac{ \cos{a(k_\nu^z\delta_{\sigma,\parallel}+k_0^x\delta_{\sigma,\perp})}
         +2\cos{a(k_\nu^z\delta_{\sigma,\parallel}+k_{\pm}^x\delta_{\sigma,\perp})}
         }{2JS}
   \right].
   \label{SE:ORDERP}
\end{align}
At $T=0$, MHFISWs are no different from modified Wick-decomposition-based-interacting SWs (MWDISWs)
\cite{N034714,Y094412}.
\end{widetext}

\subsection{Double-constraint modification}

   SC MSWs in noncollinear antiferromagnets, whether MLSWs [Figs. 2(a1) and 2(a2)] or
MWDISWs [Figs. 2($\mathrm{a}1''$) and 2($\mathrm{a}2''$)], are no longer so successful as
those in collinear antiferromagnets \cite{T1524,Y094412}.
Before pointing out that they are in total ignorance of cubic anharmonicities, we may need to
solve a more fundamental problem with them \cite{Y065004}.
The Holstein-Primakoff-boson representation in the local spin reference frame \eqref{E:HPT}
gives the correct spin magnitude but breaks the original spin rotational symmetry,
\begin{align}
   &
   \langle
    S_{\bm{r}_l}^{\tilde{x}} S_{\bm{r}_l}^{\tilde{x}}
   \rangle_T
  =S\left(
     \frac{1}{2}
    +\langle \mathcal{A} \rangle_T
    +\langle \mathcal{C} \rangle_T
   \right)
   \nonumber
   \allowdisplaybreaks \\
   &\qquad
  -\left(
    \langle \mathcal{A} \rangle_T
   +\frac{\langle \mathcal{C} \rangle_T}{2}
   \right)
   \left(
    \frac{1}{2}
   +\langle \mathcal{A} \rangle_T
   +\langle \mathcal{C} \rangle_T
   \right)
  +O(S^{-1}),
   \nonumber 
   \allowdisplaybreaks \\
   &
   \langle
    S_{\bm{r}_l}^{\tilde{y}} S_{\bm{r}_l}^{\tilde{y}}
   \rangle_T
  =S\left(
     \frac{1}{2}
    +\langle \mathcal{A} \rangle_T
    -\langle \mathcal{C} \rangle_T
   \right)
   \nonumber
   \allowdisplaybreaks \\
   &\qquad
  -\left(
    \langle \mathcal{A} \rangle_T
   -\frac{\langle \mathcal{C} \rangle_T}{2}
   \right)
   \left(
    \frac{1}{2}
   +\langle \mathcal{A} \rangle_T
   -\langle \mathcal{C} \rangle_T
   \right)
  +O(S^{-1}),
   \nonumber 
   \allowdisplaybreaks \\
   &
   \langle
    S_{\bm{r}_l}^{\tilde{z}} S_{\bm{r}_l}^{\tilde{z}}
   \rangle_T
  =S^2
  -2S\langle \mathcal{A} \rangle_T
  +  \langle \mathcal{A} \rangle_T
  +2 \langle \mathcal{A} \rangle_T^2
  +  \langle \mathcal{C} \rangle_T^2;
   \nonumber 
   \allowdisplaybreaks \\
   &
   \sum_{\lambda=\tilde{x},\tilde{y},\tilde{z}}
   \langle
    S_{\bm{r}_l}^{\lambda} S_{\bm{r}_l}^{\lambda}
   \rangle_T
  =\sum_{\lambda=x,y,z}
   \langle
    S_{\bm{r}_l}^{\lambda} S_{\bm{r}_l}^{\lambda}
   \rangle_T
  =S(S+1).
   \label{SE:XXYYZZ}
\end{align}
For collinear antiferromagnets, $\langle\mathcal{C}\rangle_T$ spontaneously vanishes to retain
a $\mathrm{U}(1)$ symmetry of each local spin operator in the rotating frame as well as
the global $\mathrm{U}(1)$ symmetry related to the conservation of the total uniform magnetization
\begin{align}
   \sum_{l=1}^L
   S_{\bm{r}_l}^z
  =\sum_{l=1}^L
   \left(
    S_{\bm{r}_l}^{\tilde{z}}\cos\bm{Q}_{\mathrm{X},\mathrm{K}}\cdot\bm{r}_l
   -S_{\bm{r}_l}^{\tilde{x}}\sin\bm{Q}_{\mathrm{X},\mathrm{K}}\cdot\bm{r}_l
   \right).
   \label{E:Mz}
\end{align}
For noncollinear antiferromagnets, any bosonic Hamiltonian, even $\mathcal{H}_{\mathrm{harm}}$,
no longer commutes with
$
   \sum_{l=1}^L
   S_{\bm{r}_l}^z
$,
yet we should expect each spin operator to possess
rotational symmetry about its local quantization axis $\tilde{z}$, which is met by demanding that
$
   \langle
    S_{\bm{r}_{l}}^{\tilde{x}}S_{\bm{r}_{l}}^{\tilde{x}}
   \rangle_T
  =\langle
    S_{\bm{r}_{l}}^{\tilde{y}}S_{\bm{r}_{l}}^{\tilde{y}}
   \rangle_T
$, i.e.,
\begin{align}
  \langle\mathcal{C}\rangle_T
  =\frac{\langle
          a_{\bm{r}_l}^\dagger a_{\bm{r}_l}^\dagger
         +a_{\bm{r}_l}         a_{\bm{r}_l}
         \rangle_T}{2}
  =0.
   \label{E:noncollinearMSWconstraintHPB2nd}
\end{align}

   Constraining SC MSWs to (\ref{E:noncollinearMSWconstraintHPB2nd}) as they are, i.e.,
together with (\ref{SE:SC-MLSW}) or (\ref{SE:SC-MHFISW}), however,
causes to increase their otherwise stable ground-state energy (cf. Fig. \ref{SF:ETuning}).
In order to solve this composite problem, we diagonalize the LSW Hamiltonian subject to
a double-constraint (DC) condition.
Namely, the DC-MLSW Hamiltonian is diagonalized as
\begin{align}
   &\!\!\!
   \mathcal{H}_{\mathrm{harm}}
  +\sum_{l=1}^L
   \left[
    \mu_l
    \left(
     S-\delta-a_{\bm{r}_l}^\dagger a_{\bm{r}_l}
    \right)
   +\eta_l
    \frac{a_{\bm{r}_l}^\dagger a_{\bm{r}_l}^\dagger
         +a_{\bm{r}_l}         a_{\bm{r}_l}        }
         {2}
   \right]
  \equiv
   \widetilde{\mathcal{H}}_{\mathrm{harm}}^{\mathrm{DC}}
   \nonumber
   \allowdisplaybreaks \\
   &\!\!\!
  =\sum_{m=1}^2
   E^{(m)}
  +\mu L
   \left(
    S-\delta+\frac{1}{2}
   \right)
  +\sum_{\sigma=0,\pm}\sum_{\nu=1}^N
   \tilde{\varepsilon}_{\bm{k}_{\nu,\sigma}}^{\mathrm{DC}}
   \alpha_{\bm{k}_{\nu,\sigma}}^\dagger
   \alpha_{\bm{k}_{\nu,\sigma}};
   \tag{8}
   \label{E:H(DC-MLSW)}
   \allowdisplaybreaks \\
   &\!\!\!
   E^{(2)}
 =-\frac{3}{2}LJS^2,\ 
   E^{(1)}
 =-\frac{3}{2}LJS
  +\sum_{\sigma=0,\pm}\sum_{\nu=1}^N
   \frac{\tilde{\varepsilon}_{\bm{k}_{\nu,\sigma}}^{\mathrm{DC}}}{2},
   \nonumber
   \allowdisplaybreaks \\
   &\!\!\!
   \frac{\tilde{\varepsilon}_{\left(k_\nu^z,k_0^x  \right)}^{\mathrm{DC}}}{JS}
  =\!\!\sqrt{2\left(1\!-\! \cos{ak_\nu^z}\!-\!\frac{\mu\!-\!\eta}{2JS}\right)\!
              \left(5\!+\!2\cos{ak_\nu^z}\!-\!\frac{\mu\!+\!\eta}{ JS}\right)},
   \nonumber
   \allowdisplaybreaks \\
   &\!\!\!
   \frac{\tilde{\varepsilon}_{\left(k_\nu^z,k_\pm^x\right)}^{\mathrm{DC}}}{JS}\!
  =\!\!\sqrt{2\left(1\!+\! \cos{ak_\nu^z}\!-\!\frac{\mu\!+\!\eta}{2JS}\right)
              \left(\frac{7}{2}\!-\!2\cos{ak_\nu^z}\!-\!\frac{\mu\!-\!\eta}{ JS}\right)}
   \tag{9}
   \label{E:ekDC-MLSW}
\end{align}
via the Bogoliubov transformation
\begin{align}
   &
   u_{\left(k_\nu^z,k_0^x\right)}
  =\sqrt{\frac{2\tilde{\varepsilon}_{\left(k_\nu^z,k_0^x\right)}^{\mathrm{DC}}+7JS-2\mu}
              {4\tilde{\varepsilon}_{\left(k_\nu^z,k_0^x\right)}^{\mathrm{DC}}}
        },
   \nonumber
   \allowdisplaybreaks \\
   &
   v_{\left(k_\nu^z,k_0^x\right)}
  =\frac{JS\left(4\cos{ak_\nu^z}+3\right)-2\eta}
        {\sqrt{4\tilde{\varepsilon}_{\left(k_\nu^z,k_0^x\right)}^{\mathrm{DC}}
               \left(
                2\tilde{\varepsilon}_{\left(k_\nu^z,k_0^x\right)}^{\mathrm{DC}}+7JS-2\mu
               \right)}},
   \nonumber 
   \allowdisplaybreaks \\
   &
   u_{\left(k_\nu^z,k_\pm^x\right)}
  =\sqrt{\frac{4\tilde{\varepsilon}_{\left(k_\nu^z,k_\pm^x\right)}^{\mathrm{DC}}+11JS-4\mu}
              {8\tilde{\varepsilon}_{\left(k_\nu^z,k_\pm^x\right)}^{\mathrm{DC}}}
        },
   \nonumber 
   \allowdisplaybreaks \\
   &
   v_{\left(k_\nu^z,k_\pm^x\right)}
  =\frac{JS\left(8\cos{ak_\nu^z}-3\right)-4\eta}
        {\sqrt{8\tilde{\varepsilon}_{\left(k_\nu^z,k_\pm^x\right)}^{\mathrm{DC}}
               \left(
                4\tilde{\varepsilon}_{\left(k_\nu^z,k_\pm^x\right)}^{\mathrm{DC}}+11JS-4\mu
               \right)}}
   \label{SE:BT-DC-MLSW}
\end{align}
on condition that
\begin{align}
   &
   S
  -\langle
    a_{\bm{r}_l}^\dagger
    a_{\bm{r}_l}
   \rangle_0^{\mathrm{DC}}
  =\delta,
   \ 
   \langle
    a_{\bm{r}_l}^\dagger
    a_{\bm{r}_l}^\dagger
   +a_{\bm{r}_l}
    a_{\bm{r}_l}
   \rangle_0^{\mathrm{DC}}
  =0
   \label{SE:DC-MLSW}
\end{align}
with $\delta$ serving to optimize the ground-state energy and
$\langle \cdots \rangle_0^{\mathrm{DC}}$ denoting a quantum average in the thus-obtained
DC-MLSW ground state.
By tuning $\delta$, we can retain the local $\mathrm{U}(1)$ symmetry with the SC-MLSW ground-state
energy remaining almost the same, as will be demonstrated later in Sect. \ref{S:DCvsSC}.

\section{Green's Function Formalism}
\label{S:GFF}

   Let us introduce Feynman's time-ordered Greens functions of Bogoliubov quasiparticles and
spin operators,
\begin{align}
   &\!\!\!
   G^{\sigma\sigma'}(\bm{k};\omega+i\epsilon)
  \equiv
   -i
   \int_{-\infty}^{\infty}
   \!\!
   \left\langle
    \mathcal{T}
     \alpha_{-\sigma \bm{k}}^\sigma(t)
     \alpha_{ \sigma'\bm{k}}^{\sigma'}
    \right\rangle
   e^{i[\omega+i\epsilon\sigma(t)]t}
   dt,
   \label{SE:Gdef}
   \allowdisplaybreaks \\
   &\!\!\!
   \mathcal{G}^{\lambda\lambda'}(\bm{q};\omega+i\epsilon)
  \equiv
  -i
   \int_{-\infty}^{\infty}
   \!\!
   \left\langle
    \mathcal{T}
    \delta S_{ \bm{q}}^{\lambda}(t)
    \delta S_{-\bm{q}}^{\lambda'}
   \right\rangle
   e^{i[\omega+i\epsilon\sigma(t)]t}
   dt,
   \label{SE:SGdef}
\end{align}
where
$\mathcal{P}(t)
\equiv
 e^{ i\widetilde{\mathcal{H}}t/\hbar}
 \mathcal{P}
 e^{-i\widetilde{\mathcal{H}}t/\hbar}$
and
$\langle\mathcal{Q}\rangle$
represent time evolution driven by $\widetilde{\mathcal{H}}$ and a quantum average in the ground
state of $\widetilde{\mathcal{H}}$, respectively, with $\widetilde{\mathcal{H}}$ being
a modified PSW (MPSW) Hamiltonian,
$\widetilde{\mathcal{H}}_{\mathrm{harm}}^{\mathrm{SC}}+\mathcal{V}$
or
$\widetilde{\mathcal{H}}_{\mathrm{harm}}^{\mathrm{DC}}+\mathcal{V}$,
while
$\alpha_{\bm{k}}^-\equiv\alpha_{\bm{k}}$ and $\alpha_{\bm{k}}^+\equiv\alpha_{\bm{k}}^\dagger$
are the annihilation and creation operators, respectively, for the corresponding harmonic
Hamiltonian
$\widetilde{\mathcal{H}}_{\mathrm{harm}}^{\mathrm{SC}}$
or
$\widetilde{\mathcal{H}}_{\mathrm{harm}}^{\mathrm{DC}}$.
$\mathcal{T}$ and $\epsilon$ shall signify time ordering operation and a positive infinitesimal
value, respectively, with $\sigma(t)=2\theta(t)-1$.
Let $\widetilde{E}_n$ and $|n\rangle$ be the eigenvalues and eigenstates of
$\widetilde{\mathcal{H}}$,
$\widetilde{\mathcal{H}}|n\rangle=\widetilde{E}_n|n\rangle$
$\left(\widetilde{E}_0\leq\widetilde{E}_1\leq\widetilde{E}_2\leq\cdots\right)$.
We shall express Green's functions of different types in terms of a Lehmann representation
\cite{L342,McGrraw-Hill1971} in the purpose of finding an identity among them.
When we consider bare magnon Green's functions replacing $\widetilde{\mathcal{H}}$ by its
harmonic part $\widetilde{\mathcal{H}}_{\mathrm{harm}}$, we use
$G_0$, $\mathcal{G}_0$,
$\mathcal{P}(t)^{\mathrm{harm}}
\equiv
 e^{ i\widetilde{\mathcal{H}}_{\mathrm{harm}}t/\hbar}
 \mathcal{P}
 e^{-i\widetilde{\mathcal{H}}_{\mathrm{harm}}t/\hbar}$,
and $\langle\mathcal{Q}\rangle_0$
instead of
$G$, $\mathcal{G}$, $\mathcal{P}(t)$, and $\langle\mathcal{Q}\rangle$, respectively,
where we let $\tilde{\varepsilon}_{\bm{k}_{\nu,\sigma}}$ be
either
$\tilde{\varepsilon}_{\bm{k}_{\nu,\sigma}}^{\mathrm{SC}}$ (\ref{E:ekSC-MLSW})
or
$\tilde{\varepsilon}_{\bm{k}_{\nu,\sigma}}^{\mathrm{DC}}$ (\ref{E:ekDC-MLSW}).
Note that the anomalous bare Green's functions $G_0^{++}$ and $G_0^{--}$ are trivially zero.

\begin{widetext}
\subsection{Spectral function $A(\bm{k},\omega)$}
\label{SS:SF}

   The ground-state spectral function $A(\bm{k},\omega)$ is calculated via the magnon
Green's function (\ref{SE:Gdef}), which is readily obtainable in the harmonic approximation as
\begin{align}
   &
   G_{0}^{-+}(\bm{k};\omega+i\epsilon)
 =-i
   \int_{-\infty}^\infty
   \left\langle
    \mathcal{T}
     \alpha_{\bm{k}}(t)^{\mathrm{harm}}
     \alpha_{\bm{k}}^\dagger
    \right\rangle_0
   e^{i[\omega+i\epsilon\sigma(t)]t}
   dt
  =-i
   \int_0^{\infty}
   \left\langle
    \alpha_{\bm{k}}
    \alpha_{\bm{k}}^\dagger
   \right\rangle_0
   e^{i(\omega-\tilde{\varepsilon}_{\bm{k}}/\hbar+i\epsilon)t}
   dt
  =\frac{\hbar}
        {\hbar\omega-\tilde{\varepsilon}_{\bm{k}}+i\hbar\epsilon}
   \label{SE:G0explicit}
\end{align}
and generally written in the interaction picture as \cite{McGrraw-Hill1971}
\begin{align}
   &
   G^{-+}  (\bm{k};\omega+i\epsilon)
  =G_0^{-+}(\bm{k};\omega+i\epsilon)
  -i\sum_{l=1}^\infty
   \frac{(-i)^l}{l!\hbar^l}
   \int_{-\infty}^\infty
   \!\!
   dt_1 
   \cdots
   \int_{-\infty}^\infty
   \!\!
   dt_l
   \int_{-\infty}^\infty
   \!\!
   \left\langle
    \mathcal{T}
    \mathcal{V}(t_1)^{\mathrm{harm}}\cdots
    \mathcal{V}(t_l)^{\mathrm{harm}}
    \alpha_{\bm{k}} (t)^{\mathrm{harm}}
    \alpha_{\bm{k}}^\dagger
   \right\rangle_0
   e^{i[\omega+i\epsilon\sigma(t)]t}
   dt.
   \label{SE:perturbedG}
\end{align}
Calculation of (\ref{SE:perturbedG}) starts with expressing $\mathcal{V}$ as well as
$\widetilde{\mathcal{H}}_{\mathrm{harm}}$ in terms of Bogoliubov bosons,
\begin{align}
   &
   \mathcal{H}^{\left(\frac{1}{2}\right)}
  =\sum_{\sigma =0,\pm}\sum_{\nu =1}^N
   \sum_{\sigma'=0,\pm}\sum_{\nu'=1}^N
   \left[
         \vphantom{\alpha_{ \bm{k}_{\nu,\sigma}}^\dagger
                   \alpha_{-\bm{k}_{\nu,\sigma}-\bm{k}_{\nu',\sigma'}}^\dagger
                   \alpha_{ \bm{k}_{\nu',\sigma'}}^\dagger}
    \varLambda_{++-}
    \left(
     \bm{k}_{\nu,\sigma},\bm{k}_{\nu',\sigma'}-\bm{k}_{\nu,\sigma},\bm{k}_{\nu',\sigma'}
    \right)
    \alpha_{\bm{k}_{\nu,\sigma}}^\dagger
    \alpha_{\bm{k}_{\nu',\sigma'}-\bm{k}_{\nu,\sigma}}^\dagger
    \alpha_{\bm{k}_{\nu',\sigma'}}
   \right.
   \nonumber
   \allowdisplaybreaks \\
   &\qquad\qquad\qquad\qquad\qquad
   \left.
   +\varLambda_{+++}
    \left(
     \bm{k}_{\nu,\sigma},-\bm{k}_{\nu',\sigma'}-\bm{k}_{\nu,\sigma},\bm{k}_{\nu',\sigma'}
    \right)
    \alpha_{ \bm{k}_{\nu,\sigma}}^\dagger
    \alpha_{-\bm{k}_{\nu,\sigma}-\bm{k}_{\nu',\sigma'}}^\dagger
    \alpha_{ \bm{k}_{\nu',\sigma'}}^\dagger
   +\mathrm{H.c.}
   \right],
   \label{SE:H(1/2)alpha}
   \allowdisplaybreaks \\
   &
   \mathcal{H}^{(0)}
  =\delta E_1^{(0)}
  +\sum_{\sigma=0,\pm}\sum_{\nu=1}^N
   \left\{
    \varLambda_{+-}
    \left(
     \bm{k}_{\nu,\sigma},\bm{k}_{\nu,\sigma}
    \right)
    \alpha_{\bm{k}_{\nu,\sigma}}^\dagger
    \alpha_{\bm{k}_{\nu,\sigma}}
   +\left[
     \varLambda_{++}
     \left(
      \bm{k}_{\nu,\sigma},-\bm{k}_{\nu,\sigma}
     \right)
     \alpha_{ \bm{k}_{\nu,\sigma}}^\dagger
     \alpha_{-\bm{k}_{\nu,\sigma}}^\dagger
    +\mathrm{H.c.}
    \right]
   \right\}
   \nonumber
   \allowdisplaybreaks \\
   &\qquad
  +\sum_{\sigma  =0,\pm}\sum_{\nu  =1}^N
   \sum_{\sigma' =0,\pm}\sum_{\nu' =1}^N
   \sum_{\sigma''=0,\pm}\sum_{\nu''=1}^N
   \left\{
    \varLambda_{++--}
    \left(
     \bm{k}_{\nu  ,\sigma  }                        ,
     \bm{k}_{\nu' ,\sigma' }                        ,
     \bm{k}_{\nu  ,\sigma  }+\bm{k}_{\nu'',\sigma''},
     \bm{k}_{\nu' ,\sigma' }-\bm{k}_{\nu'',\sigma''}
    \right)
    \alpha_{\bm{k}_{\nu  ,\sigma  }                        }^\dagger
    \alpha_{\bm{k}_{\nu' ,\sigma' }                        }^\dagger
    \alpha_{\bm{k}_{\nu  ,\sigma  }+\bm{k}_{\nu'',\sigma''}}
    \alpha_{\bm{k}_{\nu' ,\sigma' }-\bm{k}_{\nu'',\sigma''}}
   \right.
   \nonumber
   \allowdisplaybreaks \\
   &\qquad\qquad\qquad
   +\left[
     \varLambda_{+++-}
      \left(
       \bm{k}_{\nu  ,\sigma  }                        ,
       \bm{k}_{\nu' ,\sigma' }                        ,
      -\bm{k}_{\nu  ,\sigma  }+\bm{k}_{\nu'',\sigma''},
       \bm{k}_{\nu' ,\sigma' }+\bm{k}_{\nu'',\sigma''}
      \right)
     \alpha_{ \bm{k}_{\nu  ,\sigma  }                        }^\dagger
     \alpha_{ \bm{k}_{\nu' ,\sigma' }                        }^\dagger
     \alpha_{-\bm{k}_{\nu  ,\sigma  }+\bm{k}_{\nu'',\sigma''}}^\dagger
     \alpha_{ \bm{k}_{\nu' ,\sigma' }+\bm{k}_{\nu'',\sigma''}}
    \right.
   \nonumber
   \allowdisplaybreaks \\
   &\qquad\qquad\qquad\ 
   \left.
    \left.
    +\varLambda_{++++}
     \left(
      \bm{k}_{\nu  ,\sigma  }                        ,
      \bm{k}_{\nu' ,\sigma' }                        ,
     -\bm{k}_{\nu  ,\sigma  }-\bm{k}_{\nu'',\sigma''},
     -\bm{k}_{\nu' ,\sigma' }+\bm{k}_{\nu'',\sigma''}
     \right)
     \alpha_{ \bm{k}_{\nu  ,\sigma  }                        }^\dagger
     \alpha_{ \bm{k}_{\nu' ,\sigma' }                        }^\dagger
     \alpha_{-\bm{k}_{\nu  ,\sigma  }-\bm{k}_{\nu'',\sigma''}}^\dagger
     \alpha_{-\bm{k}_{\nu' ,\sigma' }+\bm{k}_{\nu'',\sigma''}}^\dagger
    +\mathrm{H.c.}
    \right]
   \right\},
   \label{SE:H(0)alpha}
\end{align}
where the ground-state energy correction
$\delta E_1^{(0)}
\equiv
 \langle
  \mathcal{H}^{(0)}
 \rangle_0
$
and vertex functions
$\varLambda_{++-}$ and $\varLambda_{+++}$ in (\ref{SE:H(1/2)alpha}) and 
$\varLambda_{+-}$, $\varLambda_{++}$, $\varLambda_{++--}$, $\varLambda_{+++-}$, and
$\varLambda_{++++}$ in (\ref{SE:H(0)alpha})
are all written in terms of the MLSW ground-state Bogoliubov parameters
$u_{\bm{k}_{\nu,\sigma}}$ and $v_{\bm{k}_{\nu,\sigma}}$.
By virtue of Wick's theorem \cite{W268},
we can rewrite (\ref{SE:perturbedG}) into a series in $G_0^{-+}$ and $\mathcal{V}$.
In the spirit of the $1/S$ expansion, we renormalize the bare magnon Green's function by
the lowest-order $O(S^0)$ corrections arising at $l=1$, to first order in $\mathcal{H}^{(0)}$,
and at $l=2$, to second order in $\mathcal{H}^{\left(\frac{1}{2}\right)}$,
\begin{align}
   &
   G^{-+}  (\bm{k};\omega+i\epsilon)
  =G_0^{-+}(\bm{k};\omega+i\epsilon)
  -\frac{1}{\hbar}
   \int_{-\infty}^\infty
   dt_1 
   \int_{-\infty}^\infty
   \left\langle
    \mathcal{T}
    \mathcal{H}^{(0)}(t_1)^{\mathrm{harm}}
    \alpha_{\bm{k}} (t)^{\mathrm{harm}}
    \alpha_{\bm{k}}^\dagger
   \right\rangle_0
   e^{i[\omega+i\epsilon\sigma(t)]t}
   dt
  +\frac{i}{2!\hbar^2}
   \nonumber
   \allowdisplaybreaks \\
   &\quad\times
   \int_{-\infty}^\infty
   dt_1 
   \int_{-\infty}^\infty
   dt_2 
   \int_{-\infty}^\infty
   \left\langle
    \mathcal{T}
    \mathcal{H}^{\left(\frac{1}{2}\right)}(t_1)^{\mathrm{harm}}
    \mathcal{H}^{\left(\frac{1}{2}\right)}(t_2)^{\mathrm{harm}}
    \alpha_{\bm{k}} (t)^{\mathrm{harm}}
    \alpha_{\bm{k}}^\dagger
   \right\rangle_0
   e^{i[\omega+i\epsilon\sigma(t)]t}
   dt
  +\cdots
   \nonumber
   \allowdisplaybreaks \\
   &
  =G_0^{-+}(\bm{k};\omega+i\epsilon)
  +G_0^{-+}(\bm{k};\omega+i\epsilon)
   \
   \frac{\varSigma_1^{(0)}(\bm{k})
        +\varSigma_2^{(0)}(\bm{k};\omega+i\epsilon)}{\hbar}
   G_0^{-+}(\bm{k};\omega+i\epsilon)
  +\cdots
   \nonumber
   \allowdisplaybreaks \\
   &
  \simeq
   G_0^{-+}(\bm{k};\omega+i\epsilon)
   \sum_{l=0}^\infty
   \left[
    \frac{\varSigma_1^{(0)}(\bm{k})
         +\varSigma_2^{(0)}(\bm{k};\omega+i\epsilon)}{\hbar}
    G_0^{-+}(\bm{k};\omega+i\epsilon)
   \right]^l
  =G_0^{-+}(\bm{k};\omega+i\epsilon)
   \left[
    1
   +\frac{\varSigma^{(0)}(\bm{k};\omega+i\epsilon)}{\hbar}
    G^{-+}  (\bm{k};\omega+i\epsilon)
   \right],
   \label{SE:normGFDysonEq}
\end{align}
where the primary self-energies
$
   \varSigma^{(0)}(\bm{k};\omega+i\epsilon)
  \equiv
   \varSigma_1^{(0)}(\bm{k})
  +\varSigma_2^{(0)}(\bm{k};\omega+i\epsilon)
$
are given by
\begin{align}
   \varSigma_1^{(0)}(\bm{k})
  =\varLambda_{+-}
    \left(
     \bm{k},\bm{k}
    \right),\ 
   \varSigma_2^{(0)}(\bm{k};\omega+i\epsilon)
  =\sum_{\sigma=0,\pm}\sum_{\nu=1}^{N}
   \frac{2|
    \varLambda_{++-}
    \left(
     \bm{k},\bm{k}_{\nu,\sigma}-\bm{k},\bm{k}_{\nu,\sigma}
    \right)|^{2}}
        {\hbar\omega
        -\tilde{\varepsilon}_{\bm{k}_{\nu,\sigma}}
        -\tilde{\varepsilon}_{{\bm{k}}-\bm{k}_{\nu,\sigma}}
        +i\hbar\epsilon}
  -\sum_{\sigma=0,\pm}\sum_{\nu=1}^{N}
   \frac{18|
    \varLambda_{+++}
    \left(
     \bm{k},-\bm{k}_{\nu,\sigma}-\bm{k},\bm{k}_{\nu,\sigma}
    \right)|^{2}}
        {\hbar\omega
         +\tilde{\varepsilon}_{\bm{k}_{\nu,\sigma}}
         +\tilde{\varepsilon}_{\bm{k}+\bm{k}_{\nu,\sigma}}
         +i\hbar\epsilon}.
   \label{SE:normSE}
\end{align}
$\varSigma_1^{(0)}(\bm{k})$ has no frequency dependence and therefore no imaginary part,
$\varSigma_1^{(0)}(\bm{k})
=\displaystyle
 \mathop{\varSigma}^{\,\protect\substack{-\!\!\!-\\[-1.6mm]}}\!{}_1^{(0)}(\bm{k})$.
$\varSigma_2^{(0)}(\bm{k};\omega+i\epsilon)$ may have both real and imaginary parts,
which we shall denote by
$\displaystyle
 \mathop{\varSigma}^{\,\protect\substack{-\!\!\!-\\[-1.6mm]}}\!{}_2^{(0)}
 (\bm{k};\omega+i\epsilon)$
and
$\displaystyle
 \mathop{\varSigma}^{\,\protect\substack{\\[-0.6mm]-\!\!\!- \\[-1.4mm] -\!\!\!-\\[-1.4mm]}}
               \!{}_2^{(0)}
 (\bm{k};\omega+i\epsilon)$,
respectively.
The Dyson formulation (\ref{SE:normGFDysonEq}) gives
\begin{align}
   G^{-+}(\bm{k};\omega+i\epsilon)
  =\frac{1}
   {\left[G_0^{-+}(\bm{k};\omega+i\epsilon)\right]^{-1}
   -\varSigma^{(0)}(\bm{k};\omega+i\epsilon)/\hbar
   +i\epsilon}
  =\frac{\hbar}
   {\hbar\omega
   -\tilde{\varepsilon}_{\bm{k}}
   -\varSigma^{(0)}(\bm{k};\omega+i\epsilon)
   +i\hbar\epsilon}.
   \label{SE:GfromDysonEq}
\end{align}
The bare and renormalized spectral functions
\begin{align}
   A(\bm{k},\omega)
  =\left\{
   \begin{array}{ll}
   \displaystyle
  -\frac{1}{\pi\hbar}
   \mathrm{Im}
   \left[
    G_0^{-+}(\bm{k};\omega+i\epsilon)
   \right]
 =-\frac{1}{\pi}
   {\mathrm{Im}}
   \frac{1}
        {\hbar\omega\!-\!\tilde{\varepsilon}_{\bm{k}}+i\hbar\epsilon}
  =\delta(\hbar\omega-\tilde{\varepsilon}_{\bm{k}})
   & \mathrm{(MLSW)} \\
   \displaystyle
  -\frac{1}{\pi\hbar}
   \mathrm{Im}
   \left[
    G^{-+}  (\bm{k};\omega+i\epsilon)
   \right]
 =-\frac{1}{\pi}
   {\mathrm{Im}}
   \frac{1}
        {\hbar\omega-\tilde{\varepsilon}_{\bm{k}}
        -\varSigma^{(0)}(\bm{k};\omega+i\epsilon)+i\hbar\epsilon}
   & \mathrm{(MPSW)} \\
   \end{array}
   \right.
   \tag{10}
   \label{E:Aqw}
\end{align}
are calculated in the SC and DC schemes and explicitly compared with each other as well as with
MLSW pair-excitation continua later in Sect. \ref{S:DCvsSC}.

\subsection{Dynamic spin structure factor $S(\bm{q},\omega)$}
\label{SS:DSF}

   The dynamic spin structure factors at temperature $T$
\begin{align}
   S^{\lambda\lambda'}(\bm{q},\omega)
  \equiv
   \frac{1}{2\pi\hbar}
   \int_{-\infty}^{\infty}
   \frac
   {\mathrm{Tr}
    \left\{
     e^{-\widetilde{\mathcal{H}}/k_{\mathrm{B}}T}
     \left[
      \delta S_{ \bm{q}}^{\lambda}(t)
      \delta S_{-\bm{q}}^{\lambda'}
     \right]
    \right\}}
   {\mathrm{Tr}
    \left\{
     e^{-\widetilde{\mathcal{H}}/k_{\mathrm{B}}T}
    \right\}}
   e^{i\omega t}dt
  =\sum_{l,l'=1}^{L}
   \frac{e^{i\bm{q}\cdot(\bm{r}_l-\bm{r}_{l'})}}{2\pi\hbar L}
   \int_{-\infty}^{\infty}
   \frac
   {\mathrm{Tr}
    \left\{
     e^{-\widetilde{\mathcal{H}}/k_{\mathrm{B}}T}
     \left[
      \delta S_{\bm{r}_{l }}^{\lambda}(t)
      \delta S_{\bm{r}_{l'}}^{\lambda'}
     \right]
    \right\}}
   {\mathrm{Tr}
    \left\{
     e^{-\widetilde{\mathcal{H}}/k_{\mathrm{B}}T}
    \right\}}
   e^{i\omega t}dt
   \label{SE:Sqw}
\end{align}
are calculated via the retarded Green's functions
\begin{align}
   \mathcal{R}^{\lambda\lambda'}(\bm{q};\omega+i\epsilon)
  \equiv
   -i
   \int_{-\infty}^{\infty}
   \frac
   {\mathrm{Tr}
    \left\{
     e^{-\widetilde{\mathcal{H}}/k_{\mathrm{B}}T}
     \left[
      \delta S_{ \bm{q}}^{\lambda}(t),
      \delta S_{-\bm{q}}^{\lambda'}
     \right]
    \right\}}
   {\mathrm{Tr}
    \left\{
     e^{-\widetilde{\mathcal{H}}/k_{\mathrm{B}}T}
    \right\}}
   e^{i(\omega+i\epsilon)t}
   \theta(t)dt.
   \label{SE:RGdef}
\end{align}
The fluctuation-dissipation theorem says that \cite{M094407}
\begin{align}
   S^{\lambda\lambda'}(\bm{q},\omega)
  =-\frac{1}{\pi\hbar}
   \frac{1}{1-e^{-\hbar\omega/k_{\mathrm{B}}T}}
   \mathrm{Im}
   \left[
    \mathcal{R}^{\lambda\lambda'}(\bm{q};\omega+i\epsilon)
   \right]
   \mathop{\longrightarrow}^{T\rightarrow 0}
   \left\{
    \begin{array}{ll}
     \displaystyle
    -\frac{1}{\pi\hbar}
     \mathop{\mathrm{lim}}_{T\rightarrow 0}
     \mathrm{Im}
     \left[
      \mathcal{R}^{\lambda\lambda'}(\bm{q};\omega+i\epsilon)
     \right]
     & (\omega>0) \\
     \ \ \ 0
     & (\omega<0) \\
    \end{array}
   \right..
   \label{SE:FDT}
\end{align}
In terms of a Lehmann representation, the retarded correlator (\ref{SE:RGdef}) reads
\begin{align}
   \lim_{T\rightarrow 0}
   \mathcal{R}^{\lambda\lambda'}(\bm{q};\omega+i\epsilon)
  =-i
   \int_{-\infty}^{\infty}
   \left\langle
    \left[
    \delta S_{ \bm{q}}^{\lambda}(t),
    \delta S_{-\bm{q}}^{\lambda'}
    \right]
   \right\rangle
   e^{i(\omega+i\epsilon)t}
   \theta(t)dt
  =-\sum_{n=0}^\infty
   \left[
    \frac{\langle 0|\delta S_{ \bm{q}}^{\lambda }|n\rangle
          \langle n|\delta S_{-\bm{q}}^{\lambda'}|0\rangle}
         {\omega-\left(\widetilde{E}_n-\widetilde{E}_0\right)/\hbar+i\epsilon}
   -\frac{\langle 0|\delta S_{-\bm{q}}^{\lambda'}|n\rangle
          \langle n|\delta S_{ \bm{q}}^{\lambda }|0\rangle}
         {\omega+\left(\widetilde{E}_n-\widetilde{E}_0\right)/\hbar+i\epsilon}
   \right],
   \label{SE:RGLEMAN}
\end{align}
while the usual Feynman correlator (\ref{SE:SGdef}) evaluates to
\begin{align}
   \mathcal{G}^{\lambda\lambda'}(\bm{q};\omega+i\epsilon)
  =-i
   \int_{-\infty}^{\infty}
   \left\langle
    \mathcal{T}
    \delta S_{ \bm{q}}^{\lambda}(t)
    \delta S_{-\bm{q}}^{\lambda'}
   \right\rangle
   e^{i[\omega+i\epsilon\sigma(t)]t}dt
  =-\sum_{\kappa=0}^\infty
   \left[
    \frac{\langle 0|\delta S_{ \bm{q}}^{\lambda }|n\rangle
          \langle n|\delta S_{-\bm{q}}^{\lambda'}|0\rangle}
         {\omega-\left(\widetilde{E}_n-\widetilde{E}_0\right)/\hbar+i\epsilon}
   -\frac{\langle 0|\delta S_{-\bm{q}}^{\lambda'}|n\rangle
          \langle n|\delta S_{ \bm{q}}^{\lambda }|0\rangle}
         {\omega+\left(\widetilde{E}_n-\widetilde{E}_0\right)/\hbar-i\epsilon}
   \right].
   \label{SE:SGLEMAN}
\end{align}
Having in mind that
\begin{align}
   \frac{1}
        {\omega+\left(\widetilde{E}_n-\widetilde{E}_0\right)/\hbar\pm i\epsilon}
  =\frac{\mathcal{P}}
        {\omega+\left(\widetilde{E}_n-\widetilde{E}_0\right)/\hbar}
  \mp i\pi\delta
          \left(
           \omega
          +\frac{\widetilde{E}_n-\widetilde{E}_0}{\hbar}
          \right)
   \label{SE:DiracRelation}
\end{align}
with $\mathcal{P}$ denoting the principal value and
$\omega+\left(\widetilde{E}_n-\widetilde{E}_0\right)/\hbar>0$ for $\omega>0$,
we learn that
\begin{align}
   S^{\lambda\lambda'}(\bm{q},\omega)
 =-\frac{1}{\pi\hbar}
   \mathrm{Im}
   \left[
    \mathcal{G}^{\lambda\lambda'}(\bm{q};\omega+i\epsilon)
   \right]\ 
   (T\rightarrow 0).
   \label{SE:Sqw(T=0)}
\end{align}
Similar to (\ref{E:TransSq}), we can construct the dynamic structure factors in
the laboratory frame from those in the rotating frame,
\begin{align}
   &
   S^{zz}(\bm{q},\omega)
  =S^{xx}(\bm{q},\omega)
  =\frac{1}{4}
   \sum_{\tau=\pm}
   \left\{
    S^{\tilde{z}\tilde{z}}(\tau\bm{Q}_{\mathrm{X},\mathrm{K}}+\bm{q},\omega)
   +S^{\tilde{x}\tilde{x}}(\tau\bm{Q}_{\mathrm{X},\mathrm{K}}+\bm{q},\omega)
   +i\tau
    \left[
     S^{\tilde{z}\tilde{x}}(\tau\bm{Q}_{\mathrm{X},\mathrm{K}}+\bm{q},\omega)
    -S^{\tilde{x}\tilde{z}}(\tau\bm{Q}_{\mathrm{X},\mathrm{K}}+\bm{q},\omega)
    \right]
   \right\}
   \nonumber
   \allowdisplaybreaks \\
   &
  =-\frac{1}{4\pi\hbar}
   \sum_{\tau=\pm}
   \mathrm{Im}
   \left\{
    \mathcal{G}^{\tilde{z}\tilde{z}}(\tau\bm{Q}_{\mathrm{X},\mathrm{K}}+\bm{q};\omega+i\epsilon)
   +\mathcal{G}^{\tilde{x}\tilde{x}}(\tau\bm{Q}_{\mathrm{X},\mathrm{K}}+\bm{q};\omega+i\epsilon)
   +\tau
    \left[
     \mathcal{G}^{\tilde{z}\tilde{x}}(\tau\bm{Q}_{\mathrm{X},\mathrm{K}}+\bm{q};\omega+i\epsilon)
    -\mathcal{G}^{\tilde{x}\tilde{z}}(\tau\bm{Q}_{\mathrm{X},\mathrm{K}}+\bm{q};\omega+i\epsilon)
    \right]
   \right\}
   \nonumber
   \allowdisplaybreaks \\
   &
%  \simeq
%   -\frac{1}{4\pi\hbar}
%   \sum_{\tau=\pm}
%   \mathrm{Im}
%   \left[
%    \mathcal{G}^{\tilde{z}\tilde{z}}(\tau\bm{Q}_{\mathrm{X},\mathrm{K}}+\bm{q};\omega+i\epsilon)
%   +\mathcal{G}^{\tilde{x}\tilde{x}}(\tau\bm{Q}_{\mathrm{X},\mathrm{K}}+\bm{q};\omega+i\epsilon)
%  \right],
%   \ 
   S^{yy}(\bm{q},\omega)
  =S^{\tilde{y}\tilde{y}}(\bm{q},\omega)
  =-\frac{1}{\pi\hbar}
   \mathrm{Im}
   \left[
    \mathcal{G}^{\tilde{y}\tilde{y}}(\bm{q};\omega+i\epsilon)
   \right].
   \label{SE:SqwLAB}
\end{align}

   First we investigate leading transverse fluctuations in (\ref{SE:SqwLAB}).
Let us approximate on-site cubic terms as
\begin{align}
   a_{\bm{r}_l}^\dagger a_{\bm{r}_l}^\dagger a_{\bm{r}_l}
  \simeq
   2\langle
    \mathcal{A}
   \rangle_0
   a_{\bm{r}_l}^{\dagger}
  +\langle
    \mathcal{C}
   \rangle_0
   a_{\bm{r}_l};\ 
   \langle a_{\bm{r}_l}^\dagger a_{\bm{r}_l} \rangle_0
  \equiv
   \langle \mathcal{A} \rangle_0
  =\sum_{\sigma=0,\pm}\sum_{\nu=1}^N
   \frac{u_{\bm{k}_{\nu,\sigma}}^{2}+v_{\bm{k}_{\nu,\sigma}}^{2}-1}{2L},\ 
   \langle
    a_{\bm{r}_l}^{\dagger}
    a_{\bm{r}_l}^{\dagger}
   \rangle_0
  \equiv
   \langle \mathcal{C} \rangle_0
  =\sum_{\sigma=0,\pm}\sum_{\nu=1}^N
   \frac{u_{\bm{k}_{\nu,\sigma}}v_{\bm{k}_{\nu,\sigma}}}{L}
   \label{SE:CubicTermsHF}
\end{align}
to write the relevant spin operators as
\begin{align}
   S_{\bm{r}_l}^{\tilde{x}}
  =\sqrt{\frac{S}{2}}
   \left[1-\frac{2\langle \mathcal{A} \rangle_0
                 +\langle \mathcal{C} \rangle_0}{4S}
   \right]
   \left(a_{\bm{r}_l}+a_{\bm{r}_l}^\dagger\right)
  +O\left(S^{-\frac{3}{2}}\right),\ 
   S_{\bm{r}_l}^{\tilde{y}}
  =-i\sqrt{\frac{S}{2}}
   \left[1-\frac{2\langle \mathcal{A} \rangle_0
                 -\langle \mathcal{C} \rangle_0}{4S}
   \right]
   \left(a_{\bm{r}_l}-a_{\bm{r}_l}^\dagger\right)
  +O\left(S^{-\frac{3}{2}}\right).
   \label{SE:RrepSxSy}
\end{align}
Via the Fourier transformation (\ref{SE:FTspin}) and (\ref{SE:FTboson}), we can express
spin Green's functions in terms of the magnon language,
\begin{align}
   &
   \mathcal{G}^{\tilde{x}\tilde{x}}(\bm{k};\omega+i\epsilon)
  =\frac{S}{2}
   \left[1-\frac{2\langle \mathcal{A} \rangle_0
                 +\langle \mathcal{C} \rangle_0}{4S}
   \right]^{2}
   \left({u}_{\bm{k}}+{v}_{\bm{k}}\right)^{2}
   \left[
    G^{-+}( \bm{k}; \omega+i\epsilon)
   +G^{-+}(-\bm{k};-\omega-i\epsilon)
   +G^{++}( \bm{k};\omega+i\epsilon)
   +G^{--}( \bm{k};\omega+i\epsilon)
   \right]
   \nonumber
   \allowdisplaybreaks \\
   &\qquad\qquad\quad\ \ 
  \simeq
   \frac{S}{2}
   \left[
    1
    -\frac{2\langle\mathcal{A}\rangle_0+\langle \mathcal{C}\rangle_0}
          {2S}
   \right]
   \left({u}_{\bm{k}}+{v}_{\bm{k}}\right)^2
   G^{-+}( \bm{k};\omega+i\epsilon),
   \label{SE:calGtxtx}
   \allowdisplaybreaks \\
   &
   \mathcal{G}^{\tilde{y}\tilde{y}}(\bm{k};\omega+i\epsilon)
  =\frac{S}{2}
   \left[
    1
   -\frac{2\langle\mathcal{A}\rangle_0-\langle \mathcal{C}\rangle_0}
         {4S}
   \right]^2
   \left({u}_{\bm{k}}-{v}_{\bm{k}}\right)^2
   \left[
    G^{-+}( \bm{k}; \omega+i\epsilon)
   +G^{-+}(-\bm{k};-\omega-i\epsilon)
   -G^{++}( \bm{k};\omega+i\epsilon)
   -G^{--}( \bm{k};\omega+i\epsilon)
   \right]
   \nonumber
   \allowdisplaybreaks \\
   &\qquad\qquad\quad\ \ 
  \simeq
   \frac{S}{2}
   \left[
    1
   -\frac{2\langle\mathcal{A}\rangle_0-\langle \mathcal{C}\rangle_0}
         {2S}
   \right]
   \left({u}_{\bm{k}}-{v}_{\bm{k}}\right)^2
   G^{-+}( \bm{k};\omega+i\epsilon),
   \label{SE:calGtyty}
\end{align}
where the terms $G^{-+}(-{\bm{k}},-\omega-i\epsilon)$ have no pole for $\omega>0$ and therefore
no contribution to the dynamic structure factor,
while compared to the normal Green's functions $G^{-+}(\bm{k};\omega+i\epsilon)$ of $O(S^{-1})$
[cf. (\ref{SE:GfromDysonEq})], the anomalous Green's functions
\begin{align}
   &
   G  ^{\sigma\sigma}(\bm{k};\omega+i\epsilon)
 =-i\sum_{l=1}^\infty
   \frac{(-i)^l}{l!\hbar^l}
   \int_{-\infty}^\infty
   dt_1
   \cdots
   \int_{-\infty}^\infty
   dt_l
   \int_{-\infty}^\infty
   \left\langle
    \mathcal{T}
    \mathcal{V}(t_1)^{\mathrm{harm}}\cdots
    \mathcal{V}(t_l)^{\mathrm{harm}}
    \alpha_{-\sigma\bm{k}}^\sigma(t)^{\mathrm{harm}}
    \alpha_{ \sigma\bm{k}}^\sigma
   \right\rangle_0
   e^{i[\omega+i\epsilon\sigma(t)]t}
   dt
   \nonumber
   \allowdisplaybreaks \\
   &\qquad\qquad\qquad
  =-\frac{1}{\hbar}
   \int_{-\infty}^\infty
   dt_1
   \int_{-\infty}^\infty
   \left\langle
    \mathcal{T}
    \mathcal{H}^{(0)}(t_1)^{\mathrm{harm}}
    \alpha_{-\sigma\bm{k}}^\sigma(t)^{\mathrm{harm}}
    \alpha_{ \sigma\bm{k}}^\sigma
   \right\rangle_0
   e^{i[\omega+i\epsilon\sigma(t)]t}
   dt
  +\frac{i}{2!\hbar^2}
   \nonumber
   \allowdisplaybreaks \\
   &
   \qquad\qquad\qquad\quad\times
   \int_{-\infty}^\infty
   dt_1
   \int_{-\infty}^\infty
   dt_2
   \int_{-\infty}^\infty
   \left\langle
    \mathcal{T}
    \mathcal{H}^{\left(\frac{1}{2}\right)}(t_1)^{\mathrm{harm}}
    \mathcal{H}^{\left(\frac{1}{2}\right)}(t_2)^{\mathrm{harm}}
    \alpha_{-\sigma\bm{k}}^\sigma(t)^{\mathrm{harm}}
    \alpha_{ \sigma\bm{k}}^\sigma
   \right\rangle_0
   e^{i[\omega+i\epsilon\sigma(t)]t}
   dt
  +\cdots
   \nonumber
   \allowdisplaybreaks \\
   &\qquad\qquad\qquad
  \simeq
   G^{-+}(\bm{k};\omega+i\epsilon)
   \frac{\varSigma'{}_{\!\!1}^{(0)}(\bm{k})
        +\varSigma'{}_{\!\!2}^{(0)}(\bm{k};\omega+i\epsilon)}
        {\hbar}
   G^{-+}(-\bm{k};-\omega-i\epsilon);
   \label{SE:anormGFDysonAppr}
   \allowdisplaybreaks \\
   &
   \varSigma'{}_{\!\!1}^{(0)}(\bm{k})
  \equiv
   \varLambda_{++}
     \left(
      \bm{k},-\bm{k}
     \right),
  \ 
   \varSigma'{}_{\!\!2}^{(0)}(\bm{k};\omega+i\epsilon)
  \equiv
   6\sum_{\sigma=0,\pm}\sum_{\nu=1}^{N}
   \nonumber
   \allowdisplaybreaks \\
   &\times
    \left[
    \frac{
     \varLambda_{+++}
     \left(
     -\bm{k},\bm{k}-\bm{k}_{\nu,\sigma},\bm{k}_{\nu,\sigma}
     \right)
     \varLambda_{++-}
     \left(
      \bm{k}-\bm{k}_{\nu,\sigma},\bm{k}_{\nu,\sigma},\bm{k}
     \right)^{*}
         }
        {\hbar\omega
        -\tilde{\varepsilon}_{\bm{k}_{\nu,\sigma}}
        -\tilde{\varepsilon}_{{\bm{k}}-\bm{k}_{\nu,\sigma}}
        +i\hbar\epsilon}
   -\frac{
     \varLambda_{+++}
     \left(
      \bm{k},-\bm{k}-\bm{k}_{\nu,\sigma},\bm{k}_{\nu,\sigma}
     \right)
     \varLambda_{++-}
     \left(
     -\bm{k}-\bm{k}_{\nu,\sigma},\bm{k}_{\nu,\sigma},-\bm{k}
     \right)^{*}
          }
         {\hbar\omega
          +\tilde{\varepsilon}_{\bm{k}_{\nu,\sigma}}
          +\tilde{\varepsilon}_{\bm{k}+\bm{k}_{\nu,\sigma}}
          +i\hbar\epsilon}
    \right],
   \label{SE:anormSE}
\end{align}
evaluate to $O(S^{-2})$ and are therefore discarded in the spirit of the $1/S$ expansion.
$\varSigma'{}_{\!\!1}^{(0)}$, to first order in $\mathcal{H}^{(0)}$, and
$\varSigma'{}_{\!\!2}^{(0)}$, to second order in $\mathcal{H}^{\left(\frac{1}{2}\right)}$,
both on the order of $S^0$, play a similar role to that played by
the usual self-energies $\varSigma_1^{(0)}$ and $\varSigma_2^{(0)}$
but with a change in the number of quasiparticles.

   Next we figure out leading longitudinal fluctuations in (\ref{SE:SqwLAB}).
MLSWs contribute
\begin{align}
   &
   \mathcal{G}_0^{\tilde{z}\tilde{z}}({\bm{k}};\omega+i\epsilon)
 =-i\int_{-\infty}^{\infty}
   \left\langle
    \mathcal{T}
    \delta S_{ \bm{k}}^{\tilde{z}}(t)^{\mathrm{harm}}
    \delta S_{-\bm{k}}^{\tilde{z}}
   \right\rangle_0
   e^{i[\omega+i\epsilon\sigma(t)]t}dt
  =-\frac{i}{L}
   \sum_{\sigma=0,\pm}\sum_{\nu=1}^N
   \sum_{\sigma'=0,\pm}\sum_{\nu'=1}^N
   \nonumber
   \allowdisplaybreaks \\
   &\qquad\times
   \int_{-\infty}^{\infty}
   \left\{
      \vphantom{\left\langle\mathcal{T}
                 \alpha_{ \bm{k}_{\nu',\sigma'}}
                 \alpha_{-\bm{k}_{\nu',\sigma'}+\bm{k}}^\dagger
                \right\rangle_0}
   v_{\bm{k}_{\nu ,\sigma }}u_{\bm{k}_{\nu ,\sigma }+\bm{k}}
   u_{\bm{k}_{\nu',\sigma'}}v_{\bm{k}_{\nu',\sigma'}-\bm{k}}
   \left\langle\mathcal{T}
     \alpha_{-\bm{k}_{\nu ,\sigma }}(t)^{\mathrm{harm}}
     \alpha_{ \bm{k}_{\nu ,\sigma }+\bm{k}}(t)^{\mathrm{harm}}
     \alpha_{ \bm{k}_{\nu',\sigma'}}^\dagger
     \alpha_{-\bm{k}_{\nu',\sigma'}+\bm{k}}^\dagger
    \right\rangle_0
   +v_{\bm{k}_{\nu ,\sigma }}v_{\bm{k}_{\nu ,\sigma }+\bm{k}}
    v_{\bm{k}_{\nu',\sigma'}}v_{\bm{k}_{\nu',\sigma'}-\bm{k}}
   \right.
   \nonumber
   \allowdisplaybreaks \\
   &\qquad\times
   \left.
   \left[
    \left\langle\mathcal{T}
     \alpha_{-\bm{k}_{\nu ,\sigma }}(t)^{\mathrm{harm}}
     \alpha_{ \bm{k}_{\nu ,\sigma }+\bm{k}}^\dagger(t)^{\mathrm{harm}}
     \alpha_{ \bm{k}_{\nu',\sigma'}}
     \alpha_{-\bm{k}_{\nu',\sigma'}+\bm{k}}^\dagger
    \right\rangle_0
   -\left\langle\mathcal{T}
     \alpha_{-\bm{k}_{\nu ,\sigma }}(t)^{\mathrm{harm}}
     \alpha_{ \bm{k}_{\nu ,\sigma }+\bm{k}}^\dagger(t)^{\mathrm{harm}}
    \right\rangle_0
    \left\langle\mathcal{T}
     \alpha_{ \bm{k}_{\nu',\sigma'}}
     \alpha_{-\bm{k}_{\nu',\sigma'}+\bm{k}}^\dagger
    \right\rangle_0
   \right]
   \right\}
   e^{i[\omega+i\epsilon\sigma(t)]t}dt
   \nonumber
   \allowdisplaybreaks \\
   &\quad
  =\frac{i}{L}
   \sum_{\sigma=0,\pm}\sum_{\nu=1}^N
   \int_{-\infty}^{\infty}
   d\omega'
   \left\{
    v_{\bm{k}_{\nu,\sigma}}u_{\bm{k}_{\nu,\sigma}+\bm{k}}
   (
    u_{\bm{k}_{\nu,\sigma}}v_{\bm{k}_{\nu,\sigma}+\bm{k}}
   +v_{\bm{k}_{\nu,\sigma}}u_{\bm{k}_{\nu,\sigma}-\bm{k}}
   )
    G_0^{-+}(-\bm{k}_{\nu,\sigma};\omega'+i\epsilon)
    G_0^{-+}(\bm{k}_{\nu,\sigma}+\bm{k};\omega-\omega'+i\epsilon)
   \right.
   \nonumber
   \allowdisplaybreaks \\
   &\qquad\qquad\qquad\qquad\quad
   \left.
   +v_{\bm{k}_{\nu ,\sigma }}^{2}
    v_{\bm{k}_{\nu ,\sigma }+\bm{k}}^{2}
    G_0^{-+}(-\bm{k}_{\nu,\sigma};\omega'+i\epsilon)
    G_0^{-+}(\bm{k}_{\nu,\sigma}+\bm{k};\omega+\omega'+i\epsilon)
      \vphantom{\left(
                 u_{\bm{k}_{\nu,\sigma}}v_{\bm{k}_{\nu,\sigma}+\bm{k}}
                +v_{\bm{k}_{\nu,\sigma}}u_{\bm{k}_{\nu,\sigma}-\bm{k}}
                \right)}
   \right\}
   \nonumber
   \allowdisplaybreaks \\
   &\quad
  =\frac{\hbar}{L}
   \sum_{\sigma=0,\pm}\sum_{\nu=1}^N
   \frac{v_{\bm{k}_{\nu ,\sigma }}u_{\bm{k}_{\nu ,\sigma }+{\bm{k}}}
        (u_{\bm{k}_{\nu ,\sigma }}v_{\bm{k}_{\nu ,\sigma }+{\bm{k}}}
        +v_{\bm{k}_{\nu ,\sigma }}u_{\bm{k}_{\nu ,\sigma }+{\bm{k}}})
        }
        {\hbar\omega-\tilde{\varepsilon}_{\bm{k}_{\nu ,\sigma }}
                    -\tilde{\varepsilon}_{\bm{k}_{\nu ,\sigma }+\bm{k}}
       +i\hbar\epsilon}
   \label{SE:calG0tztz}
\end{align}
to them, which prove to be of $O(S^{-1})$, and MPSWs merely give them $O(S^{-2})$ corrections.
Therefore, we calculate longitudinal contributions no further beyond (\ref{SE:calG0tztz}).

   Finally we discuss mixed transverse-longitudinal fluctuations emergent in (\ref{SE:SqwLAB}).
The corresponding correlators trivially vanish in the bare magnon ground state,
$\mathcal{G}_0^{\tilde{z}\tilde{x}}(\bm{k};\omega+i\epsilon)
=\mathcal{G}_0^{\tilde{x}\tilde{z}}(\bm{k};\omega+i\epsilon)
=0$ [cf. (\ref{SE:SqLocCLSW})].
Their first nonzero contribution arises to first order in $\mathcal{H}^{\left(\frac{1}{2}\right)}$.
Again utilizing the Hartree-Fock decomposition (\ref{SE:CubicTermsHF}) for expanding
$S_{\bm{r}_l}^{\tilde{x}}$ in $1/S$, we have
\begin{align}
   &
   \mathcal{G}^{\tilde{z}\tilde{x}}(\bm{k};\omega+i\epsilon)
 =-i\sum_{l=1}^\infty
   \frac{(-i)^l}{l!\hbar^l}
   \int_{-\infty}^\infty
   dt_1 
   \cdots
   \int_{-\infty}^\infty
   dt_l
   \int_{-\infty}^\infty
   \left\langle
    \mathcal{T}
    \mathcal{V}(t_1)^{\mathrm{harm}}\cdots
    \mathcal{V}(t_l)^{\mathrm{harm}}
    \delta S_{ \bm{k}}^{\tilde{z}}(t)^{\mathrm{harm}}
    \delta S_{-\bm{k}}^{\tilde{x}}
   \right\rangle_0
   e^{i[\omega+i\epsilon\sigma(t)]t}
   dt
   \nonumber
   \allowdisplaybreaks \\
   &\qquad
  =-\frac{1}{\hbar}
   \int_{-\infty}^\infty
   dt_1
   \int_{-\infty}^\infty
   \left\langle
    \mathcal{T}
    \mathcal{H}^{\left(\frac{1}{2}\right)}(t_1)^{\mathrm{harm}}
    \delta S_{ \bm{k}}^{\tilde{z}}(t)^{\mathrm{harm}}
    \delta S_{-\bm{k}}^{\tilde{x}}
   \right\rangle_0
   e^{i[\omega+i\epsilon\sigma(t)]t}
   dt
  +\cdots
  =-\mathcal{G}^{\tilde{x}\tilde{z}}(\bm{k};\omega+i\epsilon)
   \nonumber
   \allowdisplaybreaks \\
   &\qquad
  \simeq
   \sqrt{\frac{S}{2}}
   \left[1-\frac{2\langle \mathcal{A} \rangle_0
                 +\langle \mathcal{C} \rangle_0}{4S}
   \right]
   ({u}_{\bm{k}}+{v}_{\bm{k}})
   \left[
    \varGamma' {}_{    \!\!\!1}^{\left(-\frac{1}{2}\right)}(\bm{k};\omega+i\epsilon)
    G^{-+}(\bm{k};\omega+i\epsilon)
   +\varGamma''{}_{\!\!\!\!\!1}^{\left(-\frac{1}{2}\right)}(\bm{k};\omega+i\epsilon)
    G^{-+}(-\bm{k};-\omega-i\epsilon)
   \right];
   \nonumber
   \allowdisplaybreaks \\
   &
   \varGamma' {}_{    \!\!\!1}^{\left(-\frac{1}{2}\right)}(\bm{k};\omega+i\epsilon)
  \equiv
   \sum_{\sigma=0,\pm}\sum_{\nu=1}^N
    (u_{\bm{k}_{\nu ,\sigma }}v_{\bm{k}_{\nu ,\sigma }-{\bm{k}}}
    +v_{\bm{k}_{\nu ,\sigma }}u_{\bm{k}_{\nu ,\sigma }-{\bm{k}}})
    \left[
    \frac{i
     \varLambda_{++-}
     \left(
     -\bm{k}+\bm{k}_{\nu,\sigma},-\bm{k}_{\nu,\sigma},-\bm{k}
     \right)
         }
        {\hbar\omega
        -\tilde{\varepsilon}_{\bm{k}_{\nu,\sigma}}
        -\tilde{\varepsilon}_{{\bm{k}}-\bm{k}_{\nu,\sigma}}
        +i\hbar\epsilon}
   +\frac{3i
     \varLambda_{+++}
     \left(
     -\bm{k},\bm{k}-\bm{k}_{\nu,\sigma},\bm{k}_{\nu,\sigma}
     \right)
          }
         {\hbar\omega
          +\tilde{\varepsilon}_{\bm{k}_{\nu,\sigma}}
          +\tilde{\varepsilon}_{\bm{k}-\bm{k}_{\nu,\sigma}}
          +i\hbar\epsilon}
    \right],
   \nonumber
   \allowdisplaybreaks \\
   &
   \varGamma''{}_{\!\!\!\!\!1}^{\left(-\frac{1}{2}\right)}(\bm{k};\omega+i\epsilon)
  \equiv
  -\sum_{\sigma=0,\pm}\sum_{\nu=1}^N
    (u_{\bm{k}_{\nu ,\sigma }}v_{\bm{k}_{\nu ,\sigma }-{\bm{k}}}
    +v_{\bm{k}_{\nu ,\sigma }}u_{\bm{k}_{\nu ,\sigma }-{\bm{k}}})
   \left[
    \frac{3i
     \varLambda_{+++}
     \left(
      \bm{k},-\bm{k}+\bm{k}_{\nu,\sigma},-\bm{k}_{\nu,\sigma}
     \right)^*
         }
        {\hbar\omega
        -\tilde{\varepsilon}_{\bm{k}_{\nu,\sigma}}
        -\tilde{\varepsilon}_{{\bm{k}}-\bm{k}_{\nu,\sigma}}
        +i\hbar\epsilon}
   +\frac{i
     \varLambda_{++-}
     \left(
      \bm{k}-\bm{k}_{\nu,\sigma}, \bm{k}_{\nu,\sigma}, \bm{k}
     \right)^*
          }
         {\hbar\omega
          +\tilde{\varepsilon}_{\bm{k}_{\nu,\sigma}}
          +\tilde{\varepsilon}_{\bm{k}-\bm{k}_{\nu,\sigma}}
          +i\hbar\epsilon}
    \right].
   \label{SE:MixGreen}
   \\[-7.5mm]\nonumber
\end{align}
Such evaluated
$\mathcal{G}^{\tilde{z}\tilde{x}}(\bm{k};\omega+i\epsilon)$ and
$\mathcal{G}^{\tilde{x}\tilde{z}}(\bm{k};\omega+i\epsilon)$
turn out still extremely small in value.
Indeed, they merely make an invisible contribution to the total dynamic structure factor
$S(\bm{q},\omega)
\equiv
 \sum_{\lambda=z,x,y}
 S^{\lambda\lambda}(\bm{q},\omega)$
shown in Fig. 2.
$\varGamma' {}_{    \!\!\!1}^{\left(-\frac{1}{2}\right)}$ and
$\varGamma''{}_{\!\!\!\!\!1}^{\left(-\frac{1}{2}\right)}$,
both to first order in $\mathcal{H}^{\left(\frac{1}{2}\right)}$ and on the order of
$S^{-\frac{1}{2}}$, bear some resemblance to $\varSigma'{}_{\!\!2}^{(0)}$
but are dimensionless in energy.
For the triangular-lattice antiferromagnet,
Mourigal \textit{et al.} \cite{M094407} demonstrate that the mixed transverse-longitudinal
correlators (\ref{SE:MixGreen}) do not affect the leading quasiparticle-like part of
the spectrum but sometimes cause an unphysical overcompensation, and therefore conclude that
they, as well as the anomalous Green's functions (\ref{SE:anormGFDysonAppr}),
can be neglected.

   Thus and thus, the MPSW-ground-state dynamic structure factor is evaluated as
\vspace{-2mm}
\begin{align}
   &
   S(\bm{q},\omega)
  \equiv
   \sum_{\lambda=x,y,z}
   S^{\lambda\lambda}(\bm{q},\omega)
  =-\frac{1}{2\pi\hbar}\sum_{\tau=\pm}
    \mathrm{Im}
    \left[
     \mathcal{G}_0^{\tilde{z}\tilde{z}}(\tau\bm{Q}_{\mathrm{X},\mathrm{K}}+\bm{q};\omega+i\epsilon)
    +\mathcal{G}^{\tilde{x}\tilde{x}}(\tau\bm{Q}_{\mathrm{X},\mathrm{K}}+\bm{q};\omega+i\epsilon)
    \right]
   \nonumber
   \allowdisplaybreaks \\
   &\qquad\qquad
   -\frac{1}{\pi\hbar}
    \mathrm{Im}
    \,
    \mathcal{G}^{\tilde{y}\tilde{y}}(\bm{q};\omega+i\epsilon)
   -\frac{1}{2\pi\hbar}\sum_{\tau=\pm}
    \tau
    \mathrm{Im}
    \left[
    \mathcal{G}^{\tilde{z}\tilde{x}}(\tau\bm{Q}_{\mathrm{X},\mathrm{K}}+\bm{q};\omega+i\epsilon)
   -\mathcal{G}^{\tilde{x}\tilde{z}}(\tau\bm{Q}_{\mathrm{X},\mathrm{K}}+\bm{q};\omega+i\epsilon)
    \right]
   \nonumber
   \allowdisplaybreaks \\
   &\qquad\quad
  =\frac{1}{2L}
   \sum_{\tau=\pm}
   \sum_{\sigma=0,\pm}
   \sum_{\nu=1}^N
   v_{\bm{k}_{\nu ,\sigma }}u_{\bm{k}_{\nu ,\sigma }+\tau\bm{Q}_{\mathrm{X},\mathrm{K}}+\bm{q}}
   ( u_{\bm{k}_{\nu ,\sigma }}
     v_{\bm{k}_{\nu ,\sigma }+\tau\bm{Q}_{\mathrm{X},\mathrm{K}}+\bm{q}}
    +v_{\bm{k}_{\nu ,\sigma }}
     u_{\bm{k}_{\nu ,\sigma }+\tau\bm{Q}_{\mathrm{X},\mathrm{K}}+\bm{q}}
    )
   \delta
   \left(
    \hbar\omega
   -\tilde{\varepsilon}_{\bm{k}_{\nu,\sigma}}
   -\tilde{\varepsilon}_{\bm{k}_{\nu,\sigma}
                        +\tau\bm{Q}_{\mathrm{X},\mathrm{K}}+\bm{q}}
   \right)
   \nonumber
   \allowdisplaybreaks \\
   &\qquad\qquad
  -\frac{S}{4\pi}
    \left[
     1
    -\frac{2\langle \mathcal{A}\rangle_0+\langle\mathcal{C}\rangle_0}
          {2S}
    \right]
   \sum_{\tau=\pm}
    (
     {u}_{\tau\bm{Q}_{\mathrm{X},\mathrm{K}}+\bm{q}}
    +{v}_{\tau\bm{Q}_{\mathrm{X},\mathrm{K}}+\bm{q}}
    )^{2}
   \mathrm{Im}
    \frac{1}
         {\hbar\omega-\tilde{\varepsilon}_{\tau\bm{Q}_{\mathrm{X},\mathrm{K}}+\bm{q}}
         -\varSigma^{(0)}(\tau\bm{Q}_{\mathrm{X},\mathrm{K}}+\bm{q};\omega+i\epsilon)
         +i\hbar\epsilon}
   \nonumber
   \allowdisplaybreaks \\
   &\qquad\qquad
  -\frac{S}{2\pi}
    \left[
     1
    -\frac{2\langle \mathcal{A}\rangle_0-\langle\mathcal{C}\rangle_0}
          {2S}
    \right]
    ({u}_{\bm{q}}-{v}_{\bm{q}})^{2}
    \mathrm{Im}
    \frac{1}
        {\hbar\omega-\tilde{\varepsilon}_{\bm{q}}
        -\varSigma^{(0)}(\bm{q};\omega+i\epsilon)+i\hbar\epsilon}
             \vphantom{\frac{1}{L}
                       \sum_{\sigma=0,\pm}\sum_{\nu=1}^N
                      }
   \nonumber
   \allowdisplaybreaks \\
   &\qquad\qquad
  -\frac{\sqrt{S}}{\sqrt{2}\pi}
   \left[1-\frac{2\langle \mathcal{A} \rangle_0
                 +\langle \mathcal{C} \rangle_0}{4S}
   \right]
   \sum_{\tau=\pm}
   (
    {u}_{\tau\bm{Q}_{\mathrm{X},\mathrm{K}}+\bm{q}}
   +{v}_{\tau\bm{Q}_{\mathrm{X},\mathrm{K}}+\bm{q}}
    )
   \nonumber
   \allowdisplaybreaks \\
   &\qquad\qquad
   \times
   \tau
   \mathrm{Im}
   \left[
    \frac{\varGamma' {}_{    \!\!\!1}^{\left(-\frac{1}{2}\right)}
           (\tau\bm{Q}_{\mathrm{X},\mathrm{K}}+\bm{q};\omega+i\epsilon)}
         {\hbar\omega-\tilde{\varepsilon}_{\tau\bm{Q}_{\mathrm{X},\mathrm{K}}+\bm{q}}
         -\varSigma^{(0)}(\tau\bm{Q}_{\mathrm{X},\mathrm{K}}+\bm{q};\omega+i\epsilon)
         +i\hbar\epsilon}
   -\frac{\varGamma''{}_{    \!\!\!1}^{\left(-\frac{1}{2}\right)}
           (\tau\bm{Q}_{\mathrm{X},\mathrm{K}}+\bm{q};\omega+i\epsilon)}
         {\hbar\omega+\tilde{\varepsilon}_{\tau\bm{Q}_{\mathrm{X},\mathrm{K}}+\bm{q}}
         +\varSigma^{(0)}(-\tau\bm{Q}_{\mathrm{X},\mathrm{K}}-\bm{q};\omega-i\epsilon)
         +i\hbar\epsilon}
   \right].
   \label{SE:SqwtotcorrEX}
\end{align}
With
$\varSigma^{(0)}$,
$\varGamma' {}_{    \!\!\!1}^{\left(-\frac{1}{2}\right)}$, and
$\varGamma''{}_{    \!\!\!1}^{\left(-\frac{1}{2}\right)}$ all away,
(\ref{SE:SqwtotcorrEX}) reduces to the MLSW-ground-state dynamic structure factor
\vspace{-3mm}
\begin{align}
   &
   S(\bm{q},\omega)
  =\frac{1}{2L}
   \sum_{\tau=\pm}
   \sum_{\sigma=0,\pm}
   \sum_{\nu=1}^N
   v_{\bm{k}_{\nu ,\sigma }}u_{\bm{k}_{\nu ,\sigma }+\tau\bm{Q}_{\mathrm{X},\mathrm{K}}+\bm{q}}
   ( u_{\bm{k}_{\nu ,\sigma }}
     v_{\bm{k}_{\nu ,\sigma }+\tau\bm{Q}_{\mathrm{X},\mathrm{K}}+\bm{q}}
    +v_{\bm{k}_{\nu ,\sigma }}
     u_{\bm{k}_{\nu ,\sigma }+\tau\bm{Q}_{\mathrm{X},\mathrm{K}}+\bm{q}}
    )
   \delta
   \left(
    \hbar\omega
   -\tilde{\varepsilon}_{\bm{k}_{\nu,\sigma}}
   -\tilde{\varepsilon}_{\bm{k}_{\nu,\sigma}
                        +\tau\bm{Q}_{\mathrm{X},\mathrm{K}}+\bm{q}}
   \right)
   \nonumber
   \allowdisplaybreaks \\
   &\qquad\qquad
  +\frac{S}{4}
    \left[
     1
    -\frac{2\langle\mathcal{A}\rangle_0+\langle\mathcal{C}\rangle_0}
          {2S}
    \right]
   \sum_{\tau=\pm}
   (
    u_{\tau\bm{Q}_{\mathrm{X},\mathrm{K}}+\bm{q}}
   +v_{\tau\bm{Q}_{\mathrm{X},\mathrm{K}}+\bm{q}}
   )^2
   \delta
   \left(
    \hbar\omega
   -\tilde{\varepsilon}_{\tau\bm{Q}_{\mathrm{X},\mathrm{K}}+\bm{q}}
   \right)
   \nonumber
   \allowdisplaybreaks \\
   &\qquad\qquad
  +\frac{S}{2}
    \left[
     1
    -\frac{2\langle\mathcal{A}\rangle_0-\langle\mathcal{C}\rangle_0}
          {2S}
    \right]
   (u_{\bm{q}}-v_{\bm{q}})^2
   \delta\left(\hbar\omega-\tilde{\varepsilon}_{\bm{q}}\right).
   \label{SE:SqwMLSW}
\end{align}
Note that
$\langle\mathcal{A}\rangle_0^{\mathrm{SC}}=S$,
whereas
$\langle\mathcal{A}\rangle_0^{\mathrm{DC}}=S-\delta$ and
$\langle\mathcal{C}\rangle_0^{\mathrm{DC}}=0$.

   Since SC MLSWs have such a small chemical potential and therefore excitation energy that
$\mu\simeq 0.0069JS$ and
$\tilde{\varepsilon}_{\bm{Q}_{\Gamma,\Gamma}}^{\mathrm{SC}}\simeq 0.2402JS$
[Eq. (\ref{E:ekSC-MLSW}) and Fig. 1(c1)], they overestimate a single-magnon scattering
intensity in $S(\bm{Q}_{\mathrm{X},\mathrm{K}},\omega)$ [Fig. 2(a2)] as
\vspace{-1mm}
\begin{align}
   &
   S(\bm{Q}_{\mathrm{X},\mathrm{K}},\omega)
  \simeq
   \frac{S}{4}
    \left[
     1
    -\frac{2\langle\mathcal{A}\rangle_0+\langle\mathcal{C}\rangle_0}
          {2S}
    \right]
   (
    u_{-\bm{Q}_{\mathrm{X},\mathrm{K}}+\bm{Q}_{\mathrm{X},\mathrm{K}}}
   +v_{-\bm{Q}_{\mathrm{X},\mathrm{K}}+\bm{Q}_{\mathrm{X},\mathrm{K}}}
   )^2
   \delta
   \left(
    \hbar\omega
   -\tilde{\varepsilon}_{-\bm{Q}_{\mathrm{X},\mathrm{K}}
                         +\bm{Q}_{\mathrm{X},\mathrm{K}}}^{\mathrm{SC}}
   \right)
   \nonumber
   \allowdisplaybreaks \\
   &\quad
  =\left(
     \frac{S}{4}+\frac{1}{8}
    -\sum_{\sigma=0,\pm}
     \sum_{\nu=1}^N
     \frac{
       u_{\bm{k}_{\nu,\sigma}}^2
      +v_{\bm{k}_{\nu,\sigma}}^2
      +u_{\bm{k}_{\nu,\sigma}}v_{\bm{k}_{\nu,\sigma}}
     }{8L}
   \right)
   (
    u_{\bm{Q}_{\Gamma,\Gamma}}
   +v_{\bm{Q}_{\Gamma,\Gamma}}
   )^2
   \delta
   \left(
    \hbar\omega
   -\tilde{\varepsilon}_{\bm{Q}_{\Gamma,\Gamma}}^{\mathrm{SC}}
   \right)
   \nonumber
   \allowdisplaybreaks \\
   &\quad
  =\left(
     \frac{S}{4}+\frac{1}{8}
    -\sum_{\sigma=0,\pm}
     \sum_{\nu=1}^N
     \frac{
       u_{\bm{k}_{\nu,\sigma}}^2
      +v_{\bm{k}_{\nu,\sigma}}^2
      +u_{\bm{k}_{\nu,\sigma}}v_{\bm{k}_{\nu,\sigma}}
     }{8L}
   \right)
   \left\{
    \frac{1}{4\tilde{\varepsilon}_{\bm{Q}_{\Gamma,\Gamma}}^{\mathrm{SC}}}
    \left[
     21JS-2\mu
    +\frac{(7JS)^{2}}
          {2\tilde{\varepsilon}_{\bm{Q}_{\Gamma,\Gamma}}^{\mathrm{SC}}+7JS-2\mu}
    \right]
   +\frac{1}{2}
   \right\}
   \delta
   \left(
    \hbar\omega
   -\tilde{\varepsilon}_{\bm{Q}_{\Gamma,\Gamma}}^{\mathrm{SC}}
   \right)
   \nonumber
   \allowdisplaybreaks \\
   &\quad
  \simeq
   \left(
    \frac{S}{4}
   +\frac{1}{8}
   +0.1545
   \right)
   \frac{7JS}{\tilde{\varepsilon}_{\bm{Q}_{\Gamma,\Gamma}}^{\mathrm{SC}}}
   \delta
   \left(
    \hbar\omega
   -\tilde{\varepsilon}_{\bm{Q}_{\Gamma,\Gamma}}^{\mathrm{SC}}
   \right)
  =0.4045
   \frac{7J}{2\tilde{\varepsilon}_{\bm{Q}_{\Gamma,\Gamma}}^{\mathrm{SC}}}
   \delta
   \left(
    \hbar\omega
   -\tilde{\varepsilon}_{\bm{Q}_{\Gamma,\Gamma}}^{\mathrm{SC}}
   \right)
   \ 
   \left(
    S=\frac{1}{2}
   \right).
   \label{SE:Sqw-QXK}
   \\[-7.5mm]\nonumber
\end{align}
Besides, SC MLSWs have smaller excitation energies such that
$\tilde{\varepsilon}_{\pm\bm{Q}_{\mathrm{X},\mathrm{K}}}^{\mathrm{SC}}\simeq 0.2130JS$
[Eq. (\ref{E:ekSC-MLSW}) and Fig. 1(c2)], and therefore they further overestimate a two-magnon
scattering intensity in $S(\bm{Q}_{\mathrm{X},\Gamma},\omega)$ [see Fig. 2(a1)] as
\vspace{-1mm}
\begin{align}
   &
   S(\bm{Q}_{\mathrm{X},\Gamma},\omega)
  \simeq
   v_{\bm{Q}_{\mathrm{X},\mathrm{K}}}
   u_{\bm{Q}_{\mathrm{X},\mathrm{K}}+\bm{Q}_{\mathrm{X},\mathrm{K}}+\bm{Q}_{\mathrm{X},\Gamma}}
   ( u_{\bm{Q}_{\mathrm{X},\mathrm{K}}}
     v_{\bm{Q}_{\mathrm{X},\mathrm{K}}+\bm{Q}_{\mathrm{X},\mathrm{K}}+\bm{Q}_{\mathrm{X},\Gamma}}
    +v_{\bm{Q}_{\mathrm{X},\mathrm{K}}}
     u_{\bm{Q}_{\mathrm{X},\mathrm{K}}+\bm{Q}_{\mathrm{X},\mathrm{K}}+\bm{Q}_{\mathrm{X},\Gamma}}
    )
   \delta
   \left(
    \hbar\omega
   -\tilde{\varepsilon}_{\bm{Q}_{\mathrm{X},\mathrm{K}}}^{\mathrm{SC}}
   -\tilde{\varepsilon}_{\bm{Q}_{\mathrm{X},\mathrm{K}}
                        +\bm{Q}_{\mathrm{X},\mathrm{K}}+\bm{Q}_{\mathrm{X},\Gamma}}^{\mathrm{SC}}
   \right)
   \nonumber
   \allowdisplaybreaks \\
   &\quad
  =v_{ \bm{Q}_{\mathrm{X},\mathrm{K}}}
   u_{-\bm{Q}_{\mathrm{X},\mathrm{K}}}
   ( u_{ \bm{Q}_{\mathrm{X},\mathrm{K}}}
     v_{-\bm{Q}_{\mathrm{X},\mathrm{K}}}
    +v_{ \bm{Q}_{\mathrm{X},\mathrm{K}}}
     u_{-\bm{Q}_{\mathrm{X},\mathrm{K}}}
    )
   \delta
   \left(
    \hbar\omega
   -\tilde{\varepsilon}_{ \bm{Q}_{\mathrm{X},\mathrm{K}}}^{\mathrm{SC}}
   -\tilde{\varepsilon}_{-\bm{Q}_{\mathrm{X},\mathrm{K}}}^{\mathrm{SC}}
   \right)
  =\frac{(11JS)^{2}}
   {32\tilde{\varepsilon}_{ \bm{Q}_{\mathrm{X},\mathrm{K}}}^{\mathrm{SC}}
      \tilde{\varepsilon}_{-\bm{Q}_{\mathrm{X},\mathrm{K}}}^{\mathrm{SC}}}
   \delta
   \left(
    \hbar\omega
   -\tilde{\varepsilon}_{ \bm{Q}_{\mathrm{X},\mathrm{K}}}^{\mathrm{SC}}
   -\tilde{\varepsilon}_{-\bm{Q}_{\mathrm{X},\mathrm{K}}}^{\mathrm{SC}}
   \right).
   \label{SE:Sqw-QXG}
   \\[-7.5mm]\nonumber
\end{align}

   On the other hand, the SC-MWDISW-ground-state dynamic structure factor is evaluated as
\vspace{-1mm}
\begin{align}
   &
   S^{\lambda\lambda'}(\bm{q},\omega)
  =\frac{1}{2\pi\hbar}
   \int_{-\infty}^{\infty}
   \left\langle
    \delta S_{ \bm{q}}^{\lambda}(t)
    \delta S_{-\bm{q}}^{\lambda'}
   \right\rangle_0'^{\mathrm{SC}}
   e^{i\omega t}dt
  =\frac{1}{2\pi\hbar}
   \int_{-\infty}^{\infty}
   \left\langle
    e^{ i\widetilde{\mathcal{H}}'{}^{\mathrm{SC}}t/\hbar}
    \delta S_{\bm{q}}^{\lambda}
    e^{-i\widetilde{\mathcal{H}}'{}^{\mathrm{SC}}t/\hbar}
    \delta S_{-\bm{q}}^{\lambda'}
   \right\rangle_0'^{\mathrm{SC}}
   e^{i\omega t}dt;\ 
   \delta S_{\bm{q}}^{\lambda}
  =S_{\bm{q}}^{\lambda}
  -\left\langle S_{ \bm{q}}^{\lambda} \right\rangle_0'^{\mathrm{SC}},
   \label{SE:SqwWICK}
   \allowdisplaybreaks \\[-1mm]
   &
   S(\bm{q},\omega)
  =\frac{1}{2L}
   \sum_{\tau=\pm}
   \sum_{\sigma=0,\pm}\sum_{\nu=1}^N
   v_{\bm{k}_{\nu ,\sigma }}u_{\bm{k}_{\nu ,\sigma }+\tau\bm{Q}_{\mathrm{X},\mathrm{K}}+\bm{q}}
   ( u_{\bm{k}_{\nu ,\sigma }}
     v_{\bm{k}_{\nu ,\sigma }+\tau\bm{Q}_{\mathrm{X},\mathrm{K}}+\bm{q}}
    +v_{\bm{k}_{\nu ,\sigma }}
     u_{\bm{k}_{\nu ,\sigma }+\tau\bm{Q}_{\mathrm{X},\mathrm{K}}+\bm{q}}
    )
   \delta
   \left(
    \hbar\omega
   -\tilde{\varepsilon}'{}_{\!\!\bm{k}_{\nu,\sigma}}^{\mathrm{SC}}
   -\tilde{\varepsilon}'{}_{\!\!\bm{k}_{\nu,\sigma}
                                +\tau\bm{Q}_{\mathrm{X},\mathrm{K}}+\bm{q}}^{\mathrm{SC}}
   \right)
   \nonumber
   \allowdisplaybreaks \\
   &\qquad\qquad
  +\frac{S}{4}
    \left[
     1
    -\frac{2\langle \mathcal{A}\rangle'{}_{\!\!0}^{\mathrm{SC}}
          +\langle\mathcal{C}\rangle'{}_{\!\!0}^{\mathrm{SC}}}
          {2S}
    \right]
   \sum_{\tau=\pm}
   (
    u_{\tau\bm{Q}_{\mathrm{X},\mathrm{K}}+\bm{q}}
   +v_{\tau\bm{Q}_{\mathrm{X},\mathrm{K}}+\bm{q}}
   )^2
   \delta
   \left(
    \hbar\omega
   -\tilde{\varepsilon}'{}_{\!\!\tau\bm{Q}_{\mathrm{X},\mathrm{K}}+\bm{q}}^{\mathrm{SC}}
   \right)
   \nonumber
   \allowdisplaybreaks \\
   &\qquad\qquad
  +\frac{S}{2}
    \left[
     1
    -\frac{2\langle \mathcal{A}\rangle'{}_{\!\!0}^{\mathrm{SC}}
           -\langle\mathcal{C}\rangle'{}_{\!\!0}^{\mathrm{SC}}}
          {2S}
    \right]
   (u_{\bm{q}}-v_{\bm{q}})^2
   \delta\left(\hbar\omega-\tilde{\varepsilon}'{}_{\!\!\bm{q}}^{\mathrm{SC}}\right).
   \label{SE:SqwMWDISW}
   \\[-7.5mm]\nonumber
\end{align}
Note that
$\langle\mathcal{A}\rangle'{}_{\!\!0}^{\mathrm{SC}}=S$.
The expression (\ref{SE:SqwMWDISW}) looks the same as (\ref{SE:SqwMLSW}) except for
$\langle\mathcal{A}\rangle'{}_{\!\!0}^{\mathrm{SC}}$,
$\langle\mathcal{C}\rangle'{}_{\!\!0}^{\mathrm{SC}}$, and
$\tilde{\varepsilon}'{}_{\!\!\bm{k}}^{\mathrm{SC}}$
instead of
$\langle\mathcal{A}\rangle_0^{\mathrm{SC}}$,
$\langle\mathcal{C}\rangle_0^{\mathrm{SC}}$, and
$\tilde{\varepsilon}_{\bm{k}}^{\mathrm{SC}}$, respectively.

   The $\mathcal{H}^{\left(\frac{1}{2}\right)}$-driven $O(S^0)$ self-energy
$\displaystyle
 \mathop{\varSigma}^{\,\protect\substack{-\!\!\!-\\[-1.6mm]}}\!{}_2^{(0)}
 (\bm{k};\tilde{\varepsilon}_{\bm{k}}^{\mathrm{DC}}/\hbar)$
causes a strong downward renormalization of the bare magnon energies
$\tilde{\varepsilon}_{\bm{k}}^{\mathrm{DC}}$
at $k^x=\pm Q_{\mathrm{K}}^x$ and $|k^z|\alt \frac{Q_{\mathrm{X}}^z}{2}$.
The thus-obtained DC MPSWs yield the scattering intensities
\vspace{-1mm}
\begin{align}
   &
   S(\bm{q},\omega)
  \simeq
  -\frac{S}{4\pi}
    \left[
     1
    -\frac{2\langle \mathcal{A}\rangle_0^{\mathrm{DC}}+\langle\mathcal{C}\rangle_0^{\mathrm{DC}}}
          {2S}
    \right]
   \sum_{\tau=\pm}
   \left(
    u_{\tau\bm{Q}_{\mathrm{X},\mathrm{K}}+\bm{q}}+v_{\tau\bm{Q}_{\mathrm{X},\mathrm{K}}+\bm{q}}
   \right)^2
   \mathrm{Im}
   \frac{1}
        {\hbar\omega
        -\tilde{\varepsilon}_{\tau\bm{Q}_{\mathrm{X},\mathrm{K}}+\bm{q}}^{\mathrm{DC}}
        -\varSigma^{(0)}
         (\tau\bm{Q}_{\mathrm{X},\mathrm{K}}+\bm{q};\omega+i\epsilon)
        +i\hbar\epsilon}
   \label{SE:1MShoulder}
   \\[-7.5mm]\nonumber
\end{align}
at $q^x=Q_\Gamma^x$ and $\frac{Q_{\mathrm{X}}^z}{2} \ll |q^z|\alt Q_{\mathrm{X}}^z$ and
\vspace{-1mm}
\begin{align}
   &
   S(\bm{q},\omega)
  \simeq
  -\frac{S}{2\pi}
    \left[
     1
    -\frac{2\langle \mathcal{A}\rangle_0^{\mathrm{DC}}-\langle\mathcal{C}\rangle_0^{\mathrm{DC}}}
          {2S}
    \right]
   \left(u_{\bm{q}}-v_{\bm{q}}\right)^2
   \mathrm{Im}
   \frac{1}
        {\hbar\omega-\tilde{\varepsilon}_{\bm{q}}^{\mathrm{DC}}
       -\varSigma^{(0)}
        (\bm{q};\omega+i\epsilon)
       +i\hbar\epsilon}
   \label{SE:1MDownwardDispersion}
   \\[-7.5mm]\nonumber
\end{align}
at $q^x=\pm Q_{\mathrm{K}}^x$ and $Q_{\Gamma}^z \alt |q^z| \alt \frac{Q_{\mathrm{X}}^z}{2}$
to successfully reproduce the DMRG calculations,
as are illustrated with vertical arrows
in Figs. 2($\mathrm{b}1'$) and 2($\mathrm{c}1$) and
in Figs. 2($\mathrm{b}2'$) and 2($\mathrm{c}2$), respectively.
This is not the case with SC MPSWs.
$\tilde{\varepsilon}_{\bm{k}}^{\mathrm{DC}}$ owes its strong downward renormalization
at $k^x=\pm Q_{\mathrm{K}}^x$ and $|k^z|\alt \frac{Q_{\mathrm{X}}^z}{2}$
to the cubic-interaction-driven self-energy correction
$\displaystyle
 \mathop{\varSigma}^{\,\substack{-\!\!\!-\\[-1.8mm]}}\!{}_2^{(0)}
  (\bm{k};\tilde{\varepsilon}_{\bm{k}}^{\mathrm{DC}}/\hbar)\,(<0)$
surpassing the quartic-interaction-driven one
$\displaystyle
 \mathop{\varSigma}^{\,\substack{-\!\!\!-\\[-1.8mm]}}\!{}_1^{(0)}(\bm{k})\,(>0)$
in magnitude [Fig. 3($\mathrm{a}2''$)].
These self-energy corrections cancel out each other
at $k^x=\pm Q_{\mathrm{K}}^x$ and $|k^z|\alt \frac{Q_{\mathrm{X}}^z}{2}$
under the SC condition
[Fig. 3($\mathrm{b}2''$)].
There is no downward renormalization of the bare magnon energies $\tilde{\varepsilon}_{\bm{k}}$
with MWDISWs either, whether SC or DC conditioned.
Variationally interacting SWs are totally ignorant of cubic anharmonicities.

\vspace{-5mm}
\section{Double-Constraint versus Single-Constraint Modification Schemes}
\label{S:DCvsSC}

   The DC modification scheme retains the local $\mathrm{U}(1)$ symmetry without spoiling
the ground-state energy in the SC modification scheme.
Let $\widetilde{\mathcal{H}}_{\mathrm{harm}}$ be
either
$\widetilde{\mathcal{H}}_{\mathrm{harm}}^{\mathrm{SC}}$
or
$\widetilde{\mathcal{H}}_{\mathrm{harm}}^{\mathrm{DC}}$.
Then the MSW ground-state energy $E_{\mathrm{g}}$ is written as
$\langle
  \widetilde{\mathcal{H}}_{\mathrm{harm}}
 \rangle_0
=\sum_{m=1}^2 E^{(m)}
+\mu L\left(S-\delta+\frac{1}{2}\right)$
(MLSWs)
or
$\langle
  \widetilde{\mathcal{H}}_{\mathrm{harm}}
 \rangle_0
+\delta E_1^{(0)}
+\delta E_2^{(0)}
$
(MPSWs),
where $\delta$ (\ref{SE:DC-MLSW}) serves to optimize the ground-state energy,
while the perturbative corrections to first order in $\mathcal{H}^{\left(0\right)}$
(\ref{SE:H(0)alpha}) and second order in $\mathcal{H}^{\left(\frac{1}{2}\right)}$
(\ref{SE:H(1/2)alpha}) are given by
\vspace{-1mm}
\begin{align}
   &
   \delta E_1^{(0)}
  =\langle \mathcal{H}^{(0)} \rangle_0
  =-JL\sum_{\sigma=\parallel,\perp}
   \left\{
    \left[
     \vphantom{\frac{\langle\mathcal{B}_\sigma\rangle_0
                    +\langle\mathcal{D}_\sigma\rangle_0}{2}}
     \left(\langle\mathcal{A}\rangle_0\right)^2
    +\left(\langle\mathcal{B}_\sigma\rangle_0\right)^2
    +\left(\langle\mathcal{D}_\sigma\rangle_0\right)^2
    -\left(
      2\langle\mathcal{A}\rangle_0
     + \langle\mathcal{C}\rangle_0
     \right)
     \frac{\langle\mathcal{B}_\sigma\rangle_0
          +\langle\mathcal{D}_\sigma\rangle_0}{2}
    \right]
    \cos\bm{Q}_{\mathrm{X},\mathrm{K}}\cdot{\bm{\delta}_\sigma}
   \right.
   \nonumber
   \allowdisplaybreaks \\
   &\qquad\qquad\qquad
    \left.
    -\left(
      2\langle\mathcal{A}\rangle_0
     - \langle\mathcal{C}\rangle_0
     \right)
     \frac{\langle\mathcal{B}_\sigma\rangle_0
          -\langle\mathcal{D}_\sigma\rangle_0}{2}
   -\left(
     \frac{\langle\mathcal{A}\rangle_0}{2}
    -\frac{\langle\mathcal{B}_\sigma\rangle_0
           +\langle\mathcal{D}_\sigma\rangle_0}{4}
    \right)
    \cos\bm{Q}_{\mathrm{X},\mathrm{K}}\cdot{\bm{\delta}_\sigma}
   -\frac{\langle\mathcal{B}_\sigma\rangle_0
         -\langle\mathcal{D}_\sigma\rangle_0}{4}
   \right\},
   \label{SE:delEH(0)}
   \allowdisplaybreaks \\
   &
   \delta E_2^{(0)}
  =\begin{array}{c}
   \displaystyle
   \prod_{\tau=0,\pm}\prod_{\mu=1}^{N}
   \sum_{n_{\bm{k}_{\mu,\tau}}=0}^{\infty} \\
   \scriptstyle
   \left(
   \text{$\sum_{\nu=1}^{N}\sum_{\sigma=0,\pm}n_{\bm{k}_{\nu,\sigma}}\ne0$}
   \right)
   \end{array}
   \frac{\displaystyle
    \left|
    \Biggl\langle
    \prod_{\sigma=0,\pm}\prod_{\nu=1}^{N}
    \frac{
     \left(\alpha_{\bm{k}_{\nu,\sigma}}\right)^{n_{\bm{k}_{\nu,\sigma}}}
    }{\sqrt{n_{\bm{k}_{\nu,\sigma}}!}}
    \mathcal{H}^{\left(\frac{1}{2}\right)}
    \Biggr\rangle_0
    \right|^{2}
   }
   {\displaystyle
   -\sum_{\sigma=0,\pm}\sum_{\nu=1}^{N}
    \tilde{\varepsilon}_{ \bm{k}_{\nu,\sigma}}
    n_{\bm{k}_{\nu,\sigma}}
   }
  =\sum_{\sigma'=0,\pm}\sum_{\nu'=1}^N
   \sum_{\sigma=0,\pm}\sum_{\nu=1}^N
   \frac{6|\varLambda_{+++}
   \left(
    \bm{k}_{\nu ,\sigma },
   -\bm{k}_{\nu ,\sigma }-\bm{k}_{\nu',\sigma'},
    \bm{k}_{\nu',\sigma'}
   \right)|^{2}}
   {-\tilde{\varepsilon}_{ \bm{k}_{\nu ,\sigma }}
    -\tilde{\varepsilon}_{-\bm{k}_{\nu ,\sigma }-\bm{k}_{\nu',\sigma'}}
    -\tilde{\varepsilon}_{ \bm{k}_{\nu',\sigma'}}}.
   \label{SE:delEH(1/2)}
   \\[-7.5mm]\nonumber
\end{align}
We plot $E_{\mathrm{g}}$ as a function of $\delta$ in various manners of modifying SWs
in Fig. \ref{SF:ETuning}.
While just adding the second constraint condition in (\ref{SE:DC-MLSW}) to the conventional
constraint condition (\ref{SE:SC-MLSW}) causes a rise in the ground-state energy, further tuning up
the first one in (\ref{SE:DC-MLSW}) with respect to $\delta$ cancels out this energy cost.

   We further understand the necessity of boson number tuning in the DC modification scheme
when we observe DC-MSW calculations of the structure factor in more detail.
SC MLSWs describe the scattering intensities at
$\bm{Q}_{\mathrm{X},\mathrm{K}}$ and $\bm{Q}_{\mathrm{X},\Gamma}$
as (\ref{SE:Sqw-QXK}) and (\ref{SE:Sqw-QXG}), respectively, and these descriptions turn out to be
such overestimates as Figs. 2(a1) and 2(a2), respectively.
Their DC-MLSW replacements are such improvements as Figs. 2(b1) and 2(b2)
and further developed DC-MPSW calculations are even better than them as
Figs. 2($\mathrm{b}1'$) and 2($\mathrm{b}2'$),
when we rely upon the DMRG findings Figs. 2(c1) and 2(c2).
In order to demonstrate how important the tuning of the first constraint condition is
in this context, we show in Fig. \ref{SF:S(q)Tuning} DC-MSW findings as functions of $\delta$ for
the static structure factor
$
   S(\bm{q})
  =\int_{0}^{\infty}
   S({\bm{q},\omega})
   d\omega
$
at $\bm{Q}_{\mathrm{X},\mathrm{K}}$ and $\bm{Q}_{\mathrm{X},\Gamma}$
in comparison with DMRG estimates \cite{N054421}.
SC-MSW findings are terrible overestimates.
DC-MSW estimates without tuning the total number of emergent bosons remain too large,
but they can well hit the DMRG calculations at the end, if pertinently tuned.
\begin{figure*}[t]
\centering
\begin{minipage}{0.475\textwidth}
\includegraphics[width=85mm]{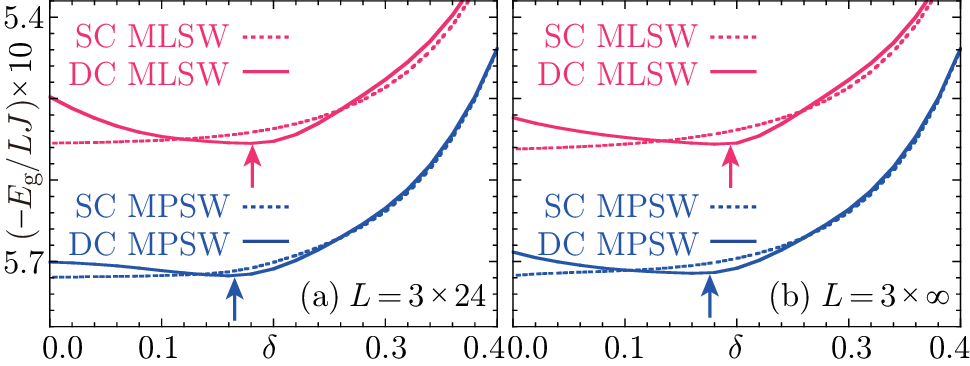}
\vspace{-6mm}
\caption{MSW ground-state energies
         $E_{\mathrm{g}}
         =\langle
           \widetilde{\mathcal{H}}_{\mathrm{harm}}
          \rangle_0
         =\sum_{m=1}^2 E^{(m)}
         +\mu L\left(S-\delta+\frac{1}{2}\right)$ (MLSW)
         and
         $E_{\mathrm{g}}
         =\langle
           \widetilde{\mathcal{H}}_{\mathrm{harm}}
          \rangle_0
         +\delta E_1^{(0)}
         +\delta E_2^{(0)}$
         (MPSW)
         under the SC (\ref{E:ekSC-MLSW}) and DC (\ref{E:ekDC-MLSW}) modification schemes
         for $S=\frac{1}{2}$ at $N=24$ (a) and in the $N \to \infty$ limit (b).
         The optimal values of $\delta$ in the DC scheme (\ref{SE:DC-MLSW}) are indicated by
         arrows, whereas those in the SC scheme are always zero.}
\label{SF:ETuning}
\end{minipage}
\hspace{5mm}
\centering
\begin{minipage}{0.475\textwidth}
\vspace{-2.5mm}
\includegraphics[width=85mm]{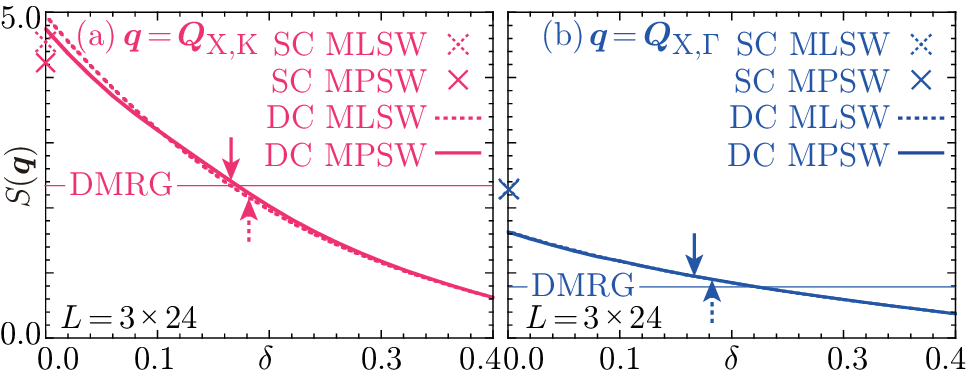}
\vspace{-6mm}
\caption{DC-MSW findings as functions of $\delta$, together with SC-MSW findings just
         for reference, for the static structure factors
         $S(\bm{Q}_{\mathrm{X,K}})$ (a) and $S(\bm{Q}_{\mathrm{X,\Gamma}})$ (b)
         in comparison with the DMRG estimates \cite{N054421} as criteria for
         $S=\frac{1}{2}$ at $N=24$.
         The optimal values of $\delta$ in the DC scheme (\ref{SE:DC-MLSW}) determined in
         Fig. \ref{SF:ETuning} are again indicated by arrows.}
\label{SF:S(q)Tuning}
\end{minipage}
\vspace*{-5mm}
\end{figure*}

   The DC modification scheme really makes subtle observations when supplemented by
perturbative corrections.
This is not the case with the SC modification scheme.
In order to contrast them in more detail, it is also worth while comparing
the single-CLSW excitation spectrum (\ref{E:ekCLSW}) and double-CLSW excitation continuum band
$
   \mathcal{E}_{\bm{k};\bm{q}}
  \equiv
   \varepsilon_{\frac{\bm{k}}{2}+\bm{q}}+\varepsilon_{\frac{\bm{k}}{2}-\bm{q}}
$
(Fig. \ref{SF:DispCLSW}), though no perturbative correction is available there.
The lower boundary of $\mathcal{E}_{\bm{k};\bm{q}}$ is analytically available as
\vspace{-2mm}
\begin{align}
   &
   \mathcal{E}_{\bm{k};\bm{q}_{\mathrm{min}}(\bm{k})}
  =\left\{
   \begin{array}{l}
   \varepsilon_{(    Q_{\mathrm{X}}^z,\mp Q_{\mathrm{K}}^x)}
  +\varepsilon_{(k^z-Q_{\mathrm{X}}^z,\pm Q_{\mathrm{K}}^x)}
  =\varepsilon_{(k^z-Q_{\mathrm{X}}^z,\pm Q_{\mathrm{K}}^x)}
   \left\{
   \begin{array}{l}
  =\varepsilon_{\bm{Q}_{\Gamma,\Gamma}}
   \hspace{4.8mm}
   \displaystyle
   \left(
    k^x=Q_\Gamma^x,\ 
    k^z=Q_\Gamma^z
   \right)
   \\[2mm]
  <\varepsilon_{(k^z,Q_\Gamma^x)}
   \hspace{2.5mm}
   \displaystyle
   \left(
    k^x=Q_\Gamma^x,\ 
    Q_\Gamma^z < |k^z| \leq Q_{\mathrm{X}}^z
   \right)\\
   \end{array}
   \right.
   \\[6mm]
   \varepsilon_{(    Q_{\mathrm{X}}^z,\mp Q_{\mathrm{K}}^x)}
  +\varepsilon_{(k^z-Q_{\mathrm{X}}^z,\mp Q_{\mathrm{K}}^x)}
  =\varepsilon_{(k^z-Q_{\mathrm{X}}^z,\mp Q_{\mathrm{K}}^x)}
  <\varepsilon_{(k^z,\pm Q_{\mathrm{K}}^x)}
   \hspace{3.1mm}
   \displaystyle
   \left(
    k^x=\pm Q_{\mathrm{K}}^x,\ 
   |k^z| < \frac{Q_{\mathrm{X}}^z}{2}
   \right) 
   \\[4mm]
   \varepsilon_{(    Q_\Gamma^z, Q_\Gamma^x)}
  +\varepsilon_{(k^z-Q_\Gamma^z,\pm Q_{\mathrm{K}}^x)}
  =\varepsilon_{(k^z-Q_\Gamma^z,\pm Q_{\mathrm{K}}^x)}
  =\varepsilon_{(k^z,\pm Q_{\mathrm{K}}^x)}
   \hspace{5.8mm}
   \displaystyle
   \left(
    k^x=\pm Q_{\mathrm{K}}^x,\ 
    \frac{Q_{\mathrm{X}}^z}{2} \leq |k^z| \le Q_{\mathrm{X}}^z
   \right)
   \end{array}
   \right..
   \label{SE:CSWCM}
\end{align}
The CLSW spectrum contains three Nambu-Goldstone modes to be buried in its pair excitation
continuum (Fig. \ref{SF:DispCLSW}).
Modifying CLSWs causes a hardening of their dispersion such that
$0
<\tilde{\varepsilon}_{\pm\bm{Q}_{\mathrm{X},\mathrm{K}}}^{\mathrm{SC}}
<\tilde{\varepsilon}_{\bm{Q}_{\Gamma,\Gamma}}^{\mathrm{SC}}
<\tilde{\varepsilon}_{\bm{Q}_{\Gamma,\Gamma}}^{\mathrm{DC}}
<\tilde{\varepsilon}_{\pm\bm{Q}_{\mathrm{X},\mathrm{K}}}^{\mathrm{DC}}$
[Figs. 3($\mathrm{a}1$) to 3($\mathrm{b}2$)],
and therefore any MLSW spectrum is no longer completely buried in its pair excitation continuum.
DC MLSWs have larger gaps than SC MLSWs in general,
$\tilde{\varepsilon}_{\bm{Q}_{\Gamma,\Gamma}}^{\mathrm{SC}}
<\tilde{\varepsilon}_{\bm{Q}_{\Gamma,\Gamma}}^{\mathrm{DC}}$
and
$\tilde{\varepsilon}_{\pm\bm{Q}_{\mathrm{X},\mathrm{K}}}^{\mathrm{SC}}
<\tilde{\varepsilon}_{\pm\bm{Q}_{\mathrm{X},\mathrm{K}}}^{\mathrm{DC}}$,
and therefore succeed in yielding such subtle observations as are illustrated with vertical arrows
in Figs. 2($\mathrm{b}1'$) and 2($\mathrm{b}2'$).
On the other hand, having in mind that
$v_{\bm{Q}_{\mathrm{X},\mathrm{K}}}=\sqrt{\frac{11}{2}}JSa
<\sqrt{7}JSa=v_{\bm{Q}_{\Gamma,\Gamma}}$,
DC MLSWs with
$\tilde{\varepsilon}_{\bm{Q}_{\Gamma,\Gamma}}^{\mathrm{DC}}
<\tilde{\varepsilon}_{\pm\bm{Q}_{\mathrm{X},\mathrm{K}}}^{\mathrm{DC}}$
are stabler than
SC MLSWs with
$\tilde{\varepsilon}_{\pm\bm{Q}_{\mathrm{X},\mathrm{K}}}^{\mathrm{SC}}
<\tilde{\varepsilon}_{\bm{Q}_{\Gamma,\Gamma}}^{\mathrm{SC}}$,
i.e.,
the single-DC (SC)-MLSW branch comes below the lower edge of the double-DC (SC) -MLSW continuum
in a major (minor) part of the Brillouin zone (Table \ref{ST:SMD}).
That is why under the successful DC modification scheme,
the inequality condition of bare magnons being kinematically unstable towards decays into pairs of
other magnons occupies a part of the Brillouin zone far smaller than
the region of the lowest-order self-energy corrections to bare magnons having
an imaginary part, implying a novel \textit{mixed instability}.

\begin{figure}
\begin{tabular}{cc}
\begin{minipage}{0.475\textwidth}
\centering
\includegraphics[width=85mm]{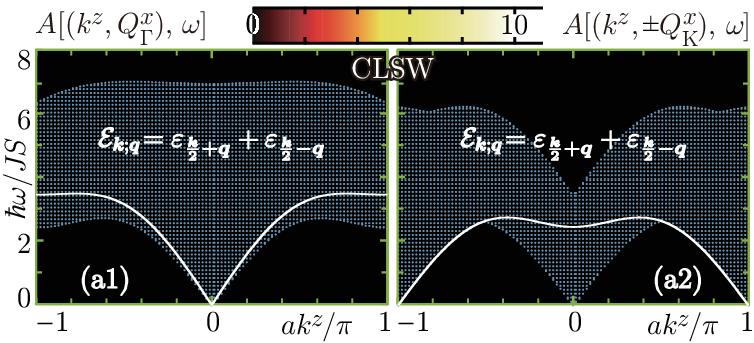}
\vspace{-8mm}
\figcaption{CLSW dispersion (\ref{E:ekCLSW}) in comparison with its double-excitation continuum
         $
            \mathcal{E}_{\bm{k};\bm{q}}
           \equiv
            \varepsilon_{\frac{\bm{k}}{2}+\bm{q}}+\varepsilon_{\frac{\bm{k}}{2}-\bm{q}}
         $
         for $S=\frac{1}{2}$ in the $N\rightarrow\infty$ limit,
         which is shown in the same format as
         Figs. 3(a1) to 3($\mathrm{b}2'$)
         so as to be easy to compare.
         The CLSW spectral function reads
         $A(\bm{k},\omega)=\delta(\hbar\omega-\varepsilon_{\bm{k}})$.}
\label{SF:DispCLSW}
\end{minipage}
\hspace{5mm}
\begin{minipage}{0.475\textwidth}
\begin{center}
\tblcaption{The shaded regions in Figs. 3($\mathrm{a}1''$) to 3($\mathrm{b}2''$),
         where every MLSW excitation is unstable with its energy $\tilde{\varepsilon}_{\bm{k}}$
         exceeding the minimum of the two-magnon continuum
         $\widetilde{\mathcal{E}}_{\bm{k};\bm{q}}
         \equiv
          \tilde{\varepsilon}_{\frac{\bm{k}}{2}+\bm{q}}
         +\tilde{\varepsilon}_{\frac{\bm{k}}{2}-\bm{q}}$
         [see Figs. 3($\mathrm{a}1$) to 3($\mathrm{b}2$)].}
\label{ST:SMD}
\begin{tabular}{ccc}
\hline\noalign{\vskip 0.5mm}\hline
 & $k^x=    Q_\Gamma^x      $\ \ \
 & $k^x=\pm Q_{\mathrm{K}}^x$\\[0.5mm]
\hline\noalign{\vskip 1.0mm}
 \ SC MLSW\ \ 
 & \ \ $\displaystyle 0.2083 <    \left|\frac{ak^z}{\pi}\right| \leq 1     $\ \ 
 & \ \ $\displaystyle             \left|\frac{ak^z}{\pi}\right| <    0.4479$\ \ \\[2.5mm]
\hline\noalign{\vskip 1.0mm}
 \ DC MLSW\ \ 
 & \ \ $\displaystyle 0.6771 <    \left|\frac{ak^z}{\pi}\right| \leq 1     $\ \ 
 & \ \ $\displaystyle             \left|\frac{ak^z}{\pi}\right| <    0.2812$\ \ \\[2.5mm]
\hline\noalign{\vskip 0.5mm}\hline
\end{tabular}
\end{center}
\end{minipage}
\end{tabular}
\end{figure}

   In sum, only when doubly constrained and perturbed, SWs succeed in revealing such complicated
dynamics $S(\bm{q},\omega)$ as a consequence of cubic dominant anharmonicities at the zone
boundaries $(k^z,\pm Q_{\mathrm{K}}^x)$ with $|k^z|\alt\frac{Q_{\mathrm{X}}^z}{2}$
[Fig. 3($\mathrm{a}2''$)].
MLSWs miss anharmonicities.
MWDISWs are still ignorant of cubic anharmonicities.
SC MPSWs leave their cubic and quartic anharmonicities to cancel out each other
[Fig. 3($\mathrm{b}2''$)].
Besides these,
SC MSWs all fail to give reasonable gaps at the zone center and corners and therefore
end up with too strong $S(\bm{q},\omega)$ intensities at the zone boundaries
[Figs. 2($\mathrm{a}1$) to 2($\mathrm{a}1''$) and
       2($\mathrm{a}2$) to 2($\mathrm{a}2''$)].
DC MSWs are generally much better than SC MSWs at describing magnon hardening effects
[Figs. 2($\mathrm{b}1$),     2($\mathrm{b}1'$),
       2($\mathrm{b}2$), and 2($\mathrm{b}2'$)].
In Table \ref{T:VariousMSWs} we summarize to what extent various MSW schemes are successful in
describing three major properties.
\begin{table*}[h]
\centering
\vspace*{0mm}
\caption{Assessment of various MSW descriptions.}
\label{T:VariousMSWs}
\begin{tabular}{cccc}
\hline\noalign{\vskip 0.5mm}\hline\noalign{\vskip -0.5mm}
  $\begin{array}{c}
    \textrm{Scheme (to which order in what manner)}
   \end{array}$
  &
  $\begin{array}{c}
    \textrm{Zone-center}      \\[-1mm]
    \textrm{magnon hardening} \\
   \end{array}$
  &
  $\begin{array}{c}
    \textrm{Zone-corner}      \\[-1mm]
    \textrm{magnon hardening} \\
   \end{array}$
  &
  $\begin{array}{c}
    \textrm{Zone-boundary}   \\[-1mm]
    \textrm{anharmonicities} \\
   \end{array}$
  \\[0.5mm]
  \hline\noalign{\vskip 0.0mm}
  SC\ \ \ \ 
  $\begin{array}{l}
    \textrm{MLSW\ \ \ \ \ \ \ \ (harmonic) } \\
    \textrm{MPSW\ \ \ \ \ \ \ \ (perturbed)} \\
    \textrm{MWDISW\ \ (variational)}         \\
   \end{array}$
  &
  $\begin{array}{c}
    \textrm{insufficient}  \\
    \textrm{insufficient}  \\
    \textrm{underestimate} \\
   \end{array}$
  &
  $\begin{array}{c}
    \textrm{insufficient} \\
    \textrm{insufficient} \\
    \textrm{overestimate} \\
   \end{array}$
  &
  $\begin{array}{c}
    \textrm{missing}      \\
    \textrm{cancel out}   \\
    \textrm{insufficient} \\
   \end{array}$
  \\[0.5mm]
  \hline\noalign{\vskip 0.0mm}
  \hspace{ 0.0mm}\!DC\ \ \ \ \hspace{-1.5mm}
  $\begin{array}{l}
    \textrm{MLSW\ \ \ \ \ \ \ \ (harmonic) } \\
    \textrm{MPSW\ \ \ \ \ \ \ \ (perturbed)} \\
   \end{array}$
  &
  $\begin{array}{c}
    \textrm{sufficient} \\
    \textrm{sufficient} \\
   \end{array}$
  &
  $\begin{array}{c}
    \textrm{sufficient} \\
    \textrm{sufficient} \\
   \end{array}$
  &
  $\begin{array}{c}
    \textrm{missing}    \\
    \textrm{sufficient} \\
   \end{array}$
  \\[0.5mm]
\hline\noalign{\vskip 0.5mm}\hline
\end{tabular}
\end{table*}
\end{widetext}